\documentclass[11pt]{article}

\usepackage{fontspec}
\usepackage[UTF8,fontset=fandol]{ctex}

\usepackage{microtype}

\usepackage{amsmath,amssymb,amsfonts,mathtools}
\usepackage{bm}
\usepackage{bbm}

\usepackage{graphicx}
\usepackage{subcaption}
\usepackage{placeins}
\usepackage{booktabs}
\usepackage{multirow}
\usepackage{array}
\usepackage{siunitx}

\usepackage{xcolor}
\usepackage{enumitem}

\usepackage{algorithm}
\usepackage{algpseudocode}

\usepackage[numbers,square,sort&compress]{natbib}

\usepackage{hyperref}
\usepackage[nameinlink,capitalise,noabbrev]{cleveref}

\graphicspath{{figures/pdf/}{figures/png/}}

\hypersetup{colorlinks=true,
  linkcolor=blue,
  citecolor=blue,
  urlcolor=blue,
  pdfauthor={Hao Zhen and Sheng Mao},
  pdftitle={Constrained Allen-Cahn phase-field topology optimization of flexoelectric devices}
}

\newcommand{\dd}{\,\mathrm{d}}

\newcommand{\bs}[1]{\boldsymbol{#1}}

\newcommand{\calH}{\mathcal{H}}

\newcommand{\eps}{\varepsilon}

\newcommand{\ufield}{\bs{u}}

\newcommand{\phifield}{\phi}

\usepackage[top=1in,bottom=1in,left=0.85in,right=0.85in]{geometry}
\usepackage{authblk}
\usepackage{titlesec}

\newlength{\titleTopSkip}
\newlength{\titleAbstractSkip}
\makeatletter
\renewcommand{\@maketitle}{%
  \newpage\null
  \begin{center}
    \vspace*{\titleTopSkip}
    \let\footnote\thanks
    {\LARGE\@title\par}
    \vskip 1.5em
    {\large\lineskip .5em\@author\par}
  \end{center}
}
\makeatother

\renewenvironment{abstract}{%
  \par\addvspace{\titleAbstractSkip}
  \hrule height 0.4pt
  \vspace{0.8em}
  \noindent\textbf{Abstract}\par\medskip
}{\par}
\providecommand{\keywords}[1]{%
  \par\addvspace{0.5em}
  \begingroup
  \def\and{\unskip,\space}%
  \noindent\textit{Keywords: }#1\par
  \endgroup
  \vspace{0.8em}
  \hrule height 0.4pt
  \vspace{0.5em}
}

\title{Constrained Allen-Cahn phase-field topology optimization of flexoelectric devices}

\author[a]{Hao Zhen}
\author[a,*]{Sheng Mao}
\affil[a]{School of Mechanics and Engineering Science, Peking University, Beijing 100871, P. R. China}

\date{}

\begin{document}

\maketitle
\begingroup
\renewcommand{\thefootnote}{\fnsymbol{footnote}}
\footnotetext[1]{Corresponding author.\newline
  \textit{Email address: }\href{mailto:maosheng@pku.edu.cn}{\texttt{maosheng@pku.edu.cn}} (Sheng Mao)}
\endgroup

\begin{abstract}

Topology optimization provides an effective approach to exploring performance-oriented designs for miniature devices.
In this work, a phase-field framework for constrained topology optimization of microscale flexoelectric devices is developed using a continuum model under the assumption of infinitesimal deformations.
The material distribution is optimized at the device level with the output voltage as the objective. The framework also accommodates objectives based on capacitance or stored electrostatic energy.
Manufacturing variations are incorporated through a worst-case objective over the prescribed geometric realizations, with spectral constraints imposed to maintain the structural connectivity.
The framework is applied to two-dimensional pressure and bending sensors and a three-dimensional thin-film device.
The optimized nominal output voltages reach approximately 4.8 times the initial value for the pressure sensor and 2.5-3.4 times the initial values for the bending sensors.
The voltage gains arise primarily from increased mechanically induced source charge, with capacitance reduction providing an additional contribution that varies with the device configuration.
Erosion analyses of representative designs further show that structural connectivity is well preserved when structural members become locally narrower, with the output voltage remaining close to that of the nominal configuration.
Combined with specific microfabrication processes, this optimization framework is expected to provide a numerical basis for configuration screening and prototype development of flexoelectric devices.

\end{abstract}


\section{Introduction}
\label{sec:introduction}

Flexoelectricity is a bidirectional electromechanical effect involving spatial gradients.
It is governed by coupling between strain gradients and polarization~\citep{nguyenNanoscaleFlexoelectricity2013,wangFlexoelectricitySolidsProgress2019,zhuangComputationalModelingFlexoelectricity2020}, with additional contributions from surface-related effects in finite dielectric bodies~\citep{mohammadiFlexoelectricityCausesSurface2025a,dingleContinuumComputationalModeling2025}.
Recent studies have shown the potential of this mechanism for applications in micro- and nanoelectronic devices, including energy harvesters based on polymer composite foams~\citep{yanFlexoelectricityenabledUltrahighPiezoelectric2023} and porous materials~\citep{jiangGiantFlexoelectricEffect2023}, flexible hydrogel sensors for object recognition~\citep{jiaGiantIontronicFlexoelectricity2024}, and silicon-based transistors in which electrical transport is mechanically modulated~\citep{guoSiliconFlexoelectronicTransistors2023}.
Geometric design offers a systematic route to controlling the flexoelectric response, since strain gradients depend on local deformation modes~\citep{abdollahiComputationalEvaluationFlexoelectric2014}.
Changes in film morphology, pore geometry, and microstructural rearrangement can modify the location and direction of localized bending, thereby controlling the spatial distribution of the flexoelectric response~\citep{liangAdvancementsFlexoelectricMaterials2024,jiaFlexoelectricEffectThin2025,wangMicrostructureDesignedFlexoelectric2025}.
For example, changes in wrinkle morphology can be used to control polarization patterns in thin films~\citep{shangMechanicalControlPolar2024}.
Geometrically polarized dielectric microstructures can exhibit a macroscopic apparent piezoelectric response as a result of local bending, enabling electromechanical transduction using non-piezoelectric materials~\citep{mocciGeometricallyPolarizedArchitected2021,barcelo-mercaderComputationalHomogenizationHigherorder2024}.
Applying these mechanisms of geometric control to device design requires predicting the specific response of a given configuration under prescribed loading and electrical conditions~\citep{codonyImmersedBoundaryHierarchical2019,barcelo-mercaderWeakEnforcementInterface2022}.

Topology optimization provides an effective approach to the structural design of flexoelectric devices~\citep{nanthakumarTopologyOptimizationFlexoelectric2017,hamdiaNovelDeepLearning2019}.
These methods typically optimize device performance under prescribed constraints to obtain globally or locally optimal material distributions, reducing reliance on predefined configurations and experimentation.
Major approaches include density-based methods~\citep{liIsogeometricTopologyOptimization2026}, level-set methods~\citep{ghasemiLevelsetBasedIGA2017,ghasemiComputationalFrameworkDesign2020,ghasemiMultimaterialLevelSetbased2018}, and the moving morphable void method~\citep{zhangFlexoelectricNanostructureDesign2022}.
Recent studies have optimized unit-cell geometry and its spatial variation in lattice structures to control the flexoelectric response~\citep{zhangExplicitTopologyOptimization2023}.
Multi-material designs improve energy conversion performance by optimizing the distribution of elastic and flexoelectric materials~\citep{zhangExplicitTopologyOptimization2025}.
Optimization of soft dielectrics and graded lattice energy harvesters under finite deformations has accounted for the effects of geometric nonlinearity on the electromechanical response~\citep{chenNonlinearTopologyOptimization2024,zhangExplicitTopologyOptimization2025a}.
For energy harvesters under dynamic excitation, material distributions have been optimized for energy conversion performance across a prescribed frequency band~\citep{chenMultimaterialDynamicTopology2026}.
The choice of optimization objectives is also guided by the intended function of the device. Greco et al.~\citep{Greco2024ApparentPiezoelectricity} obtained different optimal microstructures for sensing and actuation in flexoelectric metamaterials, indicating that structural design should match the intended function.
The phase-field method, which originated in phase separation dynamics~\citep{cahnFreeEnergyNonuniform1958,Cahn1961SpinodalDecomposition}, is also widely used for topology optimization.
In this approach, interfacial regularization is incorporated into the optimization functional, and the structure evolves through an energy-driven process.
Topology optimization can be expressed as the constrained minimization of a generalized energy functional, with material redistribution interpreted as an energy dissipation process.
The phase-field method has been applied to the optimization of stiffness and vibration characteristics~\citep{Takezawa2010PhaseFieldTopology}, multi-material design~\citep{zhouMultimaterialStructuralTopology2007,tavakoliMultimaterialTopologyOptimization2014a}, and the design of piezoelectric energy harvesters~\citep{takezawaDesignMethodologyPiezoelectric2014}.
The phase-field method has also been used to model ferroelectric domain evolution accounting for flexoelectricity~\citep{wangStrainGradientFinite2024} and its coupling with crack propagation~\citep{liuPhasefieldAnalysisBrittle2024}.
Ashour et al.~\citep{ashourIsogeometricAnalysisPhasefield2021a} incorporated bending, flexoelectric, and dielectric energies into a constrained phase-field model to predict the morphological evolution of vesicles under applied electric fields.
López et al.~\citep{lopezIsogeometricPhaseField2022} performed shape and topology optimization of two- and three-dimensional flexoelectric structures, with phase-field evolution driven by objective function sensitivities. They also compared electromechanical conversion performance under different loading conditions and material configurations.

The electromechanical performance of flexoelectric devices is usually sensitive to parameter uncertainties~\citep{hamdiaSensitivityUncertaintyAnalysis2018}.
Manufacturing variations alter the local dimensions of the optimized structure, which changes the local deformation and hence the electromechanical response of the device.
To evaluate the performance at the design stage, robust topology optimization formulations based on eroded, nominal, and dilated designs have been developed~\citep{Wang2011RobustProjection}.
Relationships between filter and projection parameters and the minimum feature sizes of solid and void regions have also been established to control the sizes of features during optimization~\citep{trilletAnalyticalRelationshipsImposing2021}.
For cantilevered piezoelectric energy harvesters, Miyajima and Yamada~\citep{Miyajima2024PiezoelectricHarvester} imposed manufacturing constraints on the optimization of substrate and piezoelectric film based on the etching process in microfabrication.
In addition to dimensional constraints, connectivity between material regions and supports or electrodes must be considered for proper device operation~\citep{mengIsogeometricAnalysisIGABased2024}.
For example, isolated electrodes increase the wiring complexity in piezoelectric transducers and electroactive polymer devices.
The virtual temperature method has been used to formulate the connectivity constraints of electrodes~\citep{Donoso2019ElectrodeConnectivity,hardConnectivityConstraintsEnsuring2024}.
In three-dimensional fluid device optimization, connectivity to the walls has been enforced by constraining floating solid regions~\citep{sasakiTopologyOptimization3D2024}.
As an alternative approach, connectivity can be characterized by the eigenvalues of auxiliary operators in spectral methods~\citep{Donoso2022SpectralGraph}.
They have been applied to exclude enclosed holes and constrain the connectivity in the structural design~\citep{Donoso2023ContinuousModel}.
These spectral approaches have been further extended to connectivity formulations with specific boundary conditions~\citep{Donoso2024DirichletLaplacian} and to level-set topology optimization~\citep{andradeLevelSetTopological2026}.
The applicability of these connectivity methods and their influence on optimized designs have been systematically compared~\citep{Cool2025ConnectivityReview}.
In the optimization of flexoelectric devices, narrow connections can also be disrupted by manufacturing variations, resulting in changes to the structural load paths. Therefore, performance evaluation under geometric perturbations and the preservation of structural connectivity need to be considered in device design.

\begin{figure}[!tbp]
  \centering
  \includegraphics[width=\linewidth]{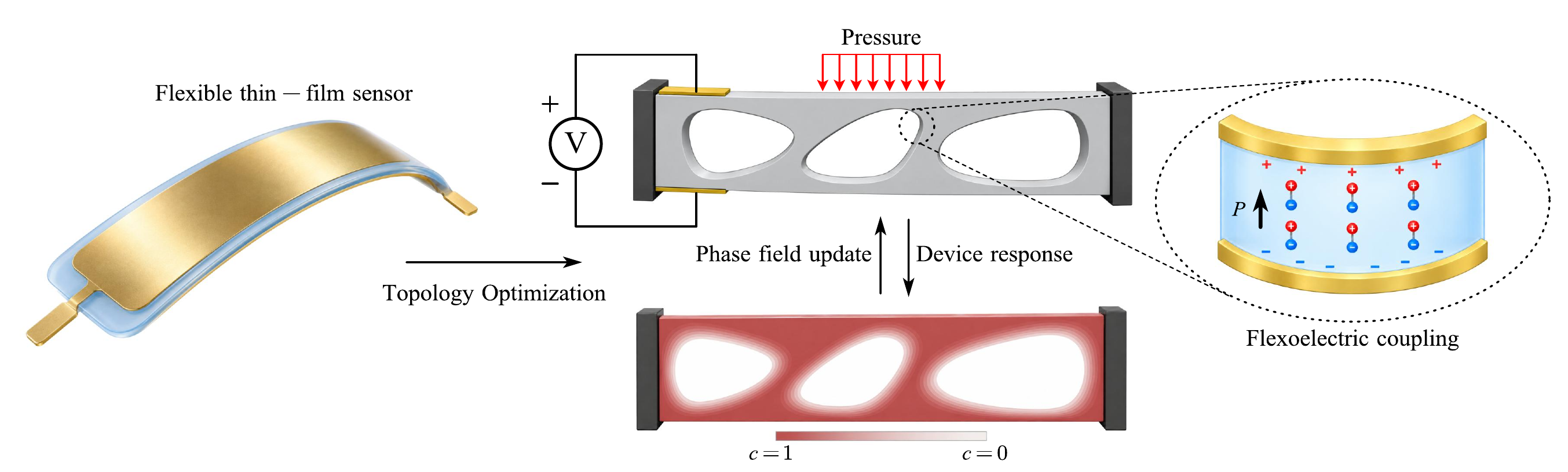}
  \caption{Schematic overview of flexoelectric sensor design and phase-field topology optimization.}
  \label{fig:sensor-design}
\end{figure}

\begingroup
\looseness=-1
With these considerations in mind, an extensible phase-field framework for constrained optimization of microscale flexoelectric devices is developed in this work.
Fig.~\ref{fig:sensor-design} provides a schematic overview of flexoelectric sensor design and the phase-field topology optimization framework.
The material distribution is optimized directly at the device level for a specified electrical performance objective.
Different performance metrics and manufacturing constraints can be incorporated into the same framework.
A robust design method is further developed in which performance and structural connectivity are maintained under manufacturing variations that can reduce electrical output and disrupt local connections.
Slender members are favored during optimization, as their local flexibility can enhance the flexoelectric response through nonuniform deformation.
Local response enhancement and connectivity preservation are balanced during optimization to maintain electrical output under prescribed geometric perturbations.
This approach can be extended to optimize electrical output under constraints imposed by specific microfabrication processes.
\par
\endgroup

The remainder of this paper is organized as follows.
In Section~\ref{sec:2-flexo-problem}, the electromechanical model of flexoelectric devices and its variational formulation are established.
Voltage readout relations for multiple electrodes are presented, and a device performance functional is defined in terms of output voltage.
In Section~\ref{sec:3-pf-formulation}, the general phase-field optimization model and its formulation for manufacturing robustness and structural connectivity are presented.
The solution method is presented in Section~\ref{sec:4-numerical}, including the optimality conditions, sensitivity analysis of the objective function and constraints, and a fully implicit scheme for phase-field evolution.
In Section~\ref{sec:5-numerical-studies}, two-dimensional pressure and bending sensors and a three-dimensional thin-film device are investigated numerically.
The evolution of the material distribution and the increase in output voltage are examined to verify the effectiveness of the proposed algorithm.
The conclusions are summarized in Section~\ref{sec:conclusions}.

\section{Modeling of flexoelectric devices}\label{sec:2-flexo-problem}

In this section, the state problem is formulated to determine the displacement field, electric potential, and electrical response at the electrode ports for a prescribed material distribution.
In Section~\ref{subsec:flexo-equation}, a linear flexoelectric continuum model with higher-order electromechanical coupling is formulated. The interpolation of material parameters is defined in terms of the material distribution.
In Section~\ref{subsec:flexo-variational-form}, the weak form of the governing equations for displacement and electric potential is derived from the first variation of the total free enthalpy functional.
In Section~\ref{subsec:flexo-port}, port voltages are determined using floating electrodes and a readout circuit model. Device performance measures are then defined in terms of these voltages.

\subsection{Flexoelectric model}\label{subsec:flexo-equation}

Consider a flexoelectric device occupying a domain $\Omega\subset\mathbb{R}^{d}$ ($d=2,3$).
Assume infinitesimal deformations and quasistatic electrical conditions. The electromechanical state is described by the displacement field $\ufield:\Omega\rightarrow\mathbb{R}^{d}$ and the electric potential field $\phifield:\Omega\rightarrow\mathbb{R}$.
The electrical Gibbs free energy density, also referred to as the free enthalpy density, is expressed as~\citep{Codony2021Flexoelectricity}
\begin{equation}
\begin{aligned}
  \calH
  ={}&
  \frac{1}{2}\mathbb C_{ijkl}\eps_{ij}\eps_{kl}
  +\frac{1}{2}\mathbb B_{ijklmn}\eps_{ij,k}\eps_{lm,n}
  -\frac{1}{2}\kappa_{ij}E_iE_j
  -\frac{1}{2}\mathbb M_{ijkl}E_{i,j}E_{k,l}
  +\calH_{\mathrm f},
\end{aligned}
\label{eqn:flexo-general-0}
\end{equation}
where $\mathbb C_{ijkl}$ is the elasticity tensor, $\mathbb B_{ijklmn}$ is the strain-gradient elasticity tensor, $\kappa_{ij}$ is the dielectric permittivity tensor, and $\mathbb M_{ijkl}$ is the gradient dielectricity tensor.
The strain tensor is defined as $\eps_{ij}(\ufield)=(u_{i,j}+u_{j,i})/2$, where a comma in the subscript $(,)$ denotes partial differentiation with respect to the spatial coordinate.
The electric field is given by $E_i(\phifield)=-\phi_{,i}$.
The strain gradient and electric field gradient are denoted by $\varepsilon_{ij,k}(\boldsymbol{u})$ and $E_{i,j}(\phi)$, respectively.
The flexoelectric coupling term $\calH_{\mathrm f}$ is written in the Lifshitz form~\citep{Codony2021Flexoelectricity}
\begin{equation}
  \calH_{\mathrm f}
  =
  -\frac{1}{2}\mathbb{F}_{ijkl}E_i\eps_{jk,l}
  +\frac{1}{2}\mathbb{F}_{ijkl}E_{i,l}\eps_{jk},
\label{eqn:flexo-general-1}
\end{equation}
where $\mathbb{F}_{ijkl}$ is the flexoelectric tensor.
Direct and converse flexoelectric effects are consistently described by the same free enthalpy density.
For simplicity, the isotropic form of the free enthalpy density is adopted
\begin{equation}
\begin{aligned}
\mathcal{H}
&=\frac{1}{2}\Lambda\varepsilon_{jj}\varepsilon_{kk}
+\mu\varepsilon_{jk}\varepsilon_{jk}
+\frac{1}{2}l_m^2(\Lambda\varepsilon_{jj,i}\varepsilon_{kk,i}
+2\mu\varepsilon_{jk,i}\varepsilon_{jk,i})-\frac{1}{2}\kappa E_iE_i-\frac{1}{2}\kappa l_e^2E_{i,j}E_{i,j}\\
&-\frac{1}{2}(f_1E_i\varepsilon_{jj,i}+2f_2\varepsilon_{ij,i}E_j)
+\frac{1}{2}(f_1E_{i,i}\varepsilon_{jj}+2f_2E_{i,j}\varepsilon_{ij}).
\end{aligned}
\label{eqn:flexo-isotropic}
\end{equation}
Here, $\Lambda$ and $\mu$ are the Lamé parameters, and $\kappa$ is the dielectric permittivity.
The flexoelectric coefficients $f_1$ and $f_2$ are obtained from the isotropic reduction of $\mathbb{F}$ in Eq.~\eqref{eqn:flexo-general-1}.
The characteristic lengths $l_m$ and $l_e$ correspond to the strain gradient and electric field gradient terms, respectively.
The contributions of the mechanical and electrical gradients are controlled by the corresponding intrinsic length scales. For geometrically similar devices under comparable loading conditions, higher-order effects become less pronounced as the device size increases relative to the intrinsic length scales.

The material distribution within the device is described by the design variable $c(\boldsymbol{x})$, with $c=1$ denoting solid material and $c=0$ denoting void.
Void regions are represented by a weak ersatz material to avoid ill-conditioning of the stiffness matrix as material parameters approach zero.
The design domain is denoted by $\Omega_d$. The non-design domain $\Omega_f=\Omega\setminus\Omega_d$ is assumed to include the supports, electrodes, and load application regions.
In this domain, $c$ is fixed at 1, so the material parameters remain unchanged during optimization.
The material parameters are interpolated in terms of $c$ using the following power-law form~\citep{Andreassen2011EfficientTopology}
\begin{equation}
  A(c) = A_- + \left(A_0-A_-\right)c^{p_A}, \qquad
  A\in \{ Y,f,\kappa \},
\label{eqn:mat_interp}
\end{equation}
where $Y$ and $f$ represent the Young's modulus and the flexoelectric coefficients, respectively.
$A_0$ and $A_-$ are the values of $A$ in the solid phase and the weak ersatz material, respectively. $p_A$ is the interpolation exponent and can be chosen independently for each material parameter.
For brevity, the dependence of the material coefficients on $c$ is omitted from the notation hereafter.

\subsection{Variational formulation}\label{subsec:flexo-variational-form}

The weak form of the governing equations is derived from the first variation of the total free enthalpy functional.
Let $\Gamma_u,\Gamma_\phi\subset\partial\Omega$ denote the Dirichlet boundaries for displacement and electric potential, respectively.
The normal derivatives are prescribed on $\Gamma_{\xi}$ and $\Gamma_{\zeta}$, respectively.
Since the free enthalpy density in Eq.~\eqref{eqn:flexo-general-0} depends on the gradients of strain and electric field, the displacement and electric potential are taken in $[H^2(\Omega)]^d$ and $H^2(\Omega)$, respectively. This regularity ensures that the volume integrals in the weak form are well defined for bounded material coefficients.
The trial spaces for displacement and electric potential are defined as
\begin{equation}
  \begin{aligned}
    \mathcal U &= \left\{\boldsymbol u\in[H^2(\Omega)]^d:\ 
    \boldsymbol u=\bar{\boldsymbol u}\ \mathrm{on}\ \Gamma_u,\ \partial_n\boldsymbol u=\bar{\boldsymbol \xi}\ \mathrm{on}\ \Gamma_\xi
    \right\},\\
    \mathcal V &= \left\{\phi\in H^2(\Omega):\ 
    \phi=\bar{\phi}\ \mathrm{on}\ \Gamma_\phi,\
    \partial_n\phi = \bar{\zeta}\ \mathrm{on}\ \Gamma_\zeta
    \right\}.
  \end{aligned}
\label{eqn:trial-space}
\end{equation}
The corresponding test spaces $\mathcal U_0$ and $\mathcal V_0$ consist of functions satisfying the homogeneous counterparts of the essential boundary conditions defining $\mathcal U$ and $\mathcal V$, respectively.

The total free enthalpy functional is defined as
\begin{equation}
  \Pi(\ufield,\phifield;c) =
  \int_\Omega \calH(\ufield,\phifield;c) \dd\Omega -\ell_u(\ufield) +\ell_\phi(\phifield),
\label{eqn:flexo-total-energy}
\end{equation}
in which $\ell_u$ and $\ell_\phi$ are the linear functionals associated with the external mechanical and electrical loads, respectively.
Let $y=(\boldsymbol{u},\phi)$ denote the state variables.
Setting the first variation of $\Pi$ to zero for all admissible variations $(\delta\boldsymbol u,\delta\phi)\in\mathcal U_0\times\mathcal V_0$ yields the weak formulation. Find $(\boldsymbol u,\phi)\in\mathcal U\times\mathcal V$ such that
\begin{equation}
\begin{aligned}
\mathcal S_u(y,c)[\delta\boldsymbol u] &= a(\ufield,\delta\boldsymbol u;c) + b(\phi,\delta\boldsymbol u;c) - \ell_u(\delta\boldsymbol u) = 0,
  \\
\mathcal S_\phi(y,c)[\delta\phi] &= b(\delta\phi,\ufield;c) - d(\phi,\delta\phi;c) + \ell_\phi(\delta\phi) = 0.
\end{aligned}
\label{eqn:flexo-weak-system}
\end{equation}
For a prescribed design variable $c$, the bilinear forms $a$, $b$, and $d$ are given by
\begin{equation}
\begin{aligned}
a(\boldsymbol u,\delta\boldsymbol u;c) &= \int_\Omega {\bigl[\Lambda\varepsilon_{jj}(\boldsymbol u) \varepsilon_{kk}(\delta\boldsymbol u) + 2\mu\varepsilon_{jk}(\boldsymbol u) \varepsilon_{jk}(\delta\boldsymbol u)\bigr]\mathrm d\Omega }\\ &+
\int_\Omega {l_m^2 \bigl[\Lambda\varepsilon_{jj,i}(\boldsymbol u) \varepsilon_{kk,i}(\delta\boldsymbol u) + 2\mu\varepsilon_{jk,i}(\boldsymbol u) \varepsilon_{jk,i}(\delta\boldsymbol u)\bigr]\mathrm d\Omega },
\\
b(\phi,\delta\boldsymbol u;c) &= \frac12 \int_\Omega {\bigl[f_1 E_{i,i}(\phi)\varepsilon_{jj}(\delta\boldsymbol u) + 2f_2 E_{i,j}(\phi)\varepsilon_{ij}(\delta\boldsymbol u)\bigr] \mathrm d\Omega}\\ &-
\frac12 \int_\Omega {\bigl[ f_1 E_i(\phi)\varepsilon_{jj,i}(\delta\boldsymbol u) + 2f_2 E_j(\phi)\varepsilon_{ij,i}(\delta\boldsymbol u) \bigr]\mathrm d\Omega},
\\
d(\phi,\delta\phi;c) &= \int_\Omega {\kappa E_i(\phi)E_i(\delta\phi)\dd\Omega } +
\int_\Omega {\kappa l_e^2 E_{i,j}(\phi)E_{i,j}(\delta\phi)\dd\Omega}.
\end{aligned}
\label{eqn:flexo-weak-bilinear}
\end{equation}
Equations~\eqref{eqn:flexo-weak-system} and \eqref{eqn:flexo-weak-bilinear} define the weak form of the state problem for displacement and electric potential.

With the test functions collected in $\delta y=(\delta\boldsymbol{u},\delta\phi)$, Eq.~\eqref{eqn:flexo-weak-system} is written as
\begin{equation}
  \mathcal S(y,c)[\delta y] \coloneqq \mathcal S_u(y,c)[\delta \boldsymbol{u}] + \mathcal S_\phi (y,c)[\delta\phi] = 0, \quad \forall \delta y\in \mathcal U_0\times\mathcal V_0.
\label{eqn:flexo-weak-system-unified}
\end{equation}
For fixed $c$, the state problem is linear in $y$.
As a fourth-order boundary value problem, it cannot be fully specified by prescribing the boundary values of $\boldsymbol u$ and $\phi$ alone.
The weak formulation also contains higher-order boundary terms conjugate to their normal derivatives.
The full treatment of these terms is provided in Appendix~\ref{app:constitutive-boundary}.

\subsection{Electrode model}\label{subsec:flexo-port}

Measurable port quantities are determined by the electrode layout and readout circuit.
Each electrode is equipotential, with its potential determined by the charge balance and circuit conditions.
The surface charge density induced by flexoelectricity can vary in sign over the device surface.
When regions of opposite sign are covered by the same electrode, their contributions to the net electrode charge partially or completely cancel. Electrically isolated electrode segments allow these charge contributions to be collected separately. Here, voltage is considered as the readout quantity and is defined by a prescribed linear combination of the electrode potentials.
The readout voltage is formulated below for a prescribed material distribution, mechanical loading, and electrode layout.

Consider a flexoelectric device with $N_p$ floating electrodes.
Each electrode is equipotential
\begin{equation}
  \phi = V_k, \quad \mathrm{on} \ \Gamma_{\phi,k},
\label{eqn:floating-eletrode-potential}
\end{equation}
where $V_k$ is the unknown potential of electrode $k$.
The electric potential is fixed at $\phi=0$ on the reference electrode $\Gamma_{\phi,0}$.
The equipotential condition in Eq.~\eqref{eqn:floating-eletrode-potential} is a special form of the Dirichlet boundary condition, with the floating potentials $\{V_k\}_{k=1}^{N_p}$ included among the unknown state variables~\citep{amannSimulationFloatingPotentials2014}.
This condition can be imposed as linear multipoint constraints or by the Lagrange multiplier method~\citep{dwarshuisDerivationSuperelementDeformable2023}.

Under open-circuit conditions, the net free charge on each electrode is conserved, giving $Q_k=0$ for initially uncharged electrodes.
For a prescribed material distribution $c$, the state problem in Eq.~\eqref{eqn:flexo-weak-system} is linear in the state variables.
By superposition, the electrode charges can be decomposed as
\begin{equation}
  \boldsymbol{Q} = \boldsymbol{Q}_{\mathrm{src}} + \boldsymbol{C}_{\mathrm{p}}\boldsymbol{V},
\label{eqn:port-0}
\end{equation}
where $\boldsymbol{V},\boldsymbol{Q}\in\mathbb R^{N_p}$ are the potential and charge vectors of the floating electrodes, respectively.
The source charge vector $\boldsymbol{Q}_{\mathrm{src}}\in\mathbb R^{N_p}$ denotes the electrode charges induced by mechanical loading with all electrode potentials held at zero.
The matrix $\boldsymbol{C}_{\mathrm{p}}\in\mathbb R^{N_p\times N_p}$ is the invertible effective capacitance matrix.
The open-circuit electrode potentials follow from $\boldsymbol{Q}=0$ as
\begin{equation}
  \boldsymbol{V}_{\mathrm{oc}} = -\boldsymbol{C}_{\mathrm{p}}^{-1}\boldsymbol{Q}_{\mathrm{src}}.
\label{eqn:port-potential}
\end{equation}

The voltage measured between electrodes $a$ and $b$ is defined as the port voltage.
With $\boldsymbol{s}_V=\boldsymbol{e}_a-\boldsymbol{e}_b$, the port voltage is written as $v_{\mathrm{o}}=\boldsymbol{s}_V^{\top}\boldsymbol{V}$, where $\boldsymbol{e}_a, \boldsymbol{e}_b\in \mathbb R^{N_p}$ are the corresponding standard basis vectors.
The readout circuit is modeled by an input capacitance $C_{\mathrm{in}}$ connected across the port. The charge transferred to the readout circuit is denoted by $q_L$, with the corresponding electrode charge vector given by $\boldsymbol{Q}_L=\boldsymbol{s}_Vq_L$.
The charge-voltage relations are
\begin{equation}
  q_L = C_{\mathrm{in}} v_{\mathrm{o}}, \quad \boldsymbol{Q}_L = C_{\mathrm{in}} \boldsymbol{s}_V\boldsymbol{s}_V^{\top} \boldsymbol{V}.
\end{equation}
Combining these relations with the electrode charge balance $\boldsymbol{Q}+\boldsymbol{Q}_L=\boldsymbol{0}$ gives the port voltage
\begin{equation}
  v_{\mathrm{o}} = - \boldsymbol{s}_V^{\top} (\boldsymbol{C}_{\mathrm{p}}+C_{\mathrm{in}} \boldsymbol{s}_V\boldsymbol{s}_V^{\top})^{-1}
  \boldsymbol{Q}_{\mathrm{src}}.
\label{eqn:electrode-balance}
\end{equation}
The open-circuit result is recovered as $C_{\mathrm{in}}\rightarrow 0$, with $v_{\mathrm{o}}\rightarrow\boldsymbol{s}^{\top}_V\boldsymbol{V}_{\mathrm{oc}}$.

The objective functional is usually chosen according to the intended device function.
For the given external loads and electrode readout configuration, the state is uniquely determined by the material distribution and is denoted by $y=y(c)$.
A scalar objective functional $J(y,c)$ can therefore be expressed as the reduced functional $j(c) = J(y(c),c)$ of the material distribution.
For sensor design, the objective functional is defined as the negative normalized port voltage
\begin{equation}
  j(c) = J(y(c),c) = - \frac{v_{\mathrm{o}}(c)}{V_{\mathrm{ref}}},
\label{eqn:reduced-obj-function}
\end{equation}
where $V_{\mathrm{ref}}>0$ is the reference voltage used for normalization.

\section{Constrained phase-field design formulation}
\label{sec:3-pf-formulation}

In this section, the material distribution is represented by a continuous phase field $c$, which serves as the design variable.
A unified optimization formulation is developed incorporating manufacturing robustness constraints and spectral connectivity constraints.
The formation, merging, and disappearance of structural components are described by the single continuous field $c$ without additional mechanisms.
The phase-field regularization is included in the total objective functional with a weight $\gamma$, which controls the balance between regularization and the device performance objective.
The phase field and the higher-order electromechanical state problem can be discretized using the same approach.
Within this formulation, the volume constraint, manufacturing robustness constraints, and spectral connectivity constraints are expressed in terms of the phase field and its derived fields.
In Section~\ref{sec:allen-cahn-model}, the phase-field evolution equation for the design variable is defined.
In Section~\ref{sec:manufacturing}, manufacturing variations are represented by eroded, nominal, and dilated design realizations. The worst-case objective value among these realizations is used as the optimization objective.
In Section~\ref{sec:spectral-conn}, constraints on the first eigenvalue are introduced to promote connectivity to the anchoring regions. After that, the complete continuous constrained optimization problem is formulated.

\subsection{Allen-Cahn phase-field model}\label{sec:allen-cahn-model}

The phase field $c$ is restricted to $[0,1]$ and fixed in the non-design region $\Omega_f$, giving the admissible set
\begin{equation}
  \mathcal K = \left\{ c\in H^1(\Omega)\cap L^\infty(\Omega):\ 0\le c\le1,\ c|_{\Omega_f}=1 \right\}.
  \label{eqn:design-space}
\end{equation}
The pointwise bounds on $c$ alone cannot control the spatial gradients of the material distribution.
Without spatial regularization, the optimization can favor increasingly fine structural features associated with enhanced strain gradients, leading to mesh-dependent designs~\citep{Miyajima2024PiezoelectricHarvester}.
A Ginzburg-Landau regularization functional is therefore introduced
\begin{equation}
  \mathcal E(c)=\int_{\Omega_d}\left[\Psi_B(c)+\frac{l_c^2}{2}|\nabla c|^2\right]\mathrm d\Omega,
\label{eqn:phase-field-energy}
\end{equation}
where the first term is a double-well potential density that favors the phase-field values of $0$ and $1$. The second term provides interfacial regularization, with the characteristic length $l_c$ controlling the diffuse interface thickness.
The quartic double-well potential is chosen as~\citep{Takezawa2010PhaseFieldTopology}
\begin{equation}
  \Psi_B(c) = c^2 (1-c)^2.
\end{equation}
The total objective functional combines the device objective with phase-field regularization,
\begin{equation}
  \mathcal F(c)=j(c)+\gamma\mathcal E(c),
\end{equation}
where $j(c)$ is the reduced objective functional defined in Eq.~\eqref{eqn:reduced-obj-function}, and the weight $\gamma$ controls the relative contributions of phase-field regularization and device performance.

The driving force for phase-field evolution is given by the variational derivative of the total functional $\mathcal F$
\begin{equation}
  z \coloneqq \frac{\delta\mathcal F}{\delta c} = 
  \frac{\delta j}{\delta c} + \gamma(\Psi'_B(c) - l_c^2 \nabla^2 c),
\end{equation}
where $\delta j/\delta c$ is the sensitivity of the objective functional, evaluated as described in Section~\ref{sec:optimal_system}.
The nonconserved evolution follows the Allen-Cahn gradient-flow form~\citep{allencahndynamic1979}
\begin{equation}
  \partial_\tau c = -M_0 z,
\label{eqn:standard-ac}
\end{equation}
where $\tau$ is pseudo-time and $M_0>0$ is a constant mobility.
This equation defines the $L^2$ gradient flow of the total functional $\mathcal F$. In \eqref{eqn:standard-ac}, conservation of the total material volume in the design domain is not enforced. Hence, a mass-conserving Allen-Cahn equation is adopted here~\citep{nonlocalreactiondiffusion1992}.
The driving force is projected onto the zero-mean subspace of $L^2(\Omega_d)$ by subtracting its spatial mean. The resulting equation is constructed as follows
\begin{equation}
  \partial_\tau c = -M_0 \mathcal P z,
\label{eqn:projected-ac}
\end{equation}
where $\mathcal P$ is the orthogonal projection onto the zero-mean subspace, defined by
\begin{equation}
  \mathcal P z = z- |\Omega_d|^{-1} \int_{\Omega_d}z\,\mathrm d\Omega.
\end{equation}
Hence, the prescribed material volume is specified by the following constraint
\begin{equation}
  g_{\mathrm{v}}(c) \coloneqq \int_{\Omega_d} {c \mathrm{d}\Omega} - m_c |\Omega_d| = 0,
\end{equation}
where $m_c$ is the prescribed target volume fraction. This constraint is preserved by the projected evolution, provided that it is satisfied by the initial design.
Although the pure phase values $c=0$ and $c=1$ are favored by the double-well potential density $\Psi_B(c)$, the bounds $0\le c\le1$ are not automatically guaranteed by Eq.~\eqref{eqn:projected-ac}.
The optimality conditions for the bound constraints are expressed using Lagrange multipliers and enforced numerically by an active-set method.
By the self-adjointness and idempotency of $\mathcal{P}$, the following energy-dissipation law is obtained
\begin{equation}
  \frac{\mathrm d}{\mathrm d \tau} \mathcal F(c) 
  = -M_0\|\mathcal{P} z\|_{L^2(\Omega_d)}^2 \le0.
\end{equation}

\subsection{Manufacturing robustness}\label{sec:manufacturing}

Topology optimization typically yields a nominal design that does not account for manufacturing variations.
Actual fabrication inevitably introduces geometric variations. Manufacturing process conditions cause the solid boundary to shift from its nominal position, producing over-etched (eroded) or under-etched (dilated) configurations.
Erosion reduces the width of solid features, whereas dilation reduces the size of void regions. For sufficiently large geometric variations, narrow connections are eliminated and small void regions are completely closed.
In microscale devices, such geometric changes can lead to stress concentrations or electric field concentrations, or disrupt mechanical and electrical paths.
For a design to be manufacturing robust, device performance must remain acceptable across the prescribed range of geometric variations, not only in the nominal configuration.
Minimum feature sizes of solid and void regions can be imposed through explicit geometric constraints~\citep{minimumlengthscale2015}. This approach requires additional geometric operators to measure the local widths and does not readily account for overall boundary shifts caused by process conditions.

A three-field model based on erosion and dilation is adopted here~\citep{Wang2011RobustProjection}.
The three fields can be constructed by applying projection directly to the phase field $c$. However, the resulting boundary shifts are then limited by the phase-field characteristic length $l_c$.
An independent filter parameter $r_f$ is therefore introduced to control the spatial frequency response before projection.
For a fixed $l_c$, $r_f$ and the projection parameters introduced below can be chosen to produce geometric variations of a prescribed magnitude.
The filtered field is first obtained by solving the Helmholtz filter equation~\citep{Lazarov2011HelmholtzFilter}
\begin{equation} 
  \begin{cases}
  -r_f^2\nabla^2 \widetilde c+\widetilde c=c, \qquad& \boldsymbol{x}\in\Omega_d, \\
  \nabla\widetilde c\cdot \boldsymbol{n}=0, \qquad& \boldsymbol{x}\in \partial\Omega_d,
  \end{cases}
\label{eqn:helmholtz_filter}
\end{equation}
where $r_f$ is the filter radius.
The filter attenuates spatial variations more strongly at shorter wavelengths.
The nominal field is defined as the filtered field $\widetilde c$ without further projection.
The eroded and dilated fields are obtained by a smooth Heaviside projection~\citep{Wang2011RobustProjection}
\begin{equation}
  c_{\mathrm i}=\widetilde c, \quad
  c_{\mathrm e}=h_{\beta,\eta_{\mathrm e}}(\widetilde c), \quad 
  c_{\mathrm d}=h_{\beta,\eta_{\mathrm d}}(\widetilde c).
\end{equation}
Here, the subscripts $\mathrm i$, $\mathrm e$, and $\mathrm d$ denote the nominal, eroded, and dilated design realizations, respectively.
The three fields are denoted by $c_{\omega}$ and their indices are collected in $\mathcal W = \{ \mathrm e, \mathrm i, \mathrm d \}$.
The smooth Heaviside projection function $h_{\beta,\eta_\omega}$ is defined as
\begin{equation}
  h_{\beta,\eta_\omega}(s) = \frac{\tanh(\beta\eta_\omega) + \tanh\!\left(\beta(s-\eta_\omega)\right)}{\tanh(\beta\eta_\omega) + \tanh\!\left(\beta(1-\eta_\omega)\right)},\quad \omega \in \{ \mathrm e, \mathrm d \},
\end{equation}
where $\beta>0$ and $\eta_\omega$ are model parameters satisfying $0<\eta_{\mathrm d}<0.5<\eta_{\mathrm e}<1$.
For the nominal field $\widetilde{c}$, the geometric boundary is defined by the contour $\widetilde{c}=0.5$.
The eroded configuration $c_{\mathrm e}$ is obtained by projection with $\eta_{\mathrm e}>0.5$. Its boundary is shifted into the solid, and excessively narrow solid features are removed entirely.
The dilated configuration $c_{\mathrm d}$ is obtained by projection with $\eta_{\mathrm d}<0.5$. Its boundary is shifted into the void, and excessively small holes are closed.
The parameter $\beta$ controls the steepness of the projection at the contour.
Such geometric mechanisms also impose lower bounds on feature sizes.
For prescribed minimum solid and void feature sizes, the parameter set $(r_f,\beta,\eta_{\mathrm e},\eta_{\mathrm d})$ can be determined by calibration~\citep{trilletAnalyticalRelationshipsImposing2021}.

Acceptable performance is required for all three design realizations.
A worst-case criterion is adopted to evaluate performance across the three design realizations, without assigning probabilities to their occurrence.
The optimization objective is therefore defined by the worst-performing design realization.
For each design realization $\omega \in \mathcal W$, the state $y_\omega$ is determined by  $\mathcal S(y_\omega,c_\omega)=0$, and the corresponding objective functional is $j_\omega(c):=J(y_\omega(c),c_\omega(c))$.
In classical topology optimization problems such as compliance minimization, the objective value typically changes monotonically with material addition or removal, allowing the design realizations to be ranked a priori~\citep{Wang2011RobustProjection}.
In the present problem, however, the port voltage depends on both the source charge and the capacitance.
The worst-case design realization must be identified by solving the state problem for all three realizations and explicitly comparing their objective values.
The optimization objective value is then defined as the maximum of $j_\omega(c)$ over the three design realizations
\begin{equation}
  \max_{\omega\in\mathcal W} \quad j_\omega(c).
\end{equation}
With phase-field regularization included, the optimization problem can be written in the following equivalent epigraph form~\citep{steinTutorialPropertiesEpigraph2025}
\begin{equation}
  \begin{array}{r@{\hspace{1.5em}}l}
    \displaystyle \min_{c,\bar{J}}
      & \displaystyle \bar{J} + \gamma\mathcal E(c), \\[0.3em]
    \mathrm{s.t.}
      & j_\omega(c)\leq \bar{J}, \ \ g_{\mathrm{v}}(c) = 0, \\
      & c\in\mathcal K, \ \ \omega \in\mathcal W,
  \end{array}
\label{eqn:epigraph_optimal}
\end{equation}
where $\bar{J}$ is an auxiliary variable representing a common upper bound on the objective values of the three design realizations. At an optimum, $\bar{J}=\max_{\omega\in\mathcal W}j_\omega(c)$.
Hence, the reformulation \eqref{eqn:epigraph_optimal} replaces the nonsmooth maximum objective with a set of standard inequality constraints.

\subsection{Anchored spectral connectivity constraints}
\label{sec:spectral-conn}

As discussed in Sections~\ref{sec:allen-cahn-model} and~\ref{sec:manufacturing},
interfacial regularization and manufacturing robustness control the local feature sizes.
The solid region in the design domain can nevertheless consist of multiple disconnected components.
The components disconnected from the fixed boundary neither transfer loads nor provide electrical paths to the ports.
Narrow connections between solid regions can also be removed entirely in the eroded design realization.
Connectivity to the anchoring boundary is therefore imposed as a global constraint on each design realization.
A comparison of existing connectivity constraints is provided in~\citep{Cool2025ConnectivityReview}.
A spectral approach based on the grounded Laplacian operator is adopted here~\citep{Donoso2022SpectralGraph,Donoso2023ContinuousModel,Donoso2024DirichletLaplacian}.
In this approach, each solid component that is not attached to the anchoring boundary contributes one zero eigenvalue to the grounded Laplacian, and the smallest eigenvalue provides a continuous measure of the connectivity between the solid phase and the anchoring boundary.
Spectral connectivity constraints have been applied to the design of piezoelectric electrodes and electromechanical devices~\citep{Donoso2023PiezoElectrode}.
Discrete graphs have also been used to assess the structural connectivity in flexoelectric topology optimization~\citep{Greco2024ApparentPiezoelectricity}.

For the threshold $c_{\mathrm{th}}=1/2$, the binary indicator of solid material is defined as
\begin{equation}
  c_{\mathrm{bin}}(\boldsymbol{x}) = 
  \begin{cases}
    1, & c_{\omega}(\boldsymbol{x}) \geq c_{\mathrm{th}} \\
    0, & c_{\omega}(\boldsymbol{x}) < c_{\mathrm{th}}.
  \end{cases}
\end{equation}
The graph $G_S=(V_S,E_S)$ is constructed by connecting solid nodes according to mesh adjacency.
Here, $V_S=\{k:c_{\mathrm{bin}}(\boldsymbol{x}_k)=1\}$ is the set of solid nodes, and $E_S$ contains the edges between adjacent solid nodes.
Let $\Gamma_A\subset\partial\Omega_d$ denote the anchoring boundary, which typically coincides with the fixed boundary of the device.
The set of solid nodes on $\Gamma_A$ is denoted by $A\subset V_S$, and the remaining nodes are collected in $F=V_S \setminus A$.
The nodes on $\Gamma_A$ are grounded by eliminating the corresponding degrees of freedom.
Hence, the resulting grounded Laplacian matrix $L_S$ is the principal submatrix of the graph Laplacian $L$ indexed by $F$.
Let $\boldsymbol{\pi}\in\mathbb{R}^{|F|}$ denote a real-valued vector on the free nodes.
The quadratic form of $L_S$ is given by
\begin{equation}
  \boldsymbol{\pi}^{\top} L_S \boldsymbol{\pi} = \sum_{\substack{(j,k)\in E_S\\ j,k\in F}} {w_{jk}(\pi_j-\pi_k)^2} + 
  \sum_{\substack{(j,k)\in E_S\\ j\in F,\;k\in A}} {w_{jk}\pi_j^2},
\end{equation}
where $w_{jk}=w_{kj}>0$ is the weight of the undirected edge $\{j,k\}\in E_S$. The first term is the edge energy between free nodes, and the second is the contribution from edges connecting free nodes to anchored nodes. Each undirected edge is counted once in the sums. The weight is zero when the two nodes are not connected by an edge.
Let $N_S$ denote the number of solid components that are not connected to $A$.
The correspondence between unanchored components and zero modes is given as~\citep{Donoso2024DirichletLaplacian}
\begin{equation}
  N_S = \dim\ker L_S.
\end{equation}
The smallest eigenvalue satisfies $\lambda_1(L_S)\geq0$, since $L_S$ is symmetric positive semidefinite. Hence,
\begin{equation}
  N_S=0 \iff \ker L_S=\{\boldsymbol{0}\} \iff \lambda_1(L_S)>0,
\label{eqn:conn-discrete}
\end{equation}
which establishes the equivalence between connectivity to the anchoring boundary and positive definiteness of the grounded Laplacian matrix.

Since the graph-based connectivity criterion in Eq.~\eqref{eqn:conn-discrete} is not differentiable with respect to the design variable, it cannot be used directly in the gradient-based optimization.
Therefore, a continuous generalized eigenvalue problem is formulated for each design realization.
Let $\vartheta$ denote the eigenfunction associated with $\lambda$. The test space of $\vartheta$ is defined as $\mathcal X_S=\{\delta\vartheta\in H^1(\Omega_d):\delta\vartheta|_{\Gamma_A}=0\}$.
The eigenfunction is prescribed to vanish on the anchoring boundary, while a natural zero-flux condition is imposed on the remaining boundary.
The corresponding weak formulation is to find an eigenpair $(\lambda,\vartheta)\in\mathbb{R}\times(\mathcal X_S\setminus\{0\})$ such that
\begin{equation}
\int_{\Omega_d}{K_S(c_\omega)\nabla\vartheta\cdot\nabla \delta\vartheta \mathrm d\Omega}
=\lambda\int_{\Omega_d}{M_S(c_\omega)\vartheta\delta\vartheta \mathrm d\Omega},
\quad \forall \delta\vartheta\in\mathcal X_S,
\label{eqn:spectral_conn_weak}
\end{equation}
in which the coefficients $K_S(c_\omega)$ and $M_S(c_\omega)$ are continuously interpolated between their solid- and void-phase values as follows
\begin{equation}
  \begin{cases}
    K_S(c_\omega) = K_- + (1-K_-) c^{r_S}_\omega, \\
    M_S(c_\omega) = M_- + (1-M_-) c_\omega,
  \end{cases}
\end{equation}
where $0 < K_-,M_-\ll 1$ are positive lower bounds for the coefficients in the void phase.
The relative decay of the diffusion coefficient $K_S$ and the mass coefficient $M_S$ across the diffuse interface is controlled by the exponent $r_S>2$.
A value of $\lambda$ is an eigenvalue if Eq.~\eqref{eqn:spectral_conn_weak} admits a nonzero solution $\vartheta$.
These eigenvalues form a discrete positive spectrum and are ordered as $0<\lambda_1\leq\lambda_2\leq\dots$.
For each design realization, the connectivity constraint is defined in terms of the first eigenvalue $\lambda_1(c_\omega)$ as
\begin{equation}
  g_{\mathrm{s},\omega}(c) = \frac{\lambda_{\mathrm{req},\omega} - \lambda_1(c_\omega(c))}{s_{\lambda,\omega}} \leq 0, \quad
  \omega \in \mathcal W,
\label{eqn:spectral_conn_constraint}
\end{equation}
where $\lambda_{\mathrm{req},\omega}$ is the prescribed threshold for connectivity strength under the given geometry and boundary conditions. $s_{\lambda,\omega}>0$ is a normalization constant.
The connectivity constraint $g_{\mathrm s,\omega}$ is differentiable with respect to the design variable $c$ when $\lambda_1$ is a simple eigenvalue.
The corresponding sensitivity analysis is presented in Appendix~\ref{app:adjoint-phase-field}.
Therefore, the complete constrained optimization problem is formulated as follows
\begin{equation}
  \begin{array}{r@{\hspace{1.5em}}l}
    \displaystyle \min_{c,\bar{J}}
      & \displaystyle \bar{J} + \gamma\mathcal E(c), \\[0.3em]
    \mathrm{s.t.}
      & j_\omega(c)\leq \bar{J}, \ \ g_{\mathrm{v}}(c) = 0, \ \ g_{\mathrm{s},\omega}(c)\leq 0, \\
      & c\in\mathcal K, \ \ \omega \in\mathcal W.
  \end{array}
\label{eqn:optimal_format}
\end{equation}

\section{Numerical procedure}
\label{sec:4-numerical}

This section presents the numerical solution of the constrained optimization problem~\eqref{eqn:optimal_format}.
Section~\ref{subsec:spatial-discretization} describes the spatial discretization and nondimensionalization. An isogeometric discretization based on non-uniform rational B-splines (NURBS) is applied for the displacement, electric potential, and phase field. 
The higher-order state problem \eqref{eqn:flexo-weak-system-unified} requires $C^1$ continuity across the element boundaries and the NURBS basis functions can provide it naturally.
The mechanical and electrical quantities are scaled to comparable magnitudes through nondimensionalization of the governing equations. Therefore, the resulting state problem is expressed in terms of characteristic scale ratios.
In Section~\ref{sec:optimal_system}, the Lagrangian functional is constructed and the first-order stationarity conditions are derived. The first- and second-order sensitivities of the objective functional and constraints are also derived.
In Section~\ref{subsec:implicit-update}, the continuous problem is discretized in pseudo-time with a fully implicit scheme. The Newton-Krylov solution procedure and the convergence criteria for phase-field updates are then presented.

\subsection{Spatial discretization}\label{subsec:spatial-discretization}

As specified in Eq.~\eqref{eqn:trial-space}, the displacement belongs to $[H^2(\Omega)]^d$, whereas the electric potential belongs to $H^2(\Omega)$.
Accordingly, $C^1$ continuity across element boundaries is required for the discretization of the displacement and electric potential.
Isogeometric analysis is adopted to represent the geometry, state fields, and design variable by NURBS basis functions~\citep{hughesIsogeometricAnalysisCAD2005}.
For simple interior knots, basis functions of degree $p\geq 2$ provide $C^{p-1}$ continuity without introducing additional degrees of freedom.
Let $R_A(\boldsymbol{x})$ denote the basis functions, and $\boldsymbol{u}_A$ and $\phi_A$ the control-point coefficients of the displacement and electric potential, respectively. The corresponding fields and gradients are given by
\begin{equation}
  \begin{cases}
    \boldsymbol{u}_h = \sum_{A}{R_A(\boldsymbol{x})\boldsymbol{u}_A}, \\
    \boldsymbol{\varepsilon}_h = \sum_{A}{\nabla^S \bigl(R_A(\boldsymbol{x})\boldsymbol{u}_A\bigr)}, \\
    \nabla\boldsymbol{\varepsilon}_h = \sum_{A}{\nabla\nabla^S \bigl(R_A(\boldsymbol{x})\boldsymbol{u}_A\bigr)},
  \end{cases}
  \quad
  \begin{cases}
    \phi_h = \sum_{A}{R_A(\boldsymbol{x})\phi_A}, \\
    \boldsymbol{E}_h = - \sum_{A}{\nabla R_A(\boldsymbol{x})\phi_A}, \\
    \nabla\boldsymbol{E}_h = - \sum_{A}{\nabla\nabla R_A(\boldsymbol{x})\phi_A}.
  \end{cases}
\label{eqn:spatial_discrete_y}
\end{equation}
Here, $\nabla^S$ denotes the symmetric gradient.
Similarly, the design variable and its gradient are interpolated as
\begin{equation}
c_h = \sum_{A}{R_A(\boldsymbol{x})c_A}, \quad
\nabla c_h = \sum_{A}{\nabla R_A(\boldsymbol{x})c_A},
\label{eqn:spatial_discrete_c}
\end{equation}
where $\nabla R_A(\boldsymbol{x})$ and $\nabla\nabla R_A(\boldsymbol{x})$ denote the first and second spatial derivatives of the basis functions, respectively.

The coefficients in Eq.~\eqref{eqn:flexo-isotropic} are characterized by large differences in magnitude.
For the solid material used in the numerical examples, the Lamé parameters, dielectric permittivity, and flexoelectric coefficients are on the order of $10^{10}\,\mathrm{Pa}$, $10^{-7}\,\mathrm{F/m}$, and $10^{-5}\,\mathrm{C/m}$, respectively.
Direct discretization in dimensional variables can produce substantial differences in the numerical scales of the mechanical and electrical equations.
Consistent convergence criteria are also difficult to define for the nonlinear Newton iterations because the residual components are poorly scaled.
Therefore, the state and optimization problems are nondimensionalized by a common set of reference scales~\citep{Codony2021Flexoelectricity}. Let $L_0$, $Y_0$, and $\kappa_0$ denote the reference length, elastic modulus and dielectric permittivity, respectively. The dimensionless quantities for the flexoelectric model are defined as
\begin{equation}
  \widehat{\boldsymbol{u}} = \frac{\boldsymbol{u}}{L_0},\quad
  \widehat{\phi} = \frac{\phi}{V_0},\quad
  \widehat{\kappa} = \frac{\kappa}{\kappa_0},\quad
  \widehat{\Lambda} = \frac{\Lambda}{Y_0},\quad
  \widehat{\mu} = \frac{\mu}{Y_0},\quad
  \widehat{f}_1 = \frac{f_1}{L_0\sqrt{Y_0(\kappa_0-\epsilon_0)}},\quad
  \widehat{f}_2 = \frac{f_2}{L_0\sqrt{Y_0(\kappa_0-\epsilon_0)}},
\label{eqn:dimless_define}
\end{equation}
in which $\epsilon_0$ denotes the vacuum permittivity.
Note that the reference scales are not all independent, since the reference electric potential satisfies $V_0=L_0\sqrt{Y_0/\kappa_0}$.
The nondimensionalized problem is characterized by a small set of characteristic scale ratios.
Under the same boundary conditions, geometrically similar designs with different sizes and material properties are characterized by a common family of dimensionless parameters, including $l_m/L_0$, $l_e/L_0$, $l_c/L_0$, $r_f/L_0$ and the ratios of material properties between the solid and void phases. This formulation provides a common basis for design comparison.
The convergence criteria in Section~\ref{subsec:implicit-update} are also defined in this dimensionless formulation.

\subsection{Optimality system}\label{sec:optimal_system}

In this section, the continuous optimality system for the optimization problem~\eqref{eqn:optimal_format} is established to specify the stationarity and feasibility requirements for the final design.
This optimization problem is nonconvex and includes both equality and inequality constraints. The Lagrange multiplier method is adopted to express the first-order optimality conditions in a unified form.
Under the assumptions of continuous differentiability of the objective and constraint functionals and an appropriate constraint qualification~\citep{hinzeOptimizationPDEConstraints2009}, a local minimizer satisfies the Karush-Kuhn-Tucker (KKT) conditions~\citep{nocedalNumericalOptimization2006}. These conditions consist of stationarity, primal feasibility, dual feasibility, and complementarity.
The Lagrangian functional is introduced below to express the optimality system, followed by the first- and second-order sensitivities required for its numerical solution.

Define the Lagrangian functional as
\begin{equation}
  \begin{aligned}
  \mathcal L(c,\bar{J},\boldsymbol{\alpha}) = \bar{J} + \gamma\mathcal E(c) 
  &+ \sum_{\omega\in\mathcal W} \alpha_\omega (j_\omega(c)-\bar{J}) + 
  \sum_{\omega\in\mathcal W} {\alpha_{\mathrm{s},\omega} g_{\mathrm{s},\omega}(c)} \\ 
  &+ \alpha_{\mathrm{v}} g_{\mathrm{v}}(c) - \int_{\Omega_d}{\alpha_c^-c \mathrm d\Omega} + \int_{\Omega_d}{\alpha_c^+(c-1) \mathrm d\Omega},
  \end{aligned}
\label{eqn:total-Lagrangian}
\end{equation}
where the set $\boldsymbol{\alpha}=(\{ \alpha_{\omega}\}_{\omega\in\mathcal{W}},\{ \alpha_{\mathrm{s},\omega} \}_{\omega\in\mathcal{W}},\alpha_{\mathrm{v}},\alpha_c^-,\alpha_c^+)$ denotes the Lagrange multipliers. $\alpha_\omega\ge0$ and $\alpha_{\mathrm{s},\omega}\ge0$ correspond to the performance and spectral connectivity constraints for design realization $\omega$, respectively.
$\alpha_{\mathrm{v}}\in\mathbb{R}$ corresponds to the volume equality constraint, and $\alpha_c^{\pm} \ge 0$ correspond to the bound constraints.
The first-order stationarity condition with respect to the design variable $c$ is
\begin{equation}
  \begin{aligned}
    \mathrm{D} \mathcal L[\delta c] = 
    \gamma \mathrm{D}\mathcal E(c)[\delta c] & + \sum_{\omega\in\mathcal W} {\alpha_\omega \mathrm{D}j_\omega(c)[\delta c]} + \sum_{\omega\in\mathcal W} {\alpha_{\mathrm{s},\omega} \mathrm{D} g_{\mathrm{s},\omega}(c)[\delta c]}  \\
    & + \alpha_{\mathrm{v}} \int_{\Omega_d}{\delta c\mathrm d\Omega}
    - \int_{\Omega_d}{\alpha_c^- \delta c \mathrm d\Omega} 
    + \int_{\Omega_d}{\alpha_c^+\delta c \mathrm d\Omega} = 0,
    \qquad
    \forall\,\delta c.
    \end{aligned}
\end{equation}
For the auxiliary variable $\bar{J}$ introduced by the epigraph reformulation, the following condition is required
\begin{equation}
  \partial_{\bar{J}} \mathcal L = 1 - \sum_{\omega\in\mathcal W} {\alpha_\omega} = 0.
\end{equation}
This equation imposes a normalization condition on the multipliers $\{\alpha_\omega\}_{\omega\in\mathcal W}$. The directional derivative of each realization-specific objective functional $j_\omega(c)$ is weighted by the corresponding multiplier $\alpha_\omega$.
For the inequality constraints, the KKT conditions are expressed as
\begin{equation}
  \begin{gathered}
    \alpha_\omega \ge 0,\ \
    j_\omega(c)-\bar{J} \le 0,\ \
    \alpha_\omega\bigl(j_\omega(c)-\bar{J}\bigr)=0, \quad
    \omega \in \mathcal W,
    \\[0.3em]
    \alpha_{\mathrm{s},\omega} \ge 0,\ \
    g_{\mathrm{s},\omega}(c)\le 0,\ \
    \alpha_{\mathrm{s},\omega} g_{\mathrm{s},\omega}(c)=0, \quad
    \omega \in \mathcal W,
    \\[0.3em]
    \alpha_c^- \ge 0,\ \
    \alpha_c^+ \ge 0,\ \
    0\le c\le1,\ \
    \alpha_c^-c=0,\ \
    \alpha_c^+(1-c)=0.
  \end{gathered}
\end{equation}
Note that the multiplier associated with an inactive inequality constraint is zero.

\subsubsection{First-order sensitivity analysis}

Evaluation of the stationarity condition requires the directional derivatives of the phase-field energy, the worst-case objective, and the connectivity constraints.
The latter two functionals depend implicitly on the design variable through the state equations and the generalized eigenvalue problem, respectively. 
In this subsection, the corresponding directional derivatives are derived below.
Because the first derivatives of the volume and bound constraints are independent of $c$, the derivations of them are omitted here.

\textbf{a. Phase-field energy functional.} The phase-field energy is an explicit functional of the design variable. The directional derivative is
\begin{equation}
  \mathrm{D}\mathcal E(c)[\delta c] = \int_{\Omega_d} \left[\Psi_B'(c)\,\delta c + l_c^2\nabla c \cdot\nabla\delta c \right]\,\mathrm d\Omega.
\label{eqn:sens_pf_1}
\end{equation}

\textbf{b. Objective functional.} 
For a given design realization $\omega$, the first directional derivative of the objective functional with respect to $c_\omega$ is obtained by the standard adjoint method~\citep{hinzeOptimizationPDEConstraints2009} as
\begin{equation}
  \mathrm Dj(c_\omega)[\delta c_\omega] = \partial_{c_\omega}J(y_\omega,c_\omega) [\delta c_\omega] + \partial_{c_\omega}\bigl(\mathcal S(y_\omega,c_\omega)[p_\omega]\bigr)[\delta c_\omega],
\label{eqn:first-order-gradient-obj}
\end{equation}
in which $p_{\omega}$ is the adjoint variable. The detailed derivation of Eq.~\eqref{eqn:first-order-gradient-obj} is provided in Appendix~\ref{app:1st-sens}.
As discussed in Section~\ref{sec:manufacturing}, the design realizations are obtained from $c$ by filtering and projection. The filter mapping $c\mapsto\widetilde c$ is defined implicitly by the Helmholtz equation~\eqref{eqn:helmholtz_filter}.
Thus, the first directional derivative $\mathrm D\widetilde c(c)[\delta c]$ of the filtered field for an arbitrary variation $\delta c$ is obtained by implicit differentiation as
\begin{equation}
  \mathrm D\widetilde c(c)[\delta c] = (\mathrm I - r_f^2\nabla^2)^{-1}\delta c,
\label{eqn:filter_direction}
\end{equation}
where $\mathrm I$ is the identity operator. The linear inverse operator $(\mathrm I - r_f^2\nabla^2)^{-1}$ is defined with the homogeneous Neumann boundary conditions in Eq.~\eqref{eqn:helmholtz_filter}.
The first directional derivative of each design realization is given by
\begin{equation}
  \mathrm D c_{\omega}(c)[\delta c] = h_{\beta,\eta_\omega}'(\widetilde c)\,\mathrm D\widetilde c(c)[\delta c],
\label{eqn:manu_direction}
\end{equation}
where $h_{\beta,\eta_\mathrm{i}}'\equiv 1$, since no projection is applied to the nominal realization. Hence, the first directional derivative of the objective functional with respect to the design variable is obtained by the chain rule
\begin{equation}
  \mathrm D j_\omega(c)[\delta c]
  = \mathrm Dj(c_\omega) \bigl[h_{\beta,\eta_\omega}'(\widetilde c)  \mathrm D\widetilde c(c)[\delta c]\bigr].
\end{equation}

\textbf{c. Spectral connectivity constraints.} The spectral constraints depend solely on the design variable and involve no electromechanical state variables.
The smallest eigenvalue $\lambda_1$ of the generalized eigenvalue problem~\eqref{eqn:spectral_conn_weak} measures the connectivity strength. 
Let $\vartheta$ denote the corresponding eigenfunction.
The first directional derivative of $\lambda_1$ with respect to the design realization is obtained by implicit differentiation of Eq.~\eqref{eqn:spectral_conn_weak} under the assumption that $\lambda_1$ is a simple eigenvalue
\begin{equation}
  \mathrm D\lambda_1(c_\omega)[\delta c_\omega] ={}\int_{\Omega_d} K_S'(c_\omega)\,\delta c_\omega \,|\nabla\vartheta|^2\,\mathrm d\Omega - \lambda_1 \int_{\Omega_d}  M_S'(c_\omega)\,\delta c_\omega \,\vartheta^2\,\mathrm d\Omega,
\label{eqn:spectral_first_order}
\end{equation}
in which the eigenfunction is normalized according to Eq.~\eqref{eqn:spec-mass-norm}.
Therefore, the first directional derivative of the connectivity constraint with respect to the design variable is given as
\begin{equation}
  \mathrm D g_{\mathrm{s},\omega}(c)[\delta c] = - \frac{1}{s_{\lambda,\omega}} \mathrm D\lambda_1(c_\omega)\bigl[\mathrm D c_\omega(c) [\delta c] \bigr],
\end{equation}
where $\mathrm D c_\omega(c) [\delta c]$ is defined in Eq.~\eqref{eqn:manu_direction}. Further details of the derivation of Eq.~\eqref{eqn:spectral_first_order} are provided in Appendix~\ref{app:adjoint-phase-field}.

\subsubsection{Second-order sensitivity analysis}

Newton linearization of the first-order stationarity conditions requires second-order derivatives with respect to the design variable.
For the fixed values of the epigraph variable and the multipliers, the Hessian-vector product for the reduced functional $\mathcal L(c)$ is defined as
\begin{equation}
  H_{\mathcal L} \Delta c \coloneqq \mathrm{D}(\nabla_c \mathcal L(c))[\Delta c],
\end{equation}
where $H_{\mathcal L}$ is the Hessian operator.
Explicit assembly of this operator is generally expensive in large-scale topology optimization. 
For a design with $N_c$ degrees of freedom, the explicitly assembled Hessian matrix contains $\mathcal{O}(N_c^2)$ entries. The columnwise assembly of it requires solving the incremental state and adjoint equations for each of the $N_c$ design directions.
Hence, a matrix-free strategy is adopted to evaluate only the Hessian-vector products needed by the Newton-Krylov solver. The total formulation is
\begin{equation}
  H_{\mathcal L}\Delta c = \gamma H_{\mathcal E}\Delta c + 
  \sum_{\omega\in\mathcal W} {\alpha_\omega H_{j,\omega} \Delta c} + 
  \sum_{\omega\in\mathcal W} {\alpha_{\mathrm{s},\omega} H_{\mathrm{s},\omega}\Delta c},
\label{eqn:second-order-formula}
\end{equation}
where $H_{\mathcal E}$, $H_{j,\omega}$, and $H_{\mathrm{s},\omega}$ correspond to the phase-field energy, objective functionals and connectivity constraints, respectively. The second derivatives of the volume and bound constraints with respect to $c$ vanish. Detailed derivations of the Hessian-vector products are provided in Appendix~\ref{app:adjoint-phase-field}.

\subsection{Fully implicit constrained optimization}\label{subsec:implicit-update}

In this section, the Allen-Cahn gradient flow is discretized by a fully implicit Euler scheme.
The first-order optimality system for problem~\eqref{eqn:optimal_format} is approached through a pseudo-time stepping strategy.

\renewcommand{\algorithmicrequire}{\textbf{Input:}}
\renewcommand{\algorithmicensure}{\textbf{Output:}}
\begin{algorithm}[!t]
\caption{Fully implicit constrained update at pseudo-time step $n$}
\label{alg:implicit_step}
\begin{algorithmic}[1]

\Require
$x^n$, $\Delta\tau$, $\epsilon_N$, $N_{\max}$, $\rho\in(0,1)$, $\chi\in(0,1)$.

\Ensure $x^{n+1}$.

\State \textbf{Initialize} $x\gets x^n$, $m\gets0$.
\State Evaluate states, active sets, $\mathcal{R}(x)$ and $r_N$.
\While{$m<N_{\max}$ \textbf{and} $r_N\geq\epsilon_N$}
  \State Solve \eqref{eqn:newton_format} by GMRES for $\Delta x$.
  \State Line search: $\varrho\gets1$.
         \While{$\Psi_n(x+\varrho\Delta x)>(1-\chi\varrho)\Psi_n(x)$}
         \State $\varrho\gets\rho\varrho$.
         \EndWhile
  \State $x\gets x+\varrho\Delta x$.
  \State Re-evaluate states, active sets, and $\mathcal{R}(x)$.
  \State Update $r_N$ by \eqref{eqn:inner_convergence}.
  \State $m\gets m+1$.
\EndWhile
\If{$r_N<\epsilon_N$}
  \State \textbf{Accept:} $x^{n+1}\gets x$.
\Else
  \State \textbf{Retry:} reduce $\Delta\tau$ and restart from $x^n$. 
\EndIf

\end{algorithmic}
\end{algorithm}

Let $c^n$ denote the converged phase field at $\tau^n$. With $c^n$ as the reference field, the incremental KKT problem over $[\tau^n,\tau^{n+1}]$ is formulated as
\begin{equation}
  \begin{array}{r@{\hspace{1.5em}}l}
    \displaystyle \min_{c,\bar{J}}
      & \displaystyle \bar{J}+\gamma\mathcal E(c) + \frac{1}{2M_0\Delta\tau}
      \int_{\Omega_d} {(c-c^n)^2 \mathrm{d}\Omega}, \\[0.3em]
    \mathrm{s.t.}
      & j_\omega(c)\leq \bar{J}, \ \ g_{\mathrm{v}}(c) = 0, \ \ g_{\mathrm{s},\omega}(c)\leq 0,  \\
      & \omega \in\mathcal W,\ \ c\in\mathcal K,
  \end{array}
\label{eqn:optimal_format_incre}
\end{equation}
where $\Delta\tau=\tau^{n+1}-\tau^n$ is the time increment. Define the primal-dual variable tuple $x=(c,\bar{J},\boldsymbol{\alpha})$. The incremental Lagrangian is given as
\begin{equation}
  \mathcal{L}_n(x) = \mathcal{L}(c,\bar{J},\boldsymbol{\alpha})
  + \frac{1}{2M_0\Delta\tau}\int_{\Omega_d}{(c-c^n)^2\,\mathrm{d}\Omega},
\label{eqn:incremental-lagrangian}
\end{equation}
where $\mathcal L$ is defined in Eq.~\eqref{eqn:total-Lagrangian}. The residual of the incremental problem~\eqref{eqn:optimal_format_incre} is written as $\mathcal R(x)[\delta x] \coloneqq \mathrm{D} \mathcal{L}_n(x)[\delta x]$. Thus, the standard Newton-Raphson linearization is
\begin{equation}
  \mathrm{D} \mathcal{R}(x)[\Delta x, \delta x] = -\mathcal R(x)[\delta x],
  \quad \forall \delta x.
\label{eqn:newton_format}
\end{equation}
Equation~\eqref{eqn:newton_format} is discretized using the approximations in Eqs.~\eqref{eqn:spatial_discrete_y} and~\eqref{eqn:spatial_discrete_c}.
The resulting linear subproblem is solved by the generalized minimal residual method (GMRES). For each Krylov search direction, only the application of the linearized KKT operator to that direction is required.
The primal-dual vector is then updated by the following damped Newton step
\begin{equation}
  x \leftarrow x + \varrho \Delta x, \quad 0 < \varrho \leq 1.
\end{equation}
The step length $\varrho$ is determined by a backtracking line search based on the KKT merit function $\Psi_n(x)=\frac{1}{2}\|\mathcal{R}(x)\|^2$.
At each pseudo-time step, convergence of the Newton-Krylov iterations and acceptance of the design update are determined by the following incremental KKT residual criterion
\begin{equation}
  r_N \coloneqq
  \frac{\|\mathcal{R}(x)\|}{1+\|\mathcal{R}(x^n)\|} < \epsilon_N.
\label{eqn:inner_convergence}
\end{equation}
Besides, the following two additional criteria are required for convergence of the overall optimization
\begin{equation}
  r_J^n = \frac{|\bar{J}^n-\bar{J}^{n_p}|}{\max(|\bar{J}^n|,|\bar{J}^{n_p}|)} < \epsilon_c,
  \quad
  r_{\mathcal F}^n = \frac{|\bar{\mathcal F}^n-\bar{\mathcal F}^{n_p}|}{\max(|\bar{\mathcal F}^n|,|\bar{\mathcal F}^{n_p}|)} < \epsilon_c.
\end{equation}
Here, $\bar{\mathcal F}=\bar{J}+\gamma\mathcal E(c)$, and the superscripts $n$ and $n_p$ denote the consecutive monitoring steps. 
These criteria assess whether the changes in the objectives between two monitoring steps are sufficiently small.
The complete algorithm is summarized in Algorithm~\ref{alg:implicit_step}. Further derivation details are provided in Appendix~\ref{app:iga-newton-pdas}.

\FloatBarrier

\section{Numerical studies}\label{sec:5-numerical-studies}

In this section, the proposed constrained optimization framework is examined through three numerical examples.
In Section~\ref{subsec:case-compression}, a two-dimensional trapezoidal pressure sensor is considered to examine the evolution of the material distribution and the resulting voltage response.
Section~\ref{subsec:case-bending} shows the optimal design of a two-dimensional bending sensor with one and multiple electrodes.
In Section~\ref{subsec:case-bending-3d}, the formulation is applied to a three-dimensional thin-film device under localized pressure loading.
For simplicity, only the nominal and eroded design realizations are considered in these examples. The dilated realization is omitted.

\captionsetup[figure]{name=Fig.,labelsep=period}
\captionsetup[table]{name=Table,labelfont=bf,labelsep=newline,justification=raggedright,singlelinecheck=false}

\subsection{Case 1: Two-dimensional pressure sensor}\label{subsec:case-compression}

\begin{figure}[!b]
  \centering
  \captionsetup[subfigure]{skip=3pt}
  \begin{subfigure}[b]{0.52\linewidth}
    \centering
    \includegraphics[width=\linewidth]{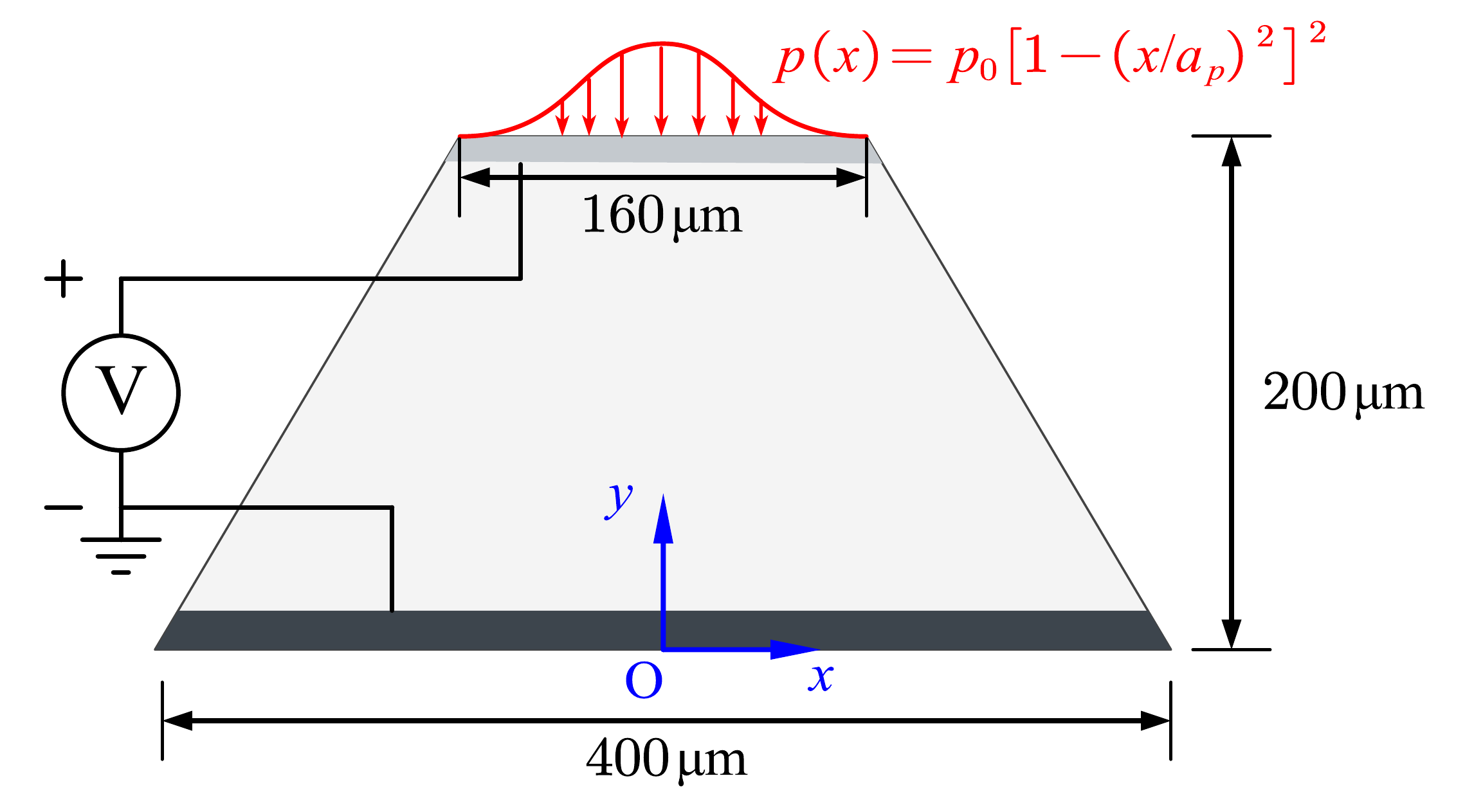}
    \caption{}
  \end{subfigure}\hspace{0.015\linewidth}%
  \begin{subfigure}[b]{0.431\linewidth}
    \centering
    \raisebox{0.090\linewidth}{\includegraphics[width=\linewidth]{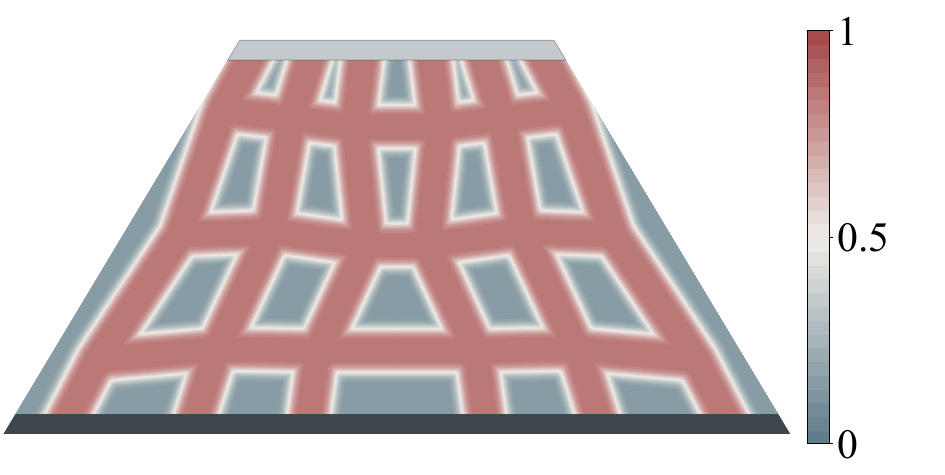}}
    \caption{}
  \end{subfigure}
  \caption{Two-dimensional pressure sensor: (a) geometry, material regions and boundary conditions; (b) initial material distribution of the nominal realization.}
  \label{fig:a1-geometry-initial}
\end{figure}

Consider the two-dimensional trapezoidal pressure sensor under plane-strain conditions, as shown in Fig.~\ref{fig:a1-geometry-initial}(a). 
The height and out-of-plane thickness are $\SI{200}{\micro\metre}$ and $\SI{10}{\micro\metre}$, respectively. The bottom and top edges are $\SI{400}{\micro\metre}$ and $\SI{160}{\micro\metre}$ wide, respectively.
The coordinate origin is located at the midpoint of the bottom edge. 
$u_x=u_y=0$ is prescribed along the bottom edge $y=0$. A pressure $p(x)=p_0[1-(x/a_p)^2]^2\ (|x|\leq a_p)$ is applied to the top edge in the negative $y$-direction. The parameters are $|x|\leq a_p=\SI{80}{\micro\metre}$ and the peak pressure $p_0=15.95\,\mathrm{MPa}$, resulting in the force of approximately $13.61\,\mathrm{mN}$ on the full device.
An equipotential sensing electrode covers the entire top edge and satisfies the ideal open-circuit condition. The bottom edge is grounded. The readout voltage is the potential of the top electrode relative to the grounded bottom edge.
Solid layers of thickness $\SI{10}{\micro\metre}$ are retained at the top and bottom as non-design regions, while the remaining region constitutes the design domain.
By symmetry of the geometry and boundary conditions, only the left half of the domain is analyzed, with the symmetry boundary conditions $u_x=0$, $\partial_xu_y=0$, and $\partial_x\phi=0$ imposed at $x=0$.
The remaining higher-order boundary conditions are prescribed in the homogeneous natural form given in Section~\ref{subsec:flexo-variational-form}.

\begin{table}[!tb]
  \centering
  \caption{Material and design parameters for the two-dimensional trapezoidal pressure sensor.}
  \label{tab:a1-parameters}
  \small
  \setlength{\tabcolsep}{5pt}
  \renewcommand{\arraystretch}{1.00}
  \begin{tabular*}{\linewidth}{@{\extracolsep{\fill}}>{\raggedright\arraybackslash}p{0.34\linewidth}lll@{}}
    \toprule
    Parameter & Symbol & Value & Unit \\
    \midrule
    Young's modulus & $(Y_0,Y_-)$ & $(1.53\times10^{11},\,3.06\times10^8)$ & $\mathrm{Pa}$ \\
    Poisson's ratio & $(\nu_0,\nu_-)$ & $(0.30,\,0.30)$ & --- \\
    Dielectric permittivity & $(\kappa_0,\kappa_-)$ & $(7.08\times10^{-8},\,1.42\times10^{-10})$ & $\mathrm{F/m}$ \\
    Flexoelectric coefficients & $(f_{1,0},f_{2,0},f_{1,-},f_{2,-})$ & $(3.0\times10^{-5},\,3.0\times10^{-5},\,0,\,0)$ & $\mathrm{C/m}$ \\
    Interpolation exponents & $(p_Y,p_f,p_\kappa)$ & $(3,\,2,\,1)$ & --- \\
    Length scales & $(l_m,l_e,l_c)$ & $(8.33,\,8.33,\,6.25)$ & $\si{\micro\metre}$ \\
    Helmholtz filter radius & $r_f$ & $10$ & $\si{\micro\metre}$ \\
    Volume fraction & $m_c$ & $0.6$ & --- \\
    Reference voltage & $V_{\mathrm{ref}}$ & 54.9 & $\mathrm{mV}$ \\
    Phase-field regularization & $\gamma L_0^2$ & $25$ & --- \\
    Erosion projection parameters & $(\beta,\eta_{\mathrm e})$ & $(5,\,0.625)$ & --- \\
    Spectral interpolation parameters & $(K_-,M_-,r_S)$ & $(10^{-3},\,10^{-3},\,6)$ & --- \\
    Spectral constraint thresholds & $(\lambda_{\mathrm{req,e}},\lambda_{\mathrm{req,i}})$ & $(0.1019,\,0.1864)$ & --- \\
    Spectral normalization scales & $(s_{\lambda,\mathrm e},s_{\lambda,\mathrm i})$ & $(0.1174,\,0.2292)$ & --- \\
    \bottomrule
  \end{tabular*}
  \par\smallskip
  \begin{minipage}{\linewidth}
    \footnotesize
    Note: Subscripts ``0'' and ``$-$'' denote the solid and void phases, respectively.
  \end{minipage}
\end{table}

The material and design parameters are listed in Table~\ref{tab:a1-parameters}.
The device is composed of barium strontium titanate ($\mathrm{Ba}_{1-x}\mathrm{Sr}_x\mathrm{TiO_3}$), with the void regions represented by the weak ersatz material interpolation in Eq.~\eqref{eqn:mat_interp}.
The reference length is $L_0=\SI{200}{\micro\metre}$ and the reference voltage is $V_0=L_0\sqrt{Y_0/\kappa_0}$.
The material volume fraction within the design domain is prescribed as $m_c=0.6$. Material in the non-design regions is excluded from this constraint.
The initial voltage of the nominal realization $V_{\mathrm{ref}}=54.9\,\mathrm{mV}$ is selected as the fixed voltage reference. Spectral connectivity constraints anchored to the bottom edge are imposed separately on the nominal and eroded realizations throughout optimization.
A mesh of $160\times160$ quadratic elements is employed, with four-point Gauss quadrature in each direction. 
The initial phase field is shown in Fig.~\ref{fig:a1-geometry-initial}(b).

\begin{figure}[!tb]
  \centering
  \captionsetup[subfigure]{skip=3pt}
  \begin{subfigure}[t]{0.23\linewidth}
    \includegraphics[width=\linewidth]{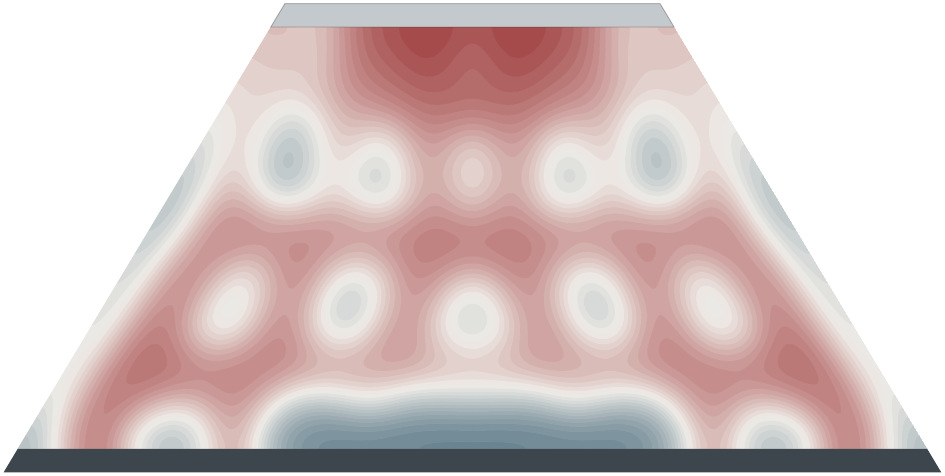}
    \par\vspace{4mm}
    \includegraphics[width=\linewidth]{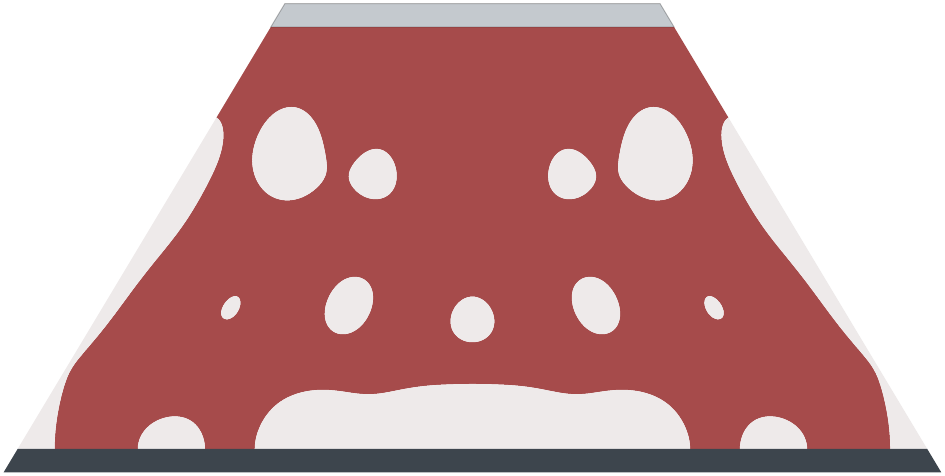}
    \caption{}
  \end{subfigure}\hfill
  \begin{subfigure}[t]{0.23\linewidth}
    \includegraphics[width=\linewidth]{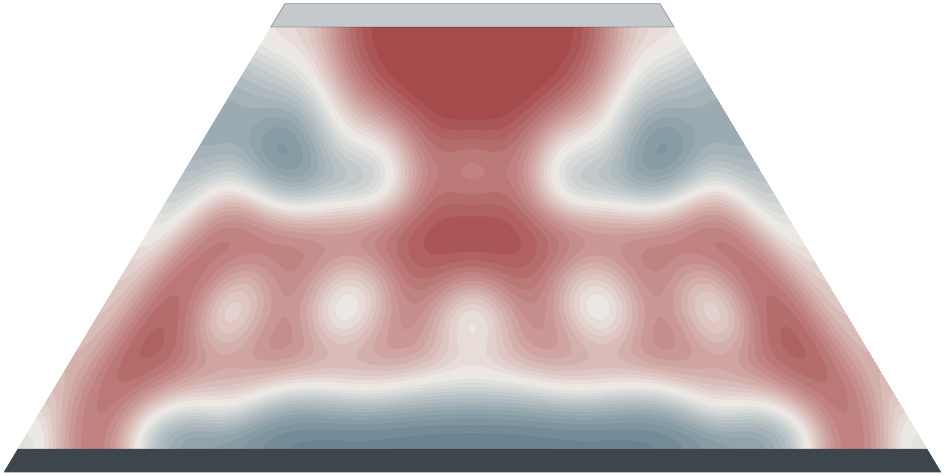}
    \par\vspace{4mm}
    \includegraphics[width=\linewidth]{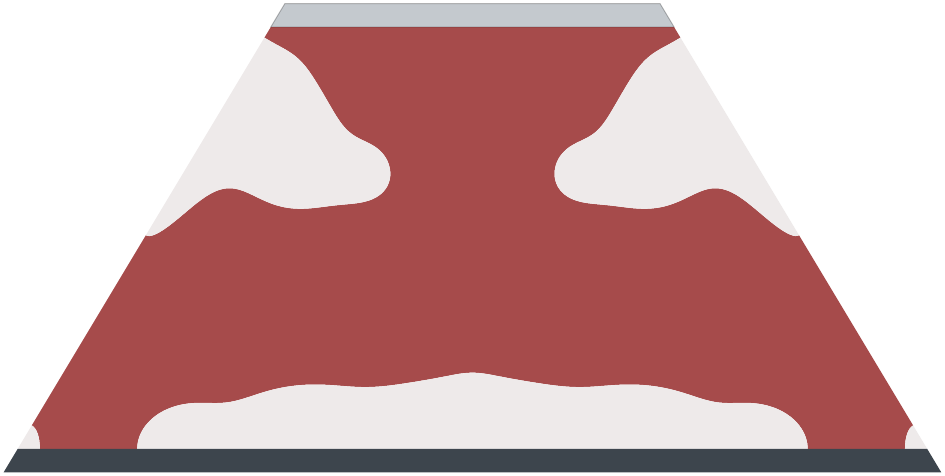}
    \caption{}
  \end{subfigure}\hfill
  \begin{subfigure}[t]{0.23\linewidth}
    \includegraphics[width=\linewidth]{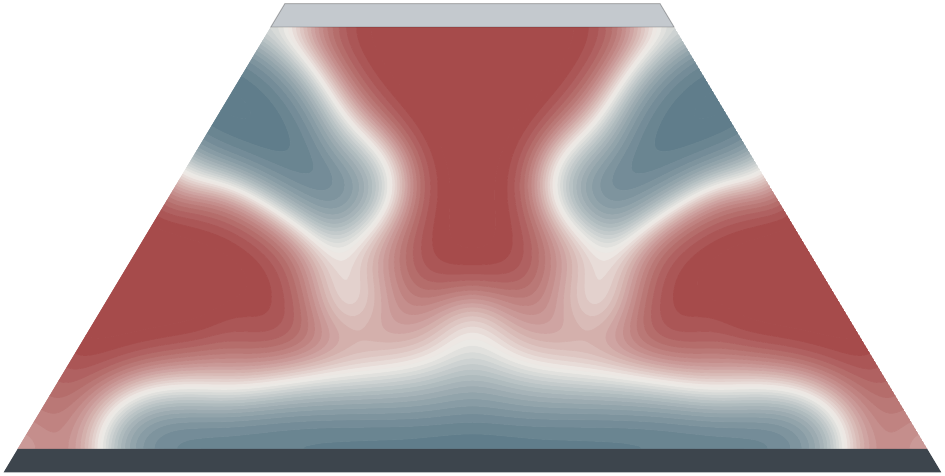}
    \par\vspace{4mm}
    \includegraphics[width=\linewidth]{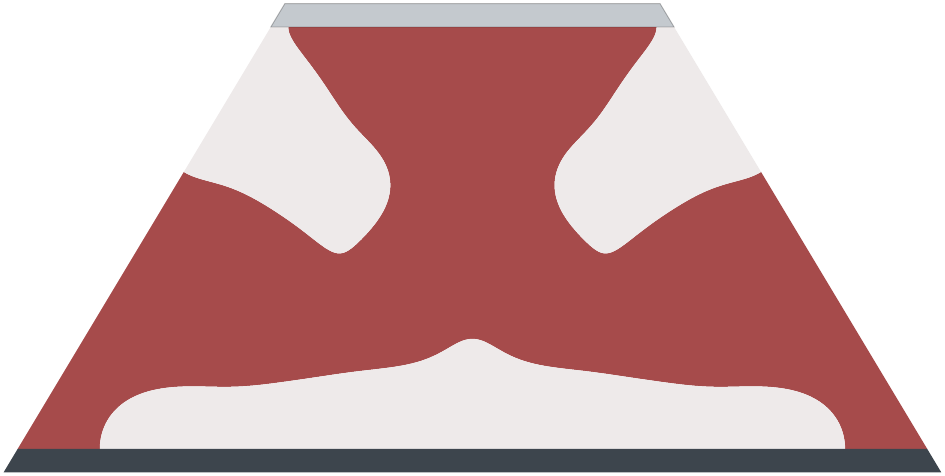}
    \caption{}
  \end{subfigure}\hfill
  \begin{subfigure}[t]{0.23\linewidth}
    \includegraphics[width=\linewidth]{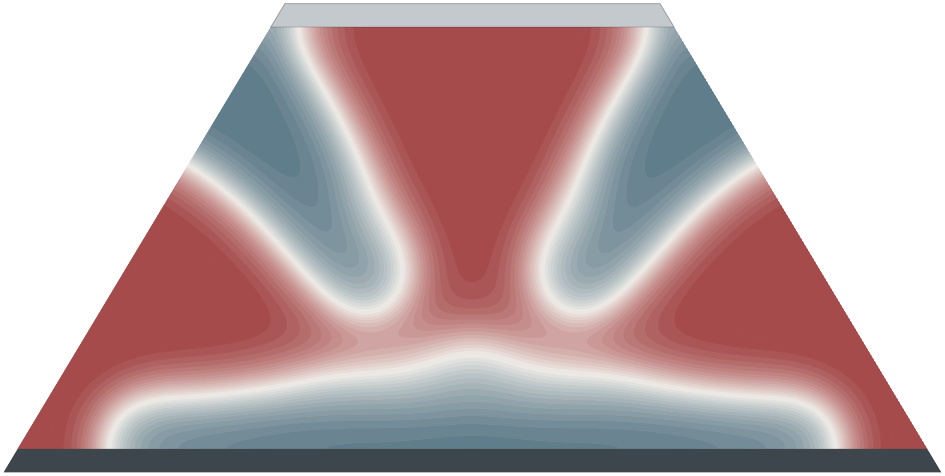}
    \par\vspace{4mm}
    \includegraphics[width=\linewidth]{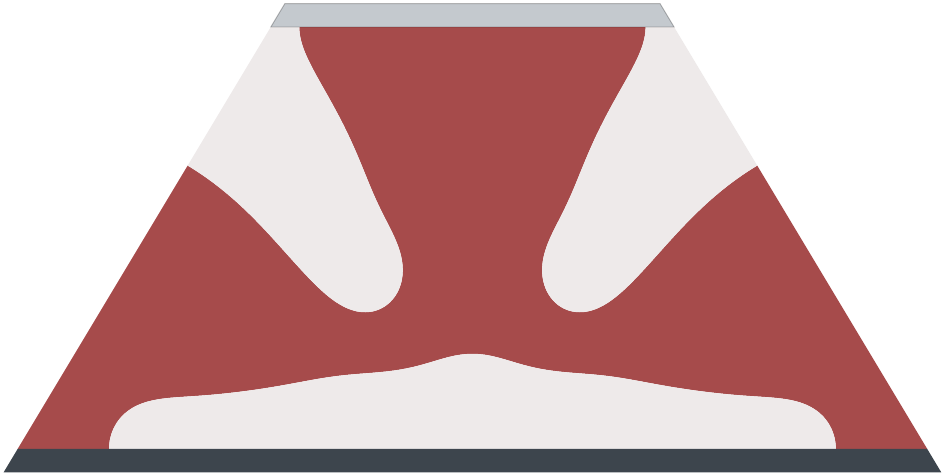}
    \caption{}
  \end{subfigure}
  \caption{Evolution of the material distribution in the two-dimensional trapezoidal pressure sensor. The top row shows the continuous phase field of the nominal realization, and the bottom row shows the corresponding binary configurations obtained with a threshold of 0.5. Panels (a) to (d) correspond to accepted steps 50, 80, 130, and 600, respectively. Regions subject to prescribed boundary conditions are highlighted in separate colors.}
  \label{fig:a1-phase-evolution}
\end{figure}

The evolution of the material distribution in the nominal realization is shown in Fig.~\ref{fig:a1-phase-evolution}.
As interior voids progressively merge, the overall geometry becomes defined by the lateral indentations and the bottom void. Material is retained in the central region and in the connections extending toward both sides of the base.
The main configuration is established by step 130. 
Further evolution occurs primarily through gradual changes in the local geometry.
The open-circuit voltage of the nominal realization increases from an initial value of $54.9\,\mathrm{mV}$ to $235.7\,\mathrm{mV}$ at step 130. It reaches $260.9\,\mathrm{mV}$ in the final design, approximately 4.75 times the initial value. These results show that the subsequent local adjustments can still provide a further increase in port voltage.
The voltage responses of the nominal and eroded realizations also become increasingly similar. At step 50, the voltages of the two realizations are $106.7\,\mathrm{mV}$ and $107.2\,\mathrm{mV}$, respectively. The nominal realization is the worst-case design realization at this step.
Therefore, the optimization improves the voltage response of both the nominal and eroded realizations.

\begin{figure}[!tb]
  \centering
  \captionsetup[subfigure]{skip=2pt}
  \begin{subfigure}[t]{0.49\linewidth}
    \centering
    \includegraphics[width=\linewidth]{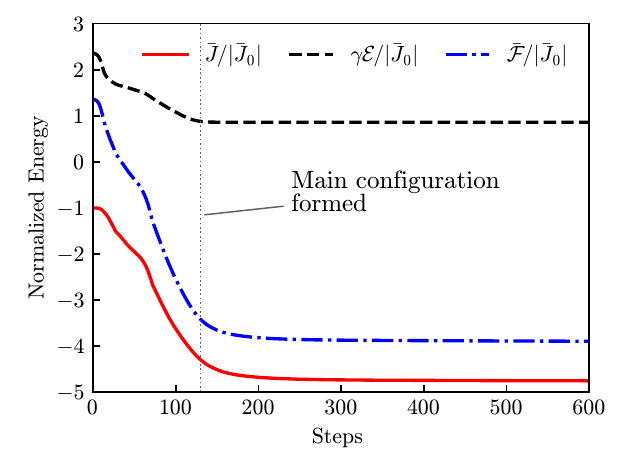}
    \caption{}
  \end{subfigure}\hfill
  \begin{subfigure}[t]{0.49\linewidth}
    \centering
    \includegraphics[width=\linewidth]{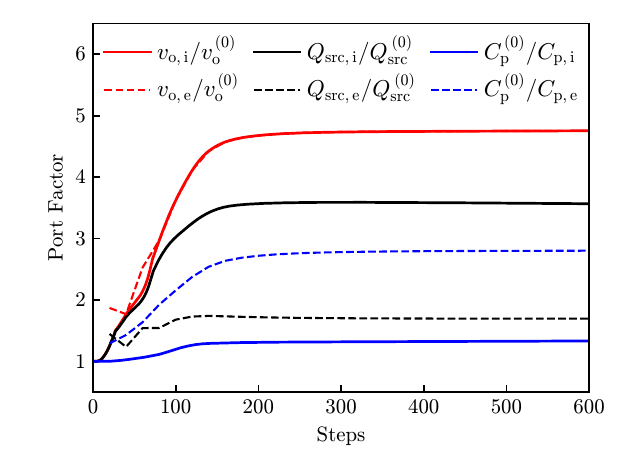}
    \caption{}
  \end{subfigure}
  \caption{Optimization history of the two-dimensional trapezoidal pressure sensor: (a) normalized voltage objective, phase-field regularization term, and total objective functional; (b) normalized open-circuit voltage, mechanically induced source charge, and inverse capacitance for the nominal and eroded realizations.}
  \label{fig:a1-optimization-port-history}
\end{figure}

The decreasing values of the three normalized functionals during optimization are illustrated in Fig.~\ref{fig:a1-optimization-port-history}(a).
The initial value of $\bar{J}$ is denoted by $\bar{J}_0$, where the subscript 0 indicates the initial design.
In the early stage, the phase-field regularization term decreases substantially as small interior voids merge. 
The double-well and gradient terms penalize the intermediate phase-field values and spatial variations, respectively. Merging of the interior voids reduces the extent of the diffuse interfaces, thereby helping lower the contributions of these terms to the regularization functional.
After step 130, the phase-field regularization term decreases only from 0.878 to 0.859, whereas the voltage objective decreases from $-4.296$ to $-4.755$. The improvement in this stage is mainly attributable to small adjustments of the existing material boundaries.
With only one electrode port considered, the open-circuit relation in Eq.~\eqref{eqn:port-potential} reduces to the scalar form $v_\mathrm{o}=-Q_{\mathrm{src}}/C_{\mathrm p}$. Thus, the voltage gain can be decomposed as
\begin{equation}
  \frac{v_{\mathrm o,\omega}}{v_{\mathrm o}^{(0)}}
  =
  \frac{Q_{\mathrm{src},\omega}}{Q_{\mathrm{src}}^{(0)}}
  \frac{C_{\mathrm p}^{(0)}}{C_{\mathrm p,\omega}},
  \quad \omega\in\{\mathrm i,\mathrm e\},
\end{equation}
where superscript $(0)$ denotes the reference value. 
Fig.~\ref{fig:a1-optimization-port-history}(b) shows that the source charge $Q_{\mathrm{src}}$ in the final design is 3.57 times its initial value, while the capacitance $C_{\mathrm{p}}$ decreases to 75.0\% of its initial value. 
The results show that the voltage gain associated with the increased source charge is further amplified by a factor of approximately 1.33 through the capacitance reduction, resulting in an overall gain of 4.75.

\begin{figure}[!tb]
  \centering
    \par\vspace{5mm}
  \begin{subfigure}[t]{0.30\linewidth}
    \centering
    \makebox[\linewidth][l]{%
      \includegraphics[width=1.21036585\linewidth,trim=1.700787bp 0bp 1.700787bp 0bp,clip]{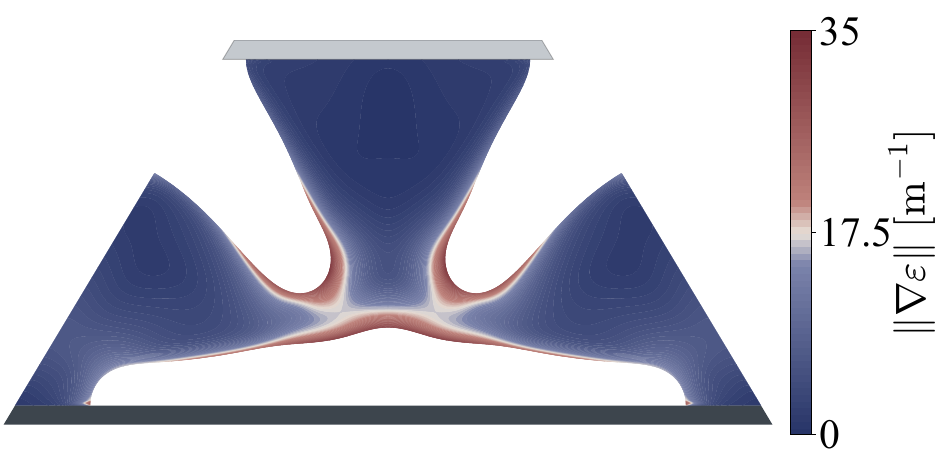}}
    \caption{}
  \end{subfigure}%
  \hspace{0.12\linewidth}%
  \begin{subfigure}[t]{0.30\linewidth}
    \centering
    \makebox[\linewidth][l]{%
      \includegraphics[width=1.21036585\linewidth,trim=1.700787bp 0bp 1.700787bp 0bp,clip]{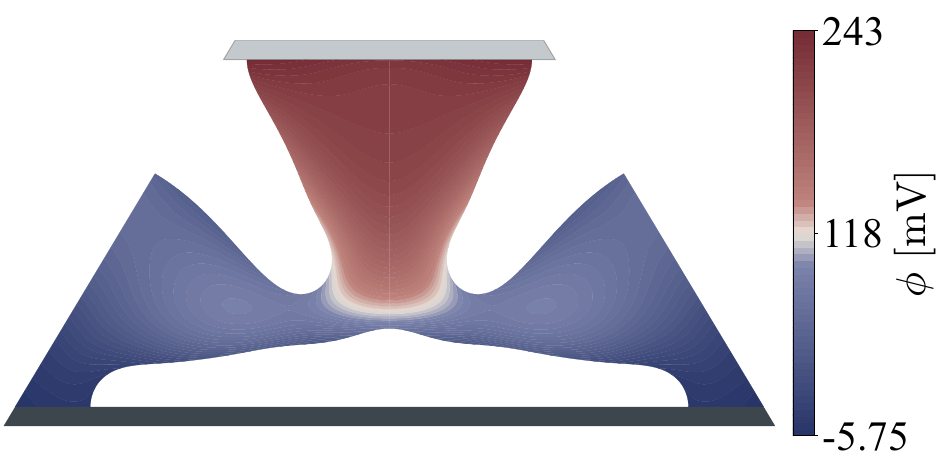}}
    \caption{}
  \end{subfigure}

  \captionsetup[subfigure]{skip=3pt}
  \begin{subfigure}[t]{0.30\linewidth}
    \centering
    \includegraphics[width=\linewidth,trim=1.700787bp 1.700787bp 1.700787bp 1.700787bp,clip]{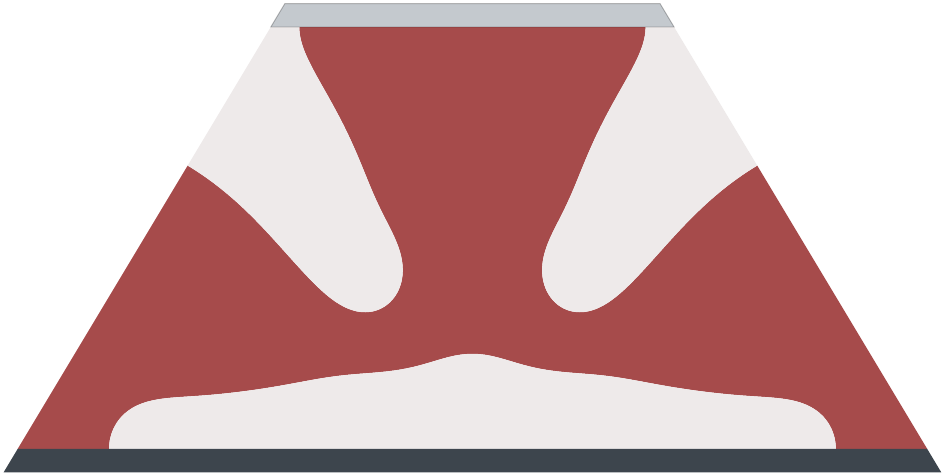}
    \caption{}
  \end{subfigure}%
  \hspace{0.12\linewidth}%
  \begin{subfigure}[t]{0.30\linewidth}
    \centering
    \includegraphics[width=\linewidth,trim=1.700787bp 1.700787bp 1.700787bp 1.700787bp,clip]{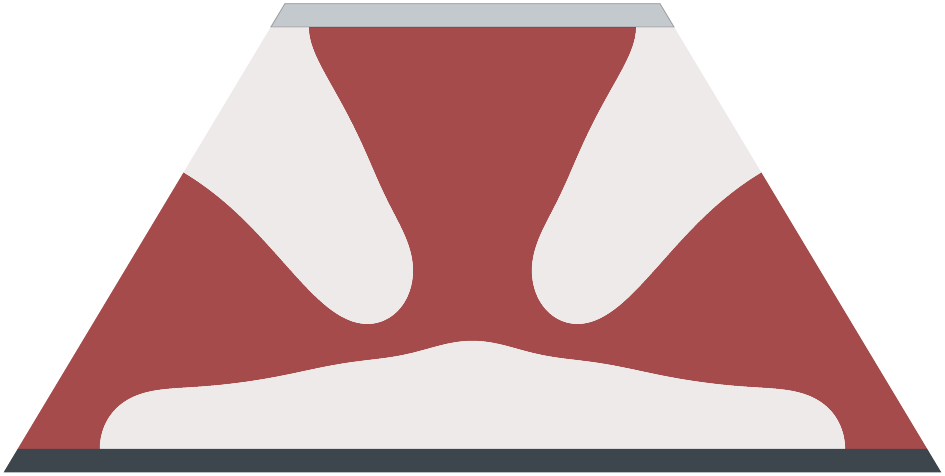}
    \caption{}
  \end{subfigure}
  \caption{Material configurations and electromechanical response at step 600: (a) strain-gradient norm distribution; (b) electric potential distribution; (c) nominal configuration; (d) eroded configuration.}
  \label{fig:a1-manufacturing-comparison}
\end{figure}

To examine the relation between nonuniform deformation and the flexoelectric response, spatial variations in strain are quantified by the strain-gradient norm $G_\varepsilon=(\varepsilon_{ij,k}\varepsilon_{ij,k})^{1/2}$.
Fig.~\ref{fig:a1-manufacturing-comparison}(a) shows that high values of $G_\varepsilon$ are concentrated near the tips of the lateral indentations and the upper boundary of the bottom void.
These regions coincide with local reductions in the width of the solid regions, indicating that deformation becomes localized in these narrow regions, with more pronounced spatial variations in strain.
In Fig.~\ref{fig:a1-manufacturing-comparison}(b), the electric potential is high in the upper region and decreases markedly toward the bottom supports. Regions of large strain gradients lie predominantly near the transition between high and low electric potentials.
Sensitivity to inward shifts of the material boundaries is examined by comparing the nominal and eroded realizations. As shown in Fig.~\ref{fig:a1-manufacturing-comparison}(c)(d), erosion enlarges the lateral indentations and the bottom void, while the solid connections in the center and on both sides remain intact. Furthermore, Fig.~\ref{fig:a1-optimization-port-history}(b) shows that the output voltages of the two configurations are similar. This indicates that the material reduction associated with manufacturing variations does not substantially degrade the voltage response of the optimized design.
These results indicate that a higher voltage output can be achieved with reduced sensitivity to inward shifts of the solid boundaries when the eroded realization is included in the optimization.

\FloatBarrier

\subsection{Case 2: Two-dimensional bending sensor}\label{subsec:case-bending}

\begin{figure}[!bp]
  \centering
  \par\medskip
  \captionsetup[subfigure]{skip=3pt}
  \begin{subfigure}[t]{0.90\linewidth}
    \centering
    \makebox[\linewidth][l]{\hspace*{0.02778\linewidth}\includegraphics[width=0.94444\linewidth]{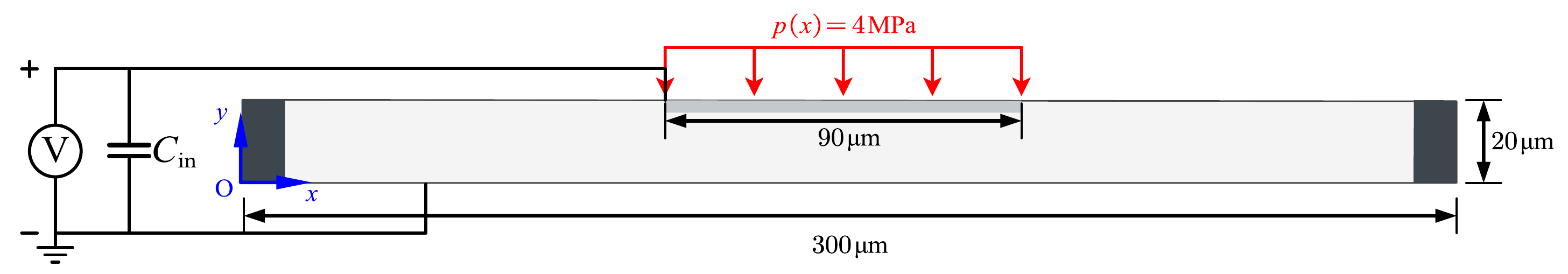}}
    \caption{}
  \end{subfigure}
  \par\medskip
  \begin{subfigure}[t]{0.90\linewidth}
    \centering
    \makebox[\linewidth][l]{\hspace*{0.0155\linewidth}\includegraphics[width=0.9623\linewidth]{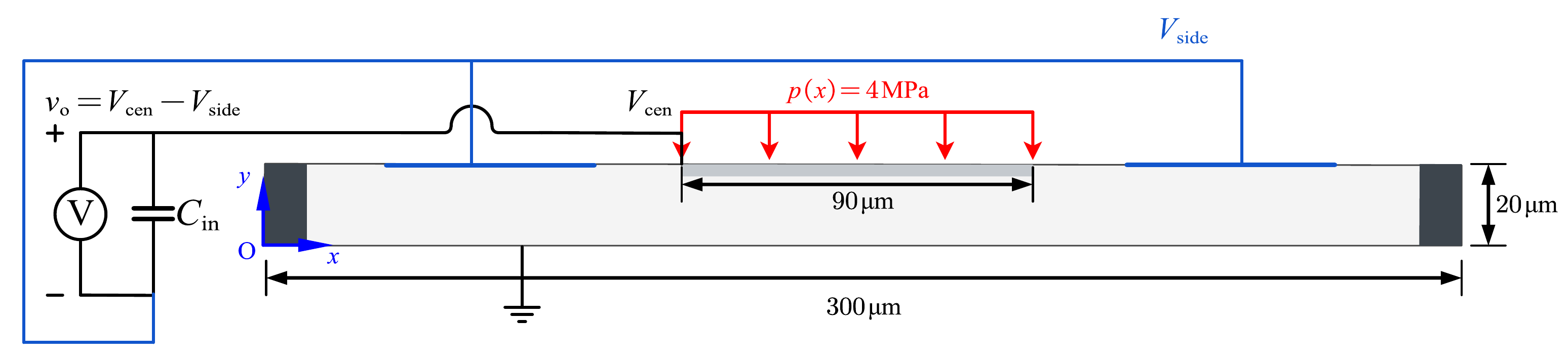}}
    \caption{}
  \end{subfigure}
  \par\medskip
  \begin{subfigure}[t]{0.90\linewidth}
    \centering
    \makebox[\linewidth][l]{\hspace*{0.1705\linewidth}\includegraphics[width=0.733\linewidth]{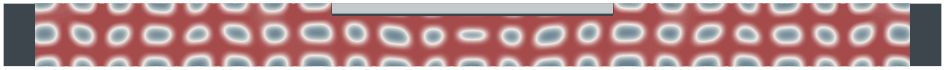}}
    \caption{}
  \end{subfigure}
  \caption{Model setup for the two-dimensional bending sensor: (a)(b) electrode layouts for E1 and E2, respectively; (c) common initial material distribution. In both configurations, the central electrode occupies $x\in[105,195]\,\si{\micro\metre}$ on the top surface. In E2, two additional electrodes occupy $x\in[30,85]\,\si{\micro\metre}$ and $x\in[215,270]\,\si{\micro\metre}$, respectively. Both configurations share the same non-design solid regions.}
  \label{fig:a2-geometry-initial}
\end{figure}

Consider the two-dimensional beam-type bending sensor in Fig.~\ref{fig:a2-geometry-initial} with the plane-strain assumption.
The device is $\SI{300}{\micro\metre}$ long and $\SI{20}{\micro\metre}$ high. 
Quantities such as energy, charge, and capacitance are expressed per unit out-of-plane thickness.
The coordinate origin is located at the lower left corner of the beam. Displacements in both the $x$- and $y$-directions are fixed at the ends $x=0$ and $x=\SI{300}{\micro\metre}$.
A uniform pressure of $4\,\mathrm{MPa}$ is applied in the negative $y$-direction over the central interval $x\in[105,195]\,\si{\micro\metre}$ of the top surface.
Solid regions of width $\SI{10}{\micro\metre}$ at both ends and a solid layer of thickness $\SI{4}{\micro\metre}$ beneath the loaded region are retained as non-design regions.
Only the left half-domain $x\in[0,150]\,\si{\micro\metre}$ is analyzed, with $u_x=0$, $\partial_xu_y=0$, and $\partial_x\phi=0$ imposed on the symmetry plane.
The remaining unspecified higher-order boundary conditions are prescribed in the homogeneous natural form given in Section~\ref{subsec:flexo-variational-form}.

The device is composed of $\mathrm{Ba}_{1-x}\mathrm{Sr}_x\mathrm{TiO_3}$, with the void regions represented by a weak ersatz material. The material parameters are interpolated according to Eq.~\eqref{eqn:mat_interp}.
The reference length is $L_0=\SI{20}{\micro\metre}$, and the reference voltage is $V_0=L_0\sqrt{Y_0/\kappa_0}$.
Two electrode configurations are shown in Fig.~\ref{fig:a2-geometry-initial}(a)(b).
Configuration E1 has a single sensing electrode at the center of the top surface. Configuration E2 has three sensing electrodes, one at the center and one on each side.
Each electrode is equipotential. In both configurations, the bottom edge is grounded. Homogeneous natural electrical boundary conditions are prescribed on the parts of the top edge not covered by electrodes.
For E1, $\boldsymbol s_V=1$. For E2, $\boldsymbol s_V=(-1,1)^{\top}$, with the entries corresponding to the lateral and central electrodes, respectively.
The parameters shared by E1 and E2 are listed in Table~\ref{tab:a2-parameters}.

\begin{table}[!tbp]
  \centering
  \caption{Common parameters for the two-dimensional bending sensor.}
  \label{tab:a2-parameters}
  \small
  \setlength{\tabcolsep}{5pt}
  \renewcommand{\arraystretch}{1.00}
  \begin{tabular*}{\linewidth}{@{\extracolsep{\fill}}>{\raggedright\arraybackslash}p{0.34\linewidth}lll@{}}
    \toprule
    Parameter & Symbol & Value & Unit \\
    \midrule
    Young's modulus & $(Y_0,Y_-)$ & $(1.53\times10^{11},\,3.06\times10^8)$ & $\mathrm{Pa}$ \\
    Poisson's ratio & $(\nu_0,\nu_-)$ & $(0.30,\,0.30)$ & --- \\
    Dielectric permittivity & $(\kappa_0,\kappa_-)$ & $(7.08\times10^{-8},\,1.42\times10^{-10})$ & $\mathrm{F/m}$ \\
    Flexoelectric coefficients & $(f_{1,0},f_{2,0},f_{1,-},f_{2,-})$ & $(3.0\times10^{-5},\,3.0\times10^{-5},\,0,\,0)$ & $\mathrm{C/m}$ \\
    Interpolation exponents & $(p_Y,p_f,p_\kappa)$ & $(3,\,2,\,1)$ & --- \\
    Length scales & $(l_m,l_e,l_c)$ & $(2,\,2,\,1.667)$ & $\si{\micro\metre}$ \\
    Helmholtz filter radius & $r_f$ & $1.927$ & $\si{\micro\metre}$ \\
    Volume fraction & $m_c$ & $0.68946$ & --- \\
    Input capacitance & $C_{\mathrm{in}}$ & $2.98655\times10^{-7}$ & $\mathrm{F/m}$ \\
    Reference voltage & $V_{\mathrm{ref}}$ & $93.9$ & $\mathrm{mV}$ \\
    Phase-field regularization & $\gamma L_0^2$ & $15$ & --- \\
    Erosion projection parameters & $(\beta,\eta_{\mathrm e})$ & $(5,\,0.625)$ & --- \\
    Spectral interpolation parameters & $(K_-,M_-,r_S)$ & $(10^{-3},\,10^{-3},\,6)$ & --- \\
    Spectral constraint thresholds & $(\lambda_{\mathrm{req,e}},\lambda_{\mathrm{req,i}})$ & $(0.002587,\,0.003248)$ & --- \\
    \bottomrule
  \end{tabular*}
  \par\smallskip
\end{table}

\begin{figure}[!tb]
  \centering
  \captionsetup[subfigure]{skip=3pt}
  \par\medskip
  \begin{subfigure}{\linewidth}
    \centering
    \includegraphics[width=0.48\linewidth]{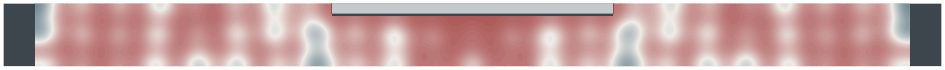}\hfill
    \includegraphics[width=0.48\linewidth]{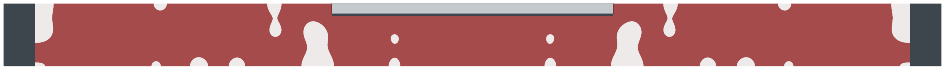}
    \caption{}
  \end{subfigure}
  \par\vspace{4mm}
  \begin{subfigure}{\linewidth}
    \centering
    \includegraphics[width=0.48\linewidth]{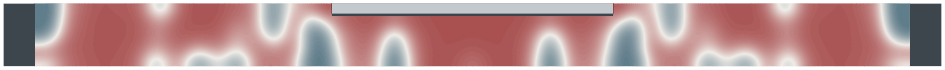}\hfill
    \includegraphics[width=0.48\linewidth]{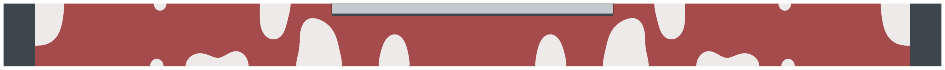}
    \caption{}
  \end{subfigure}
  \par\vspace{4mm}
  \begin{subfigure}{\linewidth}
    \centering
    \includegraphics[width=0.48\linewidth]{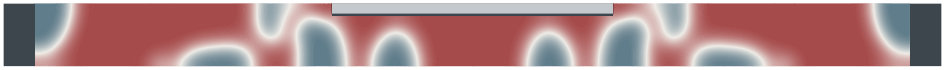}\hfill
    \includegraphics[width=0.48\linewidth]{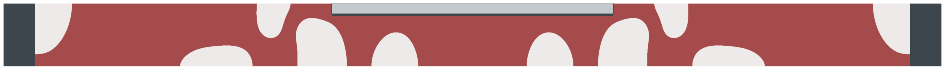}
    \caption{}
  \end{subfigure}
  \par\vspace{4mm}
  \begin{subfigure}{\linewidth}
    \centering
    \includegraphics[width=0.48\linewidth]{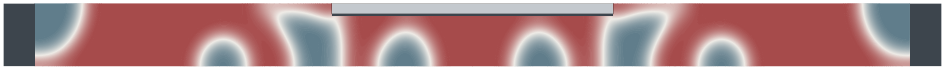}\hfill
    \includegraphics[width=0.48\linewidth]{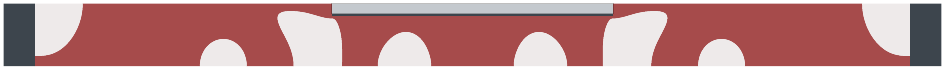}
    \caption{}
  \end{subfigure}
  \caption{Evolution of the material distribution in the two-dimensional bending sensor: configuration E1. The left column shows the continuous phase field of the nominal realization, and the right column shows the corresponding binary configurations. Panels (a) to (d) correspond to accepted steps 150, 300, 600, and 3200, respectively.}
  \label{fig:a2-e1-phase-evolution}
\end{figure}

\begin{figure}[!tb]
  \centering
  \captionsetup[subfigure]{skip=3pt}
  \par\medskip
  \begin{subfigure}{\linewidth}
    \centering
    \includegraphics[width=0.48\linewidth]{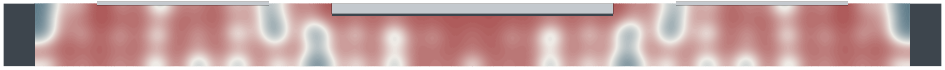}\hfill
    \includegraphics[width=0.48\linewidth]{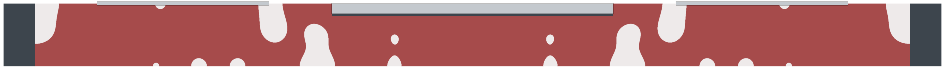}
    \caption{}
  \end{subfigure}
  \par\vspace{4mm}
  \begin{subfigure}{\linewidth}
    \centering
    \includegraphics[width=0.48\linewidth]{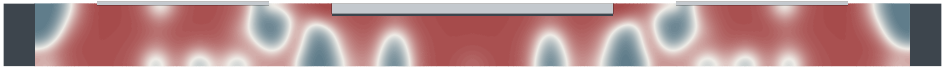}\hfill
    \includegraphics[width=0.48\linewidth]{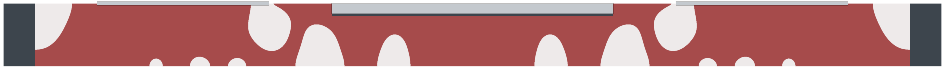}
    \caption{}
  \end{subfigure}
  \par\vspace{4mm}
  \begin{subfigure}{\linewidth}
    \centering
    \includegraphics[width=0.48\linewidth]{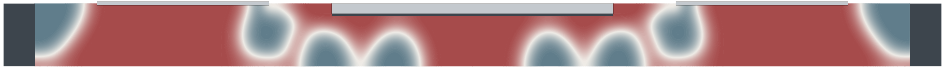}\hfill
    \includegraphics[width=0.48\linewidth]{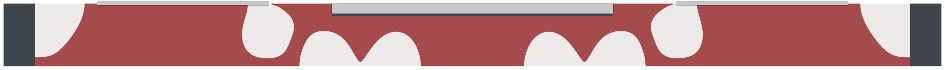}
    \caption{}
  \end{subfigure}
  \par\vspace{4mm}
  \begin{subfigure}{\linewidth}
    \centering
    \includegraphics[width=0.48\linewidth]{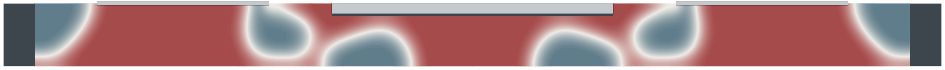}\hfill
    \includegraphics[width=0.48\linewidth]{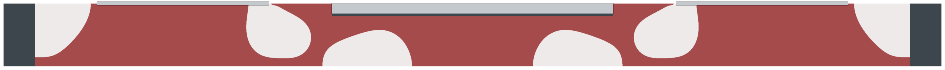}
    \caption{}
  \end{subfigure}
  \caption{Evolution of the material distribution in the two-dimensional bending sensor: configuration E2. The left column shows the continuous phase field of the nominal realization, and the right column shows the corresponding binary configurations. Panels (a) to (d) correspond to accepted steps 150, 300, 800, and 4305, respectively.}
  \label{fig:a2-e2-phase-evolution}
\end{figure}

To separate the contributions of source charge and capacitance changes to the port voltage, the equivalent capacitance and equivalent source charge are defined as
\begin{equation}
  C_{\mathrm{eq}}=\left(\boldsymbol s_V^{\top}
  \boldsymbol C_{\mathrm p}^{-1}\boldsymbol s_V\right)^{-1},\quad
  Q_{\mathrm{src,eq}}=C_{\mathrm{eq}}\boldsymbol s_V^{\top}
  \boldsymbol C_{\mathrm p}^{-1}\boldsymbol Q_{\mathrm{src}}.
\label{eqn:a2-equivalent-port-quantities}
\end{equation}
For E1, these quantities reduce to the single-port capacitance $C_{\mathrm p}$ and source charge $Q_{\mathrm{src}}$, respectively.
Hence, the port voltage is obtained by the charge balance relation~\eqref{eqn:electrode-balance}
\begin{equation}
  v_{\mathrm o}=-\frac{Q_{\mathrm{src,eq}}}{C_{\mathrm{eq}}+C_{\mathrm{in}}}.
\end{equation}
Define the following factors
\begin{equation}
  g_V = \frac{v_{\mathrm o}}{v_{\mathrm o}^{(0)}}, \quad
  g_Q = \frac{Q_{\mathrm{src,eq}}}{Q_{\mathrm{src,eq}}^{(0)}}, \quad
  g_C = \frac{C_{\mathrm{eq}}^{(0)}+C_{\mathrm{in}}}
  {C_{\mathrm{eq}}+C_{\mathrm{in}}}.
\end{equation}
Here, $g_V$ and $g_Q$ denote the voltage gain and equivalent source charge gain, respectively.
The factor $g_C$ quantifies the contribution of changes in equivalent capacitance to the voltage gain, with $g_V=g_Q g_C$.
Superscript $(0)$ denotes the initial configuration. The realization subscript $\omega$ is omitted because the analysis is restricted to the nominal realizations.
The material volume fraction constraint is imposed within the design domain of each configuration. Spectral connectivity constraints anchored to the left end of the half-domain are imposed on both the nominal and eroded realizations.
The half-domain is discretized using $450\times60$ quadratic elements, with three-point Gauss quadrature in each direction.
\begin{figure}[!tb]
  \centering
  \captionsetup[subfigure]{skip=2pt}
  \begin{subfigure}[t]{0.49\linewidth}
    \centering
    \includegraphics[width=\linewidth]{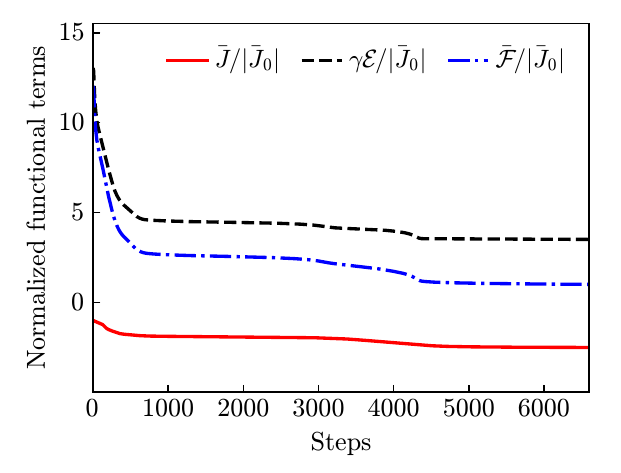}
    \caption{}
  \end{subfigure}\hfill
  \begin{subfigure}[t]{0.49\linewidth}
    \centering
    \includegraphics[width=\linewidth]{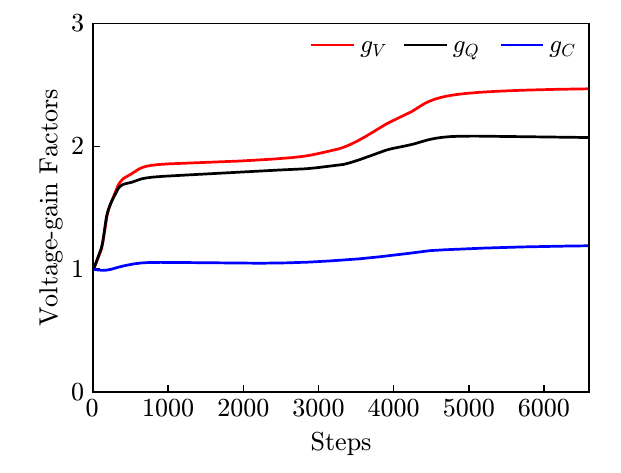}
    \caption{}
  \end{subfigure}
  \par\vspace{1mm}
  \begin{subfigure}[t]{0.49\linewidth}
    \centering
    \includegraphics[width=\linewidth]{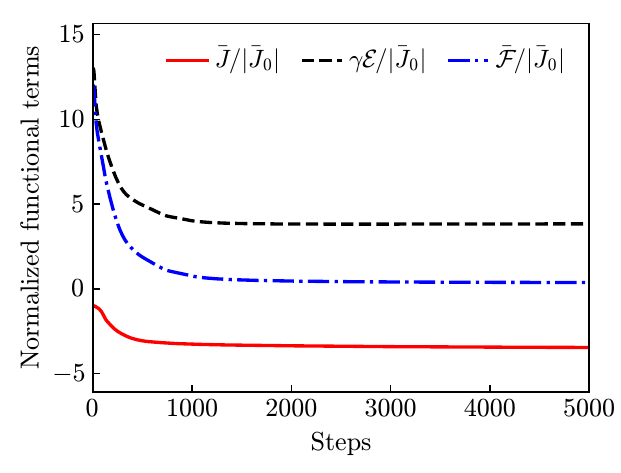}
    \caption{}
  \end{subfigure}\hfill
  \begin{subfigure}[t]{0.49\linewidth}
    \centering
    \includegraphics[width=\linewidth]{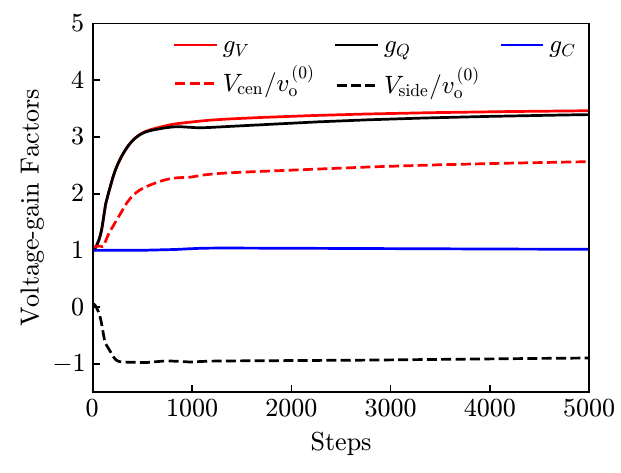}
    \caption{}
  \end{subfigure}
  \caption{Optimization history and voltage gain decomposition for the two-dimensional bending sensor. The top and bottom rows correspond to E1 and E2, respectively: (a)(c) normalized voltage objective, phase-field regularization term, and total objective functional; (b)(d) voltage gain $g_V$, equivalent source charge gain $g_Q$, and capacitive loading factor $g_C$ for the nominal realization. Panel (d) also shows the normalized central and side electrode potentials, $V_{\mathrm{cen}}/v_{\mathrm o}^{(0)}$ and $V_{\mathrm{side}}/v_{\mathrm o}^{(0)}$, respectively.}
  \label{fig:a2-optimization-port-history}
\end{figure}

To investigate the additional improvement of voltage gain using multiple electrodes, E1 and E2 are initialized with the same material distribution and equal nominal voltages.
It is usually difficult to prescribe an initial design that produces the same initial voltages for the two electrode configurations.
Therefore, a random material distribution is generated and then adjusted through a preliminary optimization to match their nominal voltages. 
This preliminary step solves the following constrained optimization problem
\begin{equation}
  \begin{aligned}
    \min_{c\in\mathcal K}\quad&
    \frac12
    \left[
    \frac{v_{\mathrm{o,i}}^{\mathrm{E2}}(c)
    -v_{\mathrm{o,i}}^{\mathrm{E1}}(c)}
    {V_{\mathrm{ref}}}
    \right]^2,\\
    \mathrm{s.t.}\quad&
    g_{\mathrm v}(c)=0, \quad g_{\mathrm{s},\omega}(c)\leq0,
    \quad \omega\in\{\mathrm i,\mathrm e\}.
  \end{aligned}
\label{eqn:case-a2-pre-opt}
\end{equation}
The resulting feasible design is illustrated in Fig.~\ref{fig:a2-geometry-initial}(c), which is adopted when the convergence criterion $|v_{\mathrm{o,i}}^{\mathrm{E2}}(c)-v_{\mathrm{o,i}}^{\mathrm{E1}}(c)|<1\,\mathrm{mV}$ for problem \eqref{eqn:case-a2-pre-opt} is satisfied.

\begin{figure}[!tb]
  \centering
  \captionsetup[subfigure]{skip=2pt}
  \begin{subfigure}[t]{0.70\linewidth}
    \centering
    \includegraphics[width=\linewidth]{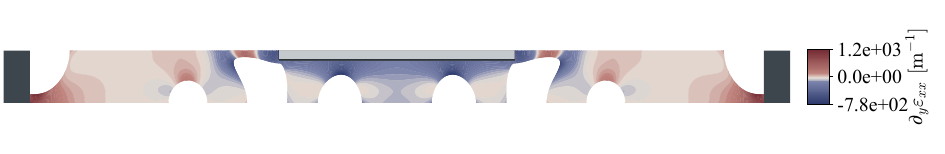}
    \caption{}
  \end{subfigure}
  \begin{subfigure}[t]{0.70\linewidth}
    \centering
    \includegraphics[width=\linewidth]{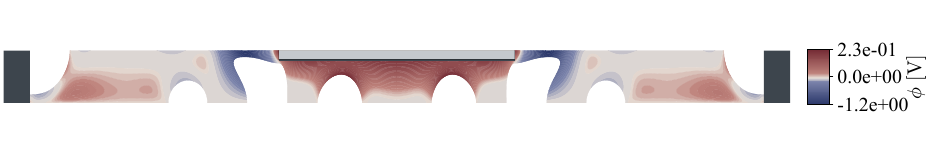}
    \caption{}
  \end{subfigure}
  \par\vspace{2mm}
  \begin{subfigure}[t]{0.70\linewidth}
    \centering
    \includegraphics[width=\linewidth]{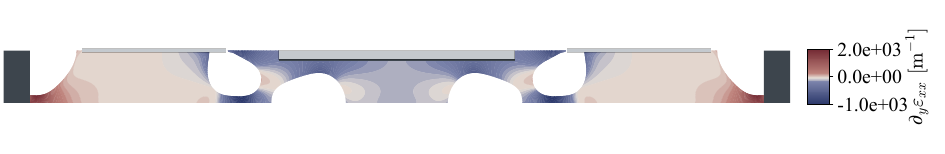}
    \caption{}
  \end{subfigure}
  \begin{subfigure}[t]{0.70\linewidth}
    \centering
    \includegraphics[width=\linewidth]{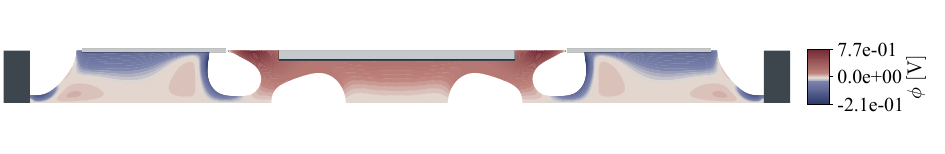}
    \caption{}
  \end{subfigure}
  \caption{Local field distributions in the nominal realizations of the two-dimensional bending sensor. Panels (a) and (b) correspond to E1 at accepted step 3200, and panels (c) and (d) to E2 at accepted step 4305. Panels (a) and (c) show the axial strain gradient $\partial_y\varepsilon_{xx}$, while panels (b) and (d) show the electric potential $\phi$.}
  \label{fig:a2-gradient-potential}
\end{figure}

The evolution of the material distribution for the two electrode configurations is shown in Figs.~\ref{fig:a2-e1-phase-evolution} and~\ref{fig:a2-e2-phase-evolution}, respectively.
In both configurations, the initially dispersed small voids gradually develop into several larger voids beneath the central loaded region, while indentations form near the clamped ends.
Later changes mainly involve adjustments to the widths of local solid connections.
The narrow solid connections promote localized bending deformation near the clamped ends, loading edges and electrodes.
Fig.~\ref{fig:a2-optimization-port-history}(a)(c) shows that the phase-field regularization term decreases rapidly as small voids merge, then declines more gradually.
Starting from the same nominal voltages, E2 reaches a final voltage of 321.4 mV, which is 38.8\% higher than the 231.6 mV achieved by E1.
The equivalent source charge gain $g_Q$ is higher in E2 than in E1. Capacitance reduction contributes more to the voltage gain in E1, whereas the capacitive loading factor remains close to unity in E2.
As shown in Fig.~\ref{fig:a2-optimization-port-history}(d), potentials of opposite signs are obtained at the central electrode and the interconnected lateral electrodes, resulting in a larger voltage difference between them.
These results indicate that optimization with multiple electrodes achieves a greater voltage gain than the single-electrode configuration.

The gradient of the axial strain in the $y$-direction $\partial_y\varepsilon_{xx}$ for both configurations is illustrated in Fig.~\ref{fig:a2-gradient-potential}(a)(c).
High values of $|\partial_y\varepsilon_{xx}|$ are concentrated near the supports and loading edges and within the narrow solid connections around the voids.
The gradient is predominantly negative beneath the loaded region but locally positive near the supports.

In Fig.~\ref{fig:a2-gradient-potential}(b)(d), positive potentials are observed beneath the central electrode in both configurations.
In E1, negative potentials occur mainly in localized regions on either side of the loaded region, whereas in E2 these regions extend beneath the lateral electrodes.
In E2, the regions of negative potential are spatially separated from the more continuous region of positive potential beneath the central electrode.
These potential distributions provide a spatial explanation for the higher voltage gain obtained for E2.

The target voltage is determined by the prescribed electrode locations and connections, which also influence the material redistribution during optimization through the electrical boundary conditions and the sensitivities of the objective functional.
Therefore, electrode locations and connections can be selected to increase the voltage difference between the measurement terminals.

\FloatBarrier

\subsection{Case 3: Three-dimensional thin-film device}\label{subsec:case-bending-3d}

\begin{figure}[!b]
  \centering
  \captionsetup[subfigure]{skip=3pt}
  \begin{subfigure}[c]{0.50\linewidth}
    \centering
    \includegraphics[width=\linewidth]{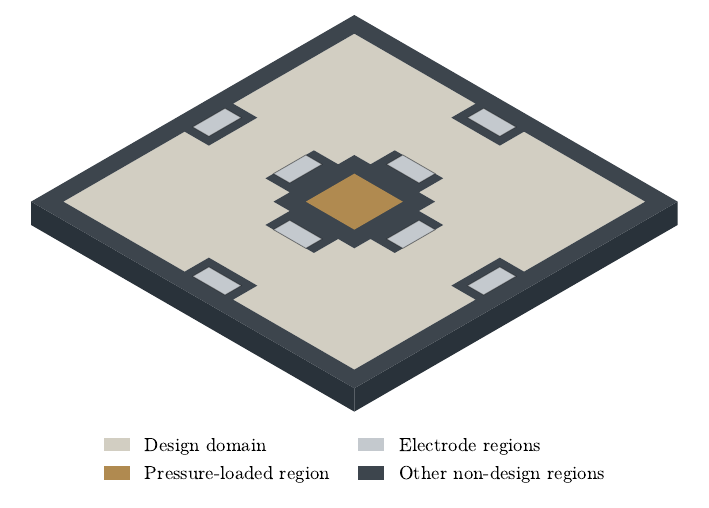}
    \caption{}
    \label{fig:a3-model-overview}
  \end{subfigure}\hspace{0.05\linewidth}%
  \begin{minipage}[c]{0.18\linewidth}
    \begin{subfigure}{\linewidth}
      \centering
      \includegraphics[width=\linewidth]{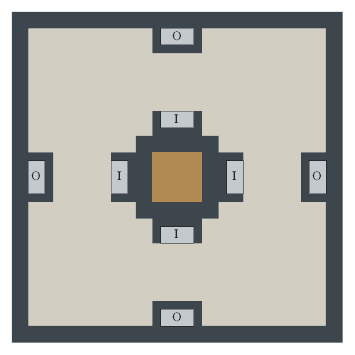}
      \caption{}
      \label{fig:a3-top-view}
    \end{subfigure}
    \par\smallskip
    \begin{subfigure}{\linewidth}
      \centering
      \includegraphics[width=\linewidth]{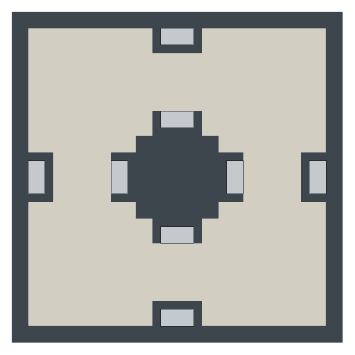}
      \caption{}
      \label{fig:a3-bottom-view}
    \end{subfigure}
  \end{minipage}
  \caption{Model setup for the three-dimensional thin-film device: (a) overall view; (b) top view showing the floating electrodes, with I and O denoting the inner and outer electrode groups, respectively. Electrodes in each group are electrically connected to form an equipotential port. (c) Bottom view showing the grounded electrodes aligned with the top electrodes. The design domain is beige, and the non-design solid regions are dark gray. Electrodes are light gray, and the pressure-loaded area on the top surface is gold. Zero displacement is prescribed on all four outer lateral faces.}
  \label{fig:a3-geometry-initial}
\end{figure}

Consider the three-dimensional thin-film device shown in Fig.~\ref{fig:a3-geometry-initial}(a).
The device occupies the domain $[-240,240]^2\times[-15,15]\,\si{\micro\metre}$. 
The length of the square midsurface is $\SI{480}{\micro\metre}$, and the thickness is $h=\SI{30}{\micro\metre}$.
Zero displacement is prescribed on all four outer lateral faces. A localized pressure $p_3(x,y)=p_0\mathfrak b(x/a_p)\mathfrak b(y/a_p)$ is applied at the center of the top surface in the negative $z$-direction, where
\begin{equation}
  \mathfrak b(\xi) =
  \begin{cases}
    (1-\xi^2)^2, & |\xi|\leq1, \\
    0, & |\xi|>1.
  \end{cases}
\end{equation}
The parameters are selected as $a_p=\SI{36}{\micro\metre}$ and $p_0=286.675\,\mathrm{MPa}$, giving a resultant force of magnitude $0.4227\,\mathrm N$ on the full device.
The non-design regions comprise an outer frame of width $\SI{24}{\micro\metre}$ and a central square region of $120\times120\,\si{\micro\metre\squared}$.
Four pairs of connection regions are also retained around the central region, with each pair consisting of an inner and an outer region.
These regions are extruded along the thickness direction to form the three-dimensional non-design domains.
One pair is formed by an inner connection region occupying $[60,96]\times[-36,36]\,\si{\micro\metre}$ and an outer connection region occupying $[180,216]\times[-36,36]\,\si{\micro\metre}$. 
The other three pairs are obtained by successive rotations of $90^\circ$ about the device center.
By symmetry, only the quarter-domain ($x\geq0,y\geq0$) is analyzed.
The conditions $u_x=0$ and $\partial_xu_y=\partial_xu_z=\partial_x\phi=0$ are imposed on $x=0$, while $u_y=0$ and $\partial_yu_x=\partial_yu_z=\partial_y\phi=0$ are imposed on $y=0$.

The elastic, dielectric and flexoelectric coefficients are the same as in Cases 1 and 2. 
The material and design parameters are listed in Table~\ref{tab:a3-parameters}.
The reference length is $L_0=\SI{30}{\micro\metre}$.
As shown in Fig.~\ref{fig:a3-geometry-initial}(b), one pair of inner and outer electrodes is placed on the top surface, occupying $[72,96]\times[-24,24]\,\si{\micro\metre}$ and $[192,216]\times[-24,24]\,\si{\micro\metre}$, respectively.
The other three pairs are obtained by rotating this pair through $90^\circ$, $180^\circ$, and $270^\circ$ about the device center.
The four inner electrodes are electrically connected to form an equipotential inner port, and the four outer electrodes are connected to form an equipotential outer port.
A grounded electrode of the same shape and size is placed on the bottom surface directly beneath each top electrode. 
Both ports satisfy the ideal open-circuit condition. 
The target voltage is defined using $\boldsymbol s_V=(-1,1)^{\top}$, with the entries ordered by the inner and outer ports.
The magnitude of the initial target voltage for the nominal realization $V_{\mathrm{ref}}=348.2\,\mathrm{mV}$ is used as the fixed normalization reference.
The design variable is constant in the thickness direction and invariant under reflection across $x=0$ and $y=0$ and interchange of $x,y$.
The quarter-domain is discretized using $20\times20\times4$ quadratic elements. Top and bottom views of the device are shown in Fig.~\ref{fig:a3-geometry-initial}(b)(c), respectively.

\begin{table}[!b]
  \centering
  \caption{Material and design parameters for the three-dimensional thin-film device.}
  \label{tab:a3-parameters}
  \small
  \setlength{\tabcolsep}{5pt}
  \renewcommand{\arraystretch}{1.00}
  \begin{tabular*}{\linewidth}{@{\extracolsep{\fill}}>{\raggedright\arraybackslash}p{0.34\linewidth}lll@{}}
    \toprule
    Parameter & Symbol & Value & Unit \\
    \midrule
    Young's modulus & $(Y_0,Y_-)$ & $(1.53\times10^{11},\,3.06\times10^8)$ & $\mathrm{Pa}$ \\
    Poisson's ratio & $(\nu_0,\nu_-)$ & $(0.30,\,0.30)$ & --- \\
    Dielectric permittivity & $(\kappa_0,\kappa_-)$ & $(7.08\times10^{-8},\,1.42\times10^{-10})$ & $\mathrm{F/m}$ \\
    Flexoelectric coefficients & $(f_{1,0},f_{2,0},f_{1,-},f_{2,-})$ & $(3.0\times10^{-5},\,3.0\times10^{-5},\,0,\,0)$ & $\mathrm{C/m}$ \\
    Interpolation exponents & $(p_Y,p_f,p_\kappa)$ & $(3,\,2,\,1)$ & --- \\
    \midrule
    Length scales & $(l_m,l_e,l_c)$ & $(10,\,10,\,8)$ & $\si{\micro\metre}$ \\
    Helmholtz filter radius & $r_f$ & $12$ & $\si{\micro\metre}$ \\
    Volume fraction & $m_c$ & $0.62504$ & --- \\
    Reference voltage & $V_{\mathrm{ref}}$ & $348.2$ & $\mathrm{mV}$ \\
    Phase-field regularization & $\gamma h L_0^2$ & $0.2$ & --- \\
    Erosion projection parameters & $(\beta,\eta_{\mathrm e})$ & $(5,\,0.625)$ & --- \\
    \midrule
    Spectral interpolation parameters & $(K_-,M_-,r_S)$ & $(10^{-3},\,10^{-3},\,6)$ & --- \\
    Spectral constraint thresholds & $(\lambda_{\mathrm{req,e}},\lambda_{\mathrm{req,i}})$ & $(0.002405,\,0.004829)$ & --- \\
    Spectral normalization scales & $(s_{\lambda,\mathrm e},s_{\lambda,\mathrm i})$ & $(0.006016,\,0.0009242)$ & --- \\
    \bottomrule
  \end{tabular*}
  \par\smallskip
\end{table}

\begin{figure}[!p]
  \centering
  \captionsetup[subfigure]{skip=3pt}
  \begin{subfigure}[t]{0.26\linewidth}
    \centering
    \includegraphics[width=\linewidth]{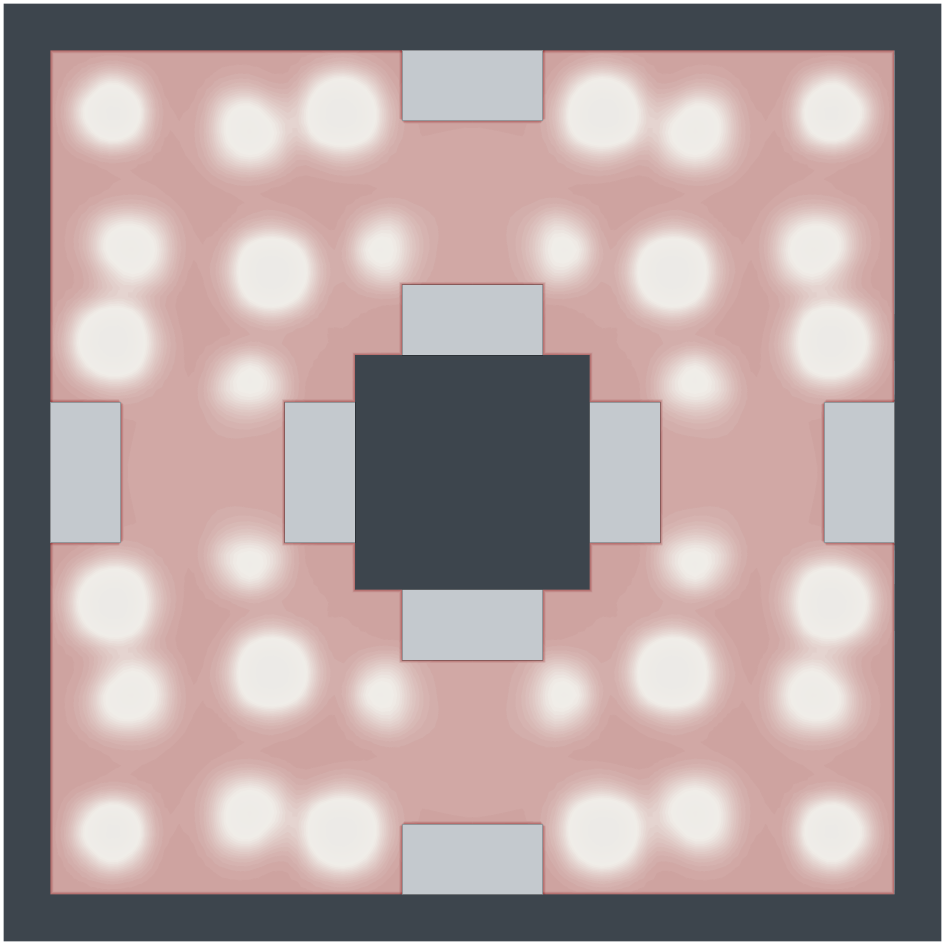}
    \caption{}
    \label{fig:a3-phase-100}
  \end{subfigure}\hspace{0.02\linewidth}%
  \begin{subfigure}[t]{0.26\linewidth}
    \centering
    \includegraphics[width=\linewidth]{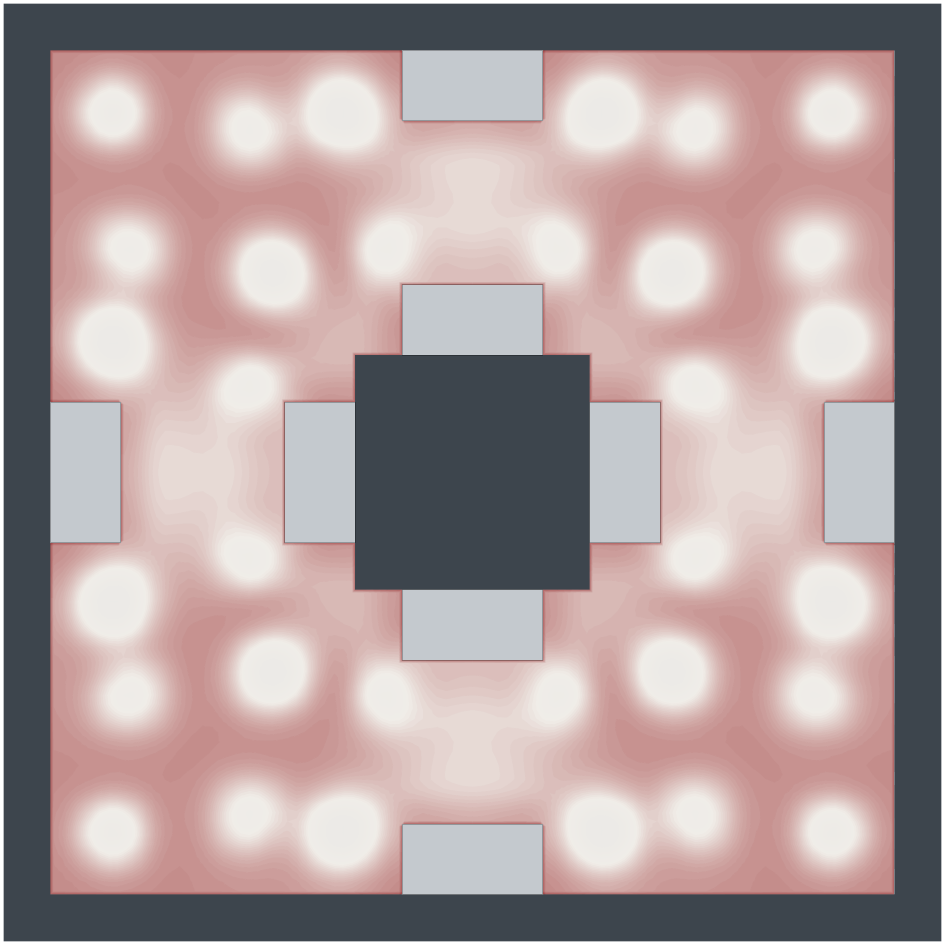}
    \caption{}
    \label{fig:a3-phase-1000}
  \end{subfigure}\hspace{0.02\linewidth}%
  \begin{subfigure}[t]{0.26\linewidth}
    \centering
    \includegraphics[width=\linewidth]{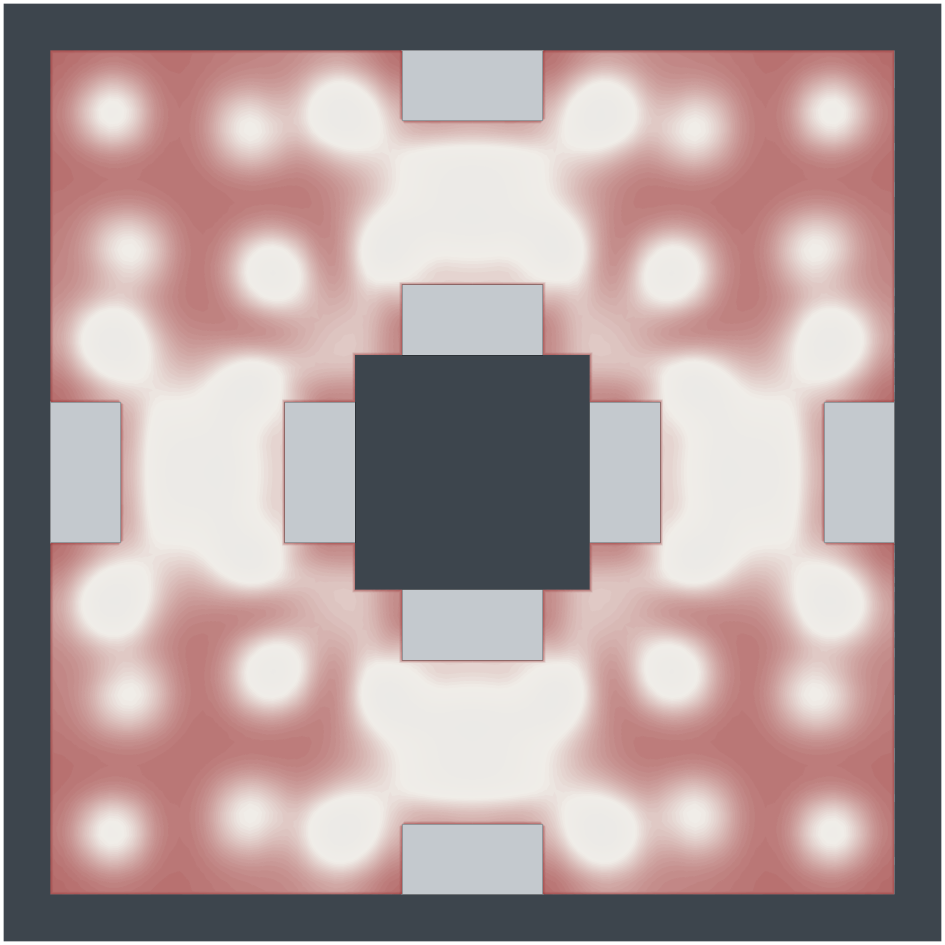}
    \caption{}
    \label{fig:a3-phase-2000}
  \end{subfigure}
  \par\medskip
  \begin{subfigure}[t]{0.26\linewidth}
    \centering
    \includegraphics[width=\linewidth]{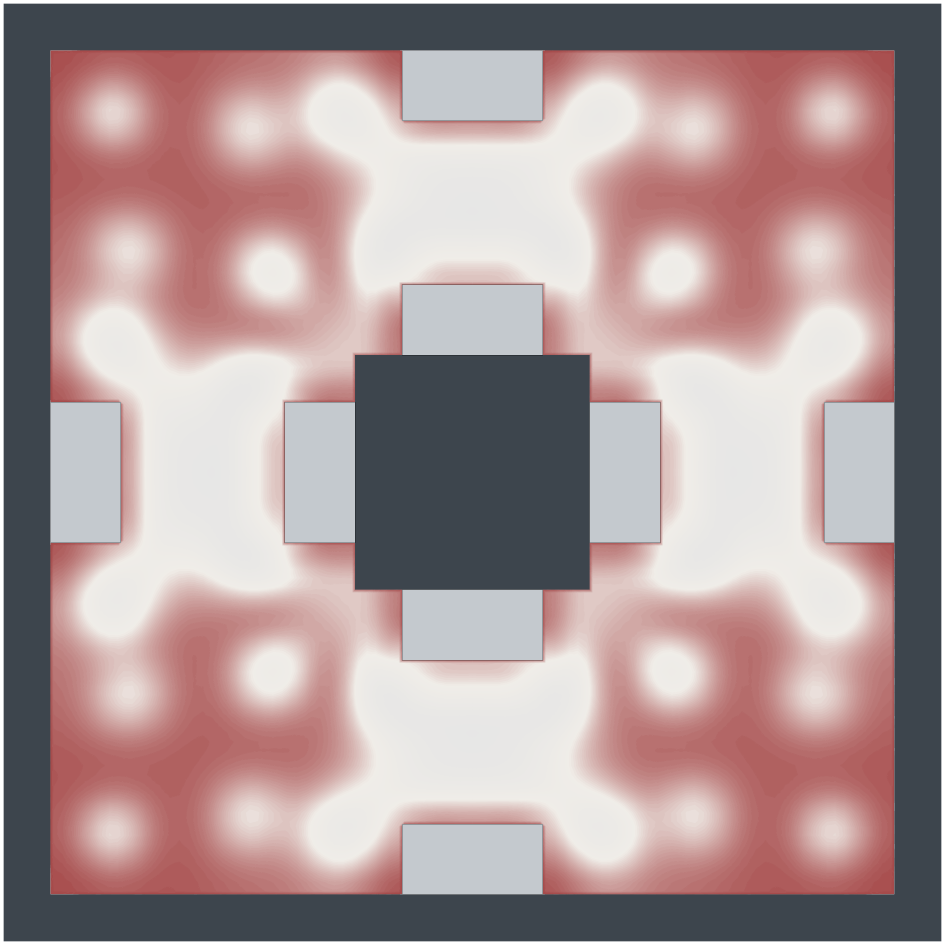}
    \caption{}
    \label{fig:a3-phase-3000}
  \end{subfigure}\hspace{0.02\linewidth}%
  \begin{subfigure}[t]{0.26\linewidth}
    \centering
    \includegraphics[width=\linewidth]{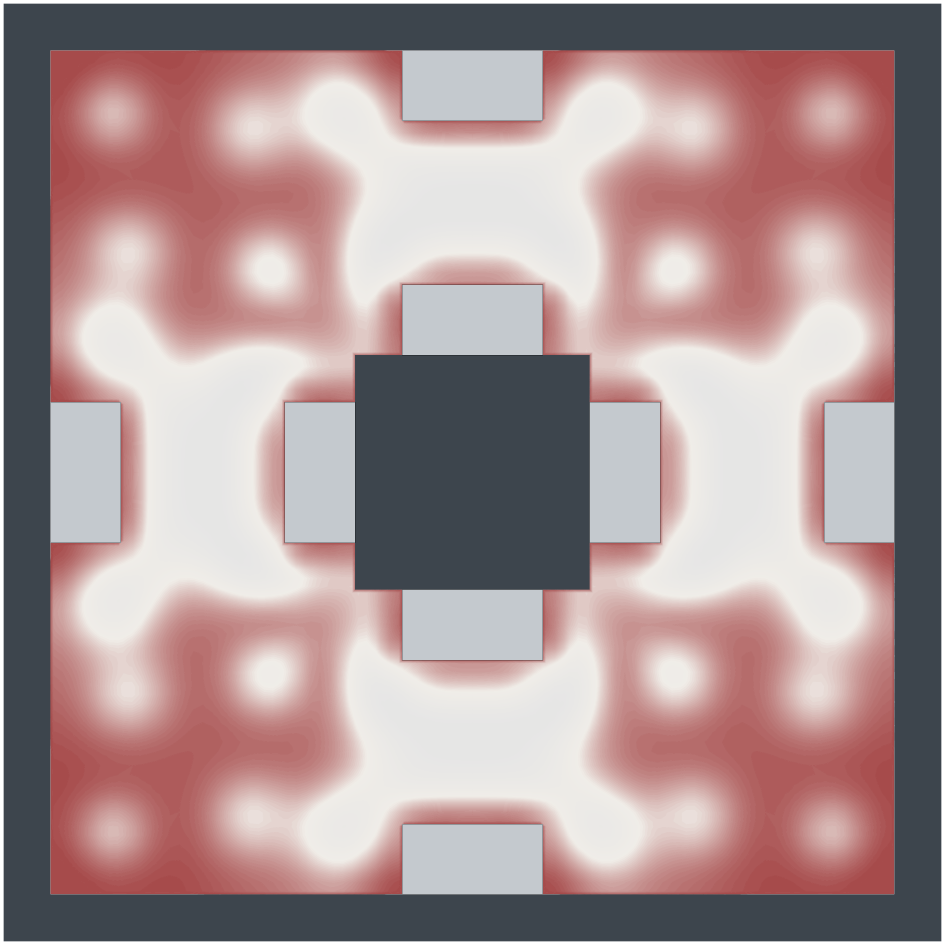}
    \caption{}
    \label{fig:a3-phase-4000}
  \end{subfigure}\hspace{0.02\linewidth}%
  \begin{subfigure}[t]{0.26\linewidth}
    \centering
    \includegraphics[width=\linewidth]{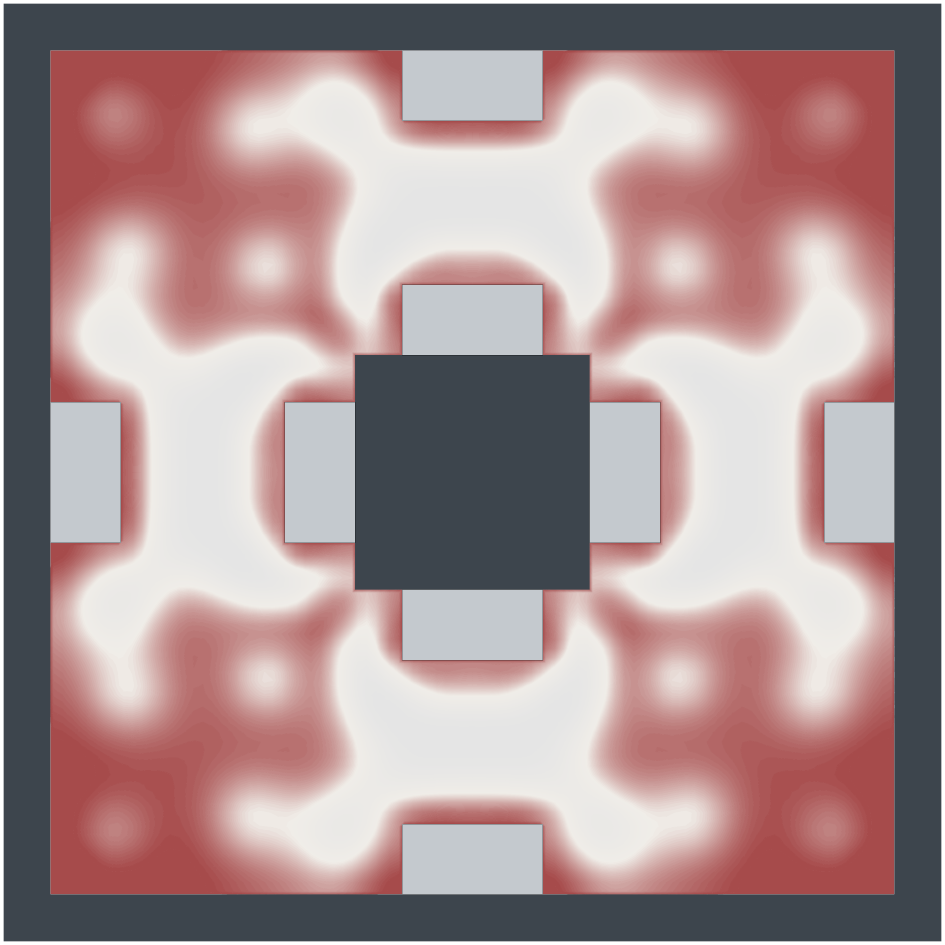}
    \caption{}
    \label{fig:a3-phase-5600}
  \end{subfigure}
  \caption{Evolution of the phase field in the three-dimensional thin-film device. Panels (a) to (f) correspond to steps 100, 1000, 2000, 3000, 4000, and 5600, respectively. Each panel shows the in-plane material distribution of the full device, with the design variable constant through the thickness.}
  \label{fig:a3-phase-evolution}
\end{figure}

\begin{figure}[!p]
  \centering
  \captionsetup[subfigure]{skip=3pt}
  \begin{subfigure}[t]{0.75\linewidth}
    \centering
    \makebox[\linewidth][l]{%
      \hspace*{0.08\linewidth}%
      \includegraphics[width=\linewidth]{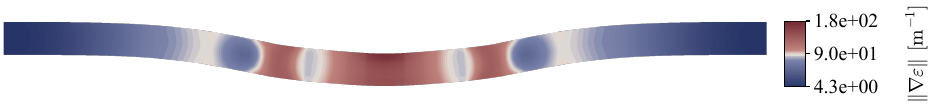}%
    }
    \caption{}
    \label{fig:a3-deformed-strain-gradient}
  \end{subfigure}

  \medskip
  \begin{subfigure}[t]{0.75\linewidth}
    \centering
    \makebox[\linewidth][l]{%
      \hspace*{0.08\linewidth}%
      \includegraphics[width=\linewidth]{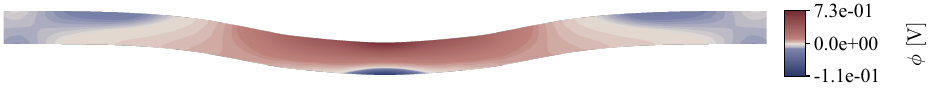}%
    }
    \caption{}
    \label{fig:a3-deformed-potential}
  \end{subfigure}

  \medskip
  \begin{subfigure}[t]{0.75\linewidth}
    \centering
    \makebox[\linewidth][l]{%
      \hspace*{0.08\linewidth}%
      \includegraphics[width=\linewidth]{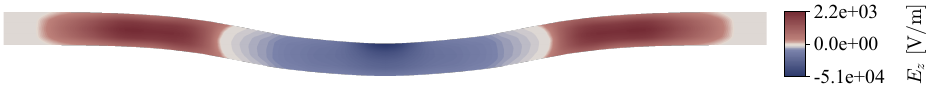}%
    }
    \caption{}
    \label{fig:a3-deformed-electric-field}
  \end{subfigure}
  \caption{Field distributions in the three-dimensional thin-film device at step 5600, shown on the deformed geometry: (a) strain-gradient norm; (b) electric potential; (c) electric field component $E_z$. Displacements are magnified by a factor of 20 for visualization.}
  \label{fig:a3-deformed-fields}
\end{figure}

\begin{figure}[!t]
  \centering
  \captionsetup[subfigure]{skip=3pt}
  \begin{subfigure}[t]{0.40\linewidth}
    \centering
    \includegraphics[width=\linewidth]{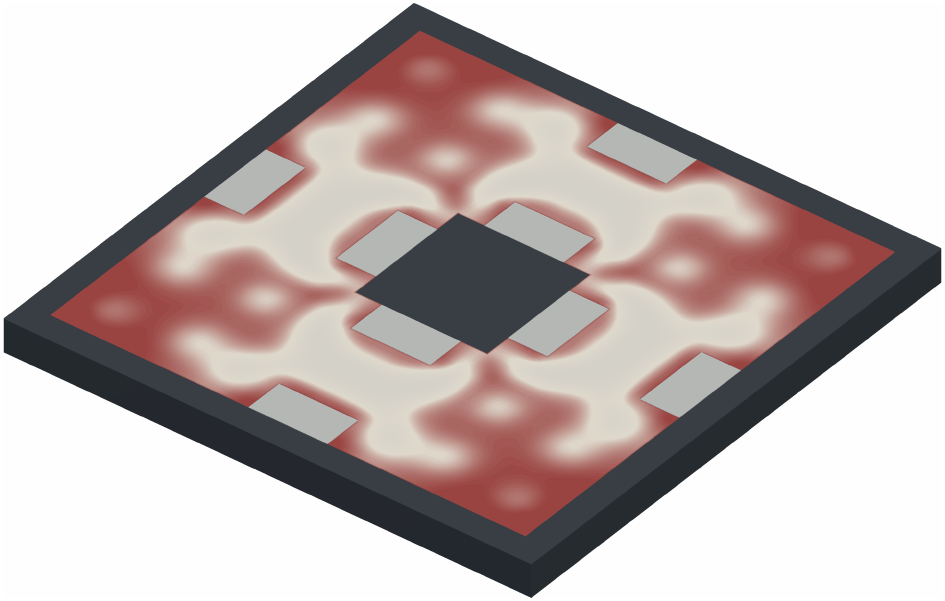}
    \caption{}
    \label{fig:a3-phase-xyz}
  \end{subfigure}\hspace{0.02\linewidth}%
  \begin{subfigure}[t]{0.40\linewidth}
    \centering
    \includegraphics[width=\linewidth]{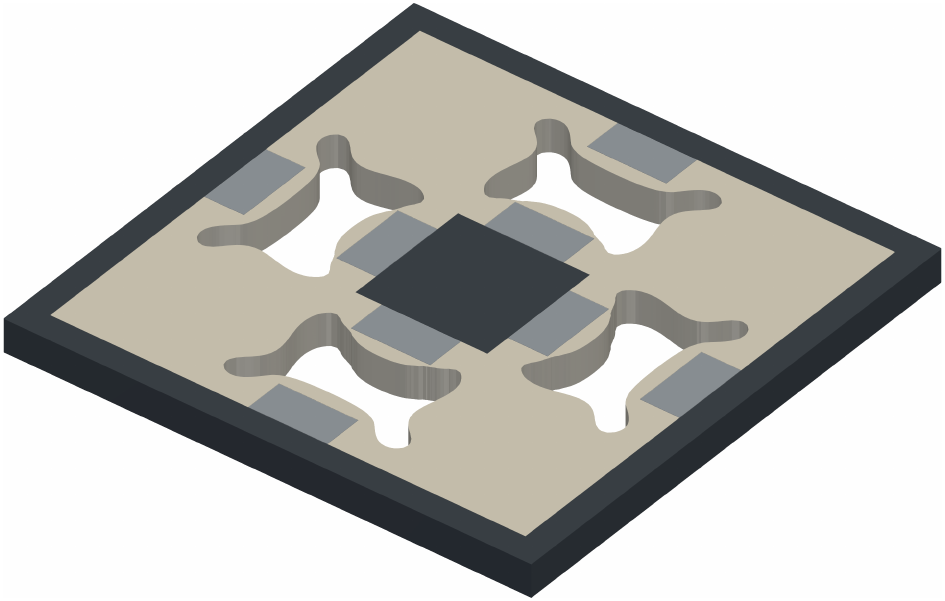}
    \caption{}
    \label{fig:a3-binary-xyz}
  \end{subfigure}
  \caption{Material distribution in the three-dimensional thin-film device at step 5600: (a) continuous phase field; (b) corresponding binary configuration. Both panels show three-dimensional views of the full device, with the non-design solid regions and electrode locations highlighted in separate colors.}
  \label{fig:a3-phase-binary-3d}
\end{figure}

\begingroup
\widowpenalty=10000
\clubpenalty=10000
Fig.~\ref{fig:a3-phase-evolution} presents $xy$-plane views of the evolving phase field in the three-dimensional thin-film device.
It can be seen that the structural layout gradually becomes well defined as the voids around the central loaded region are enlarged and merged around step 3000.
The central region is connected to the supported outer frame by several solid members, through which the applied load is transmitted to the outer boundary where zero displacement is prescribed.
Subsequent evolution mainly involves minor changes in the shapes of the openings and the widths of the solid connections.
Some small voids away from the central region persist into the later stages, indicating that material redistribution varies spatially.
The responses on a vertical section along an in-plane diagonal of the final design are shown in Fig.~\ref{fig:a3-deformed-fields}.
The central region is deflected downward under the applied pressure, while the strain-gradient norm is relatively high around the center and generally decreases toward the outer regions, as shown in Fig.~\ref{fig:a3-deformed-fields}(a).
In Fig.~\ref{fig:a3-deformed-fields}(b), the electric potential generally increases from bottom to top near the center but decreases in parts of the outer regions. This contrast contributes to the potential difference between the inner and outer electrode groups.
The corresponding electric field component $E_z$ is negative in the central region and positive in parts of the outer regions, as illustrated in Fig.~\ref{fig:a3-deformed-fields}(c).
Fig.~\ref{fig:a3-phase-binary-3d} shows the solid connections between the central loaded region and the outer frame, through which the applied load is transmitted to the outer boundary.
Hence, bending deformation in these connections depends in part on their shapes and widths.
These observations motivate further investigation of the joint optimization of structural geometry and electrode coverage.
\par
\endgroup

\FloatBarrier

\begin{samepage}
\section{Conclusions}\label{sec:conclusions}

A phase-field framework for constrained topology optimization is developed for the structural design of microscale flexoelectric devices.
The output voltage of the device under prescribed loads and electrical boundary conditions is adopted as the objective for optimizing the material distribution.
The formulation incorporates performance evaluation under manufacturing variations together with structural connectivity constraints.
The same formulation can be extended to include additional performance metrics and manufacturing constraints.
\par
\end{samepage}

The optimized nominal output voltages of the pressure and bending sensors reach approximately 4.8 and 2.5-3.4 times their respective initial values.
The voltage gains are decomposed into contributions from mechanically induced source charge and capacitance changes.
For the pressure sensor, the source charge increases to approximately 3.57 times its initial value, while capacitance reduction provides an additional multiplicative factor of approximately 1.33, yielding an overall voltage gain of about 4.75.
For the bending sensor, the equivalent source charge gains are approximately 2.07 and 3.40 in configurations E1 and E2, respectively. Capacitance changes provide a smaller additional contribution in E1 and have little effect on the voltage gain in E2.
Furthermore, the optimized geometry is influenced by the prescribed electrode pattern. Starting from the same material distribution and matched initial nominal voltages, the multiple-electrode configuration achieves a final nominal voltage 38.8\% higher than that of the single-electrode configuration in the two-dimensional bending example.
For the representative pressure-sensor design, structural connectivity is preserved under the prescribed erosion, and the output voltage remains close to that of the nominal design.
Finally, the applicability of the framework to three-dimensional device design is demonstrated by the thin-film example.
Prescribed electrode layouts are used throughout the numerical studies, with the additional restriction of a constant material distribution through the thickness in the three-dimensional example.

Potential directions for future work include joint optimization of the material distribution and electrode layout, together with three-dimensional designs in which the material distribution can vary through the thickness.
The effects of electrode position, shape, surface coverage, and electrical connections on the relationship between local deformation and output voltage could then be investigated.
Finite-deformation analysis is also needed to determine whether the voltage gains obtained under the assumption of infinitesimal deformations are retained at higher loads.
These extensions would support the development of microfabricated flexoelectric sensors for integration into microelectromechanical systems.

\section*{CRediT authorship contribution statement}
\textbf{Hao Zhen:} Writing -- original draft, Conceptualization, Methodology, Investigation, Visualization, Validation.
\textbf{Sheng Mao:} Conceptualization, Supervision, Funding acquisition, Writing -- review and editing.

\section*{Declaration of competing interest}
The authors declare that they have no known competing financial interests or personal relationships that could have appeared to influence the work reported in this paper.

\section*{Acknowledgments}

This work is supported by the National Natural Science Foundation of China (Grant No. 12272005). The authors also acknowledge the support from the high-performance computing platform of
Peking University.

\appendix
\titleformat{\section}
  {\normalfont\Large\bfseries}
  {Appendix \thesection.}
  {0.5em}
  {}

\numberwithin{equation}{section}
\section{Boundary value problem}
\label{app:constitutive-boundary}

The isotropic free enthalpy density in Eq.~\eqref{eqn:flexo-isotropic} corresponds to the following material tensors
\begin{equation}
  \begin{gathered}
    \mathbb C_{ijkl} = \Lambda\delta_{ij}\delta_{kl}
    +\mu(\delta_{ik}\delta_{jl}+\delta_{il}\delta_{jk}), \quad
    \mathbb B_{ijklmn} = l_m^2\mathbb C_{ijlm}\delta_{kn}, \quad
    \kappa_{ij} = \kappa\delta_{ij},\\
    \mathbb M_{ijkl} = \kappa l_e^2\delta_{ik}\delta_{jl}, \quad
    \mathbb F_{ijkl} = f_1\delta_{il}\delta_{jk}+f_2(\delta_{ij}\delta_{kl}+\delta_{ik}\delta_{jl}).
  \end{gathered}
\end{equation}
The material properties $\{ \Lambda,\mu,\kappa,f_1,f_2 \}$ depend only on the design variable $c$. The constitutive quantities conjugate to the strain, strain gradient, electric field, and electric field gradient are defined by
\begin{equation}
  \widehat\sigma_{ij}=\frac{\partial\mathcal H}{\partial\varepsilon_{ij}},\quad
  \widetilde\sigma_{ijk}=\frac{\partial\mathcal H}{\partial\varepsilon_{ij,k}},\quad
  \widehat D_i=-\frac{\partial\mathcal H}{\partial E_i},\quad
  \widetilde D_{ij}=-\frac{\partial\mathcal H}{\partial E_{i,j}}.
\end{equation}
The stress and higher-order stress are denoted by $\widehat\sigma_{ij}$ and $\widetilde\sigma_{ijk}$, respectively, and the electric displacement and higher-order electric displacement by $\widehat D_i$ and $\widetilde D_{ij}$. Substitution of the free enthalpy density in Eq.~\eqref{eqn:flexo-isotropic} into these constitutive definitions yields
\begin{equation}
  \begin{aligned}
    \widehat\sigma_{ij}&=\Lambda\varepsilon_{kk}\delta_{ij}+2\mu\varepsilon_{ij}
        +\frac{f_1}{2}E_{k,k}\delta_{ij}
        +\frac{f_2}{2}(E_{i,j}+E_{j,i}),\\
    \widetilde\sigma_{ijk}&=l_m^2\left(\Lambda\varepsilon_{ll,k}\delta_{ij}
       +2\mu\varepsilon_{ij,k}\right)
        -\frac{f_1}{2}E_k\delta_{ij}
        -\frac{f_2}{2}(E_i\delta_{jk}+E_j\delta_{ik}),\\
    \widehat D_i&=\kappa E_i+\frac{f_1}{2}\varepsilon_{jj,i}
        +f_2\varepsilon_{ji,j},\\
    \widetilde D_{ij}&=\kappa l_e^2E_{i,j}
        -\frac{f_1}{2}\varepsilon_{kk}\delta_{ij}
        -f_2\varepsilon_{ij}.
  \end{aligned}
\label{eqn:flexo-constitutive}
\end{equation}
The physical stress and physical electric displacement are then given by
\begin{equation}
  \sigma_{ij}=\widehat\sigma_{ij}-\partial_k\widetilde\sigma_{ijk}, \quad
  D_i=\widehat D_i-\partial_j\widetilde D_{ij}.
\label{eqn:flexo-physical-sigma-D}
\end{equation}

The first variation of the internal free enthalpy is
\begin{equation}
  \delta\Pi_{\mathrm{int}} = 
  \int_{\Omega} \delta\calH(\ufield,\phifield;c) \mathrm d\Omega = 
  \int_{\Omega} (\widehat\sigma_{ij}\delta u_{i,j}
  +\widetilde\sigma_{ijk}\delta u_{i,jk}
  +\widehat D_i\delta\phi_{,i}
  +\widetilde D_{ij}\delta\phi_{,ij})\mathrm d\Omega.
\label{eqn:flexo-internal-energy}
\end{equation}
Let $\Sigma$ denote a smooth boundary patch with outward unit normal $\boldsymbol{n}$. Integration by parts in Eq.~\eqref{eqn:flexo-internal-energy} yields
\begin{equation}
  \begin{aligned}
  \delta\Pi_{\mathrm{int}}=
  &-
  \int_\Omega \left(\partial_j\sigma_{ij}\,\delta u_i + \partial_iD_i\,\delta\phi\right)\,\mathrm d\Omega \\
  &+ 
  \sum_{\Sigma}\int_{\Sigma}
  \left(\sigma_{ij}n_j\delta u_i
       +\widetilde\sigma_{ijk}n_k\delta u_{i,j}
       +D_in_i\delta\phi
       +\widetilde D_{ij}n_j\delta\phi_{,i}\right)\,\mathrm d\Gamma.
  \end{aligned}
\label{eqn:flexo-internal-energy-1}
\end{equation}
The independent boundary traces are the field values and their normal derivatives. With the tangential projection operator $P_{ij}=\delta_{ij}-n_in_j$, the gradients of the variations can be decomposed as
\begin{equation}
  \delta u_{i,j} = n_j\partial_n\delta u_i + P_{jk}\delta u_{i,k}, \quad
  \delta \phi_{,i} = n_i\partial_n\delta \phi + P_{ik}\delta \phi_{,k}.
\end{equation}
Define the conjugate boundary quantities
\begin{equation}
  \begin{aligned}
    t_i&=\sigma_{ij}n_j - P_{\ell a}\partial_a(P_{\ell j}\widetilde\sigma_{ijk}n_k),
    & r_i&=\widetilde\sigma_{ijk}n_jn_k,
    & j_i&=\sum_{\Sigma\supset\Gamma_e}
       \widetilde\sigma_{ijk}n_k^{\Sigma}\nu_j^{\Sigma}, \\
    s_\phi&=D_in_i - P_{\ell a}\partial_a(P_{\ell i}\widetilde D_{ij}n_j),
    & r_\phi&=\widetilde D_{ij}n_in_j,
    & j_\phi&=\sum_{\Sigma\supset\Gamma_e}
       \widetilde D_{ij}n_j^{\Sigma}\nu_i^{\Sigma},
  \end{aligned}
\label{eqn:flexo-conjugate}
\end{equation}
where $\Gamma_e$ denotes the intersection of adjacent smooth boundary patches, corresponding to an edge in three dimensions or a corner in two dimensions. The outward unit conormal $\boldsymbol{\nu}^{\Sigma}$ is tangent to the boundary patch $\Sigma$ and perpendicular to its intersection with the adjacent patch.
Surface integration by parts is applied to the terms containing $\delta u_{i,j}$ and $\delta\phi_{,i}$ in Eq.~\eqref{eqn:flexo-internal-energy-1}. With Eq.~\eqref{eqn:flexo-conjugate}, the first variation of the internal free enthalpy becomes
\begin{equation}
  \begin{aligned}
    \delta\Pi_{\mathrm{int}}
    ={}&-\int_\Omega
    \left(\partial_j\sigma_{ij}\,\delta u_i+\partial_iD_i\,\delta\phi\right)
    \,\mathrm d\Omega\\
    &+\sum_{\Sigma}\int_{\Sigma}
    \left(t_i\delta u_i+r_i\partial_n\delta u_i
          +s_\phi\delta\phi+r_\phi\partial_n\delta\phi\right)
    \,\mathrm d\Gamma\\
    &+\sum_{\Gamma_e}\int_{\Gamma_e}
    \left(j_i\delta u_i+j_\phi\delta\phi\right)\,\mathrm ds.
\end{aligned}
\end{equation}
With the linear load functionals in Eq.~\eqref{eqn:flexo-total-energy} included, stationarity of the total free enthalpy with respect to all admissible variations yields the following governing equations and boundary conditions in the absence of body forces and free volume charge
\begin{equation}
  \begin{aligned}
    & \partial_j\sigma_{ij} = 0, \quad \partial_i D_i = 0, \\
    & u_i = \bar{u}_i, \ \partial_n u_i = \bar{\xi}_i, \
      \phi = \bar{\phi}, \ \partial_n \phi = \bar{\zeta}, \\
    & t_i = \bar{t}_i, \ r_i = \bar{r}_i, \
      s_{\phi} = \bar{s}_{\phi}, \ r_{\phi} = \bar{r}_{\phi}, \\
    & j_i = \bar{j}_i, \ j_{\phi} = \bar{j}_{\phi}.
  \end{aligned}
\label{eqn:flexo-equation-bc}
\end{equation}
For brevity, the boundary portions on which the respective boundary conditions are prescribed are not explicitly labeled. For each conjugate pair $(u_i,t_i)$, $(\partial_n u_i, r_i)$, $(\phi,s_{\phi})$, and $(\partial_n\phi, r_{\phi})$, either the Dirichlet or the Neumann condition is prescribed on a given boundary portion.
Equations~\eqref{eqn:flexo-constitutive}, \eqref{eqn:flexo-physical-sigma-D}, \eqref{eqn:flexo-conjugate}, and \eqref{eqn:flexo-equation-bc} specify the complete boundary value problem.

\section{First- and second-order sensitivity analysis}
\label{app:adjoint-phase-field}

\subsection{First-order sensitivity analysis}
\label{app:1st-sens}

For a given $\omega$, the objective functional is $j_\omega(c)=J(y_\omega,c_\omega)$. Differentiation with respect to the corresponding design realization $c_\omega$ gives
\begin{equation}
  \mathrm Dj(c_\omega)[\delta c_\omega] = \partial_{c_\omega}J(y_\omega,c_\omega) [\delta c_\omega] + \mathrm D_yJ(y_\omega,c_\omega) \bigl[\mathrm Dy(c_\omega)[\delta c_\omega]\bigr].
\label{eqn:first-order-partial-obj}
\end{equation}
Introduce the adjoint variable $p_\omega$ and the Lagrangian functional
\begin{equation}
  \mathcal L_{\omega}(y_\omega,c_\omega,p_\omega) \coloneqq J(y_\omega,c_\omega) + \mathcal S(y_\omega,c_\omega)[p_\omega].
\end{equation}
The stationarity condition with respect to $y_{\omega}$ is
\begin{equation}
  \mathrm D_yJ(y_\omega,c_\omega)[\delta y] + \mathcal T(c_\omega)[\delta y, p_\omega] =0, \quad \forall \delta y\in\mathcal U_0\times\mathcal V_0,
\label{eqn:adjoint-equ}
\end{equation}
where $\mathcal T(c_\omega)\coloneqq \mathrm D_y \mathcal S(y_\omega,c_\omega)$ is the bilinear form for the current design realization.
The state equation is first differentiated in the direction $\delta c_\omega$. The resulting relation and the adjoint equation~\eqref{eqn:adjoint-equ} are then substituted into Eq.~\eqref{eqn:first-order-partial-obj}, yielding the expression in Eq.~\eqref{eqn:first-order-gradient-obj}.

\subsection{Second-order sensitivity analysis}
\label{app:2nd-sens}

Let $\delta c$ denote an arbitrary variation and $\Delta c$ the design perturbation direction for evaluating the Hessian action. For each $\omega\in\mathcal W$, the corresponding increments in the design realization, state, and adjoint variable are denoted by $\Delta c_\omega$, $\Delta y_\omega$, and $\Delta p_\omega$, respectively. All expressions indexed by $\omega$ apply to each realization separately. In particular,
\begin{equation}
  \delta c_\omega = \mathrm D c_\omega(c) [\delta c], \quad
  \Delta c_\omega = \mathrm D c_\omega(c) [\Delta c].
\end{equation}
For a functional $f_\omega(c)=f(c_\omega(c))$ evaluated on a design realization, the second-order chain rule reads
\begin{equation}
  \mathrm D^2 f_\omega(c)[\Delta c,\delta c] = \mathrm D^2 f(c_\omega) [\Delta c_\omega,\delta c_\omega] + \mathrm D f(c_\omega) \bigl[ \mathrm D^2 c_\omega(c)[\Delta c, \delta c] \bigr],
\label{eqn:chain-manu-func}
\end{equation}
where
\begin{equation}
  \begin{aligned}
  \mathrm D c_{\omega}(c)[\delta c]
  &=h_{\beta,\eta_\omega}'(\widetilde c)\,\mathrm D\widetilde c(c)[\delta c],
  \\
  \mathrm D^2c_\omega(c)[\Delta c,\delta c]
  &= h_{\beta,\eta_\omega}''(\widetilde c)\,
   \mathrm D\widetilde c(c)[\Delta c]\,
   \mathrm D\widetilde c(c)[\delta c].
  \end{aligned}
\label{eqn:realization-second-order}
\end{equation}
Here, $h_{\beta,\eta_\omega}'$ and $h_{\beta,\eta_\omega}''$ denote the first and second derivatives of the Heaviside projection function, respectively. For the nominal realization, $h_{\beta,\eta_{\mathrm i}}'\equiv1$ and $h_{\beta,\eta_{\mathrm i}}''\equiv0$.

\textbf{a. Objective functional.} 
For a given design realization $\omega$, the incremental state equation is obtained by differentiating the state equation~\eqref{eqn:flexo-weak-system-unified} in the direction $\Delta c_\omega$
\begin{equation}
  \mathcal T(c_\omega)[\Delta y_\omega,\delta y] = -\mathrm D_{c_\omega}\mathcal S(y_\omega,c_\omega)[\Delta c_\omega,\delta y], \quad \forall \delta y\in\mathcal U_0\times\mathcal V_0.
\label{eqn:increment-state}
\end{equation}
The linear functionals $\ell_u$ and $\ell_\phi$, together with the vector $\boldsymbol{s}_V$ defining the port voltage, are assumed to be independent of the design variable.
For a given $c_\omega$, the state equation is linear in $y_\omega$. Eq.~\eqref{eqn:increment-state} can be written as
\begin{equation}
  \mathcal T(c_\omega)[\Delta y_\omega,\delta y] = - \mathrm D\mathcal T_\omega[\Delta c_\omega;y_\omega,\delta y],
\label{eqn:increment-state-linear}
\end{equation}
where the directional derivative of the bilinear form with respect to $c_\omega$ in the direction $\Delta c_\omega$ is defined by
\begin{equation}
  \mathrm D\mathcal T_\omega[\Delta c_\omega;y_\omega,\delta y] := \left. \frac{\mathrm d}{\mathrm dt} \mathcal T(c_\omega+t\Delta c_\omega)[y_\omega,\delta y] \right|_{t=0}.
\end{equation}
The objective functional is also linear in $y_\omega$. Differentiation of the adjoint equation~\eqref{eqn:adjoint-equ} in the same direction $\Delta c_\omega$ yields the incremental adjoint equation
\begin{equation}
  \mathcal T(c_\omega)[\delta y,\Delta p_\omega] = -\mathrm D\mathcal T_\omega[\Delta c_\omega;\delta y,p_\omega].
\label{eqn:increment-adjoint}
\end{equation}
The increments $\Delta y_\omega$ and $\Delta p_\omega$ associated with $\Delta c_\omega$ are obtained from Eqs.~\eqref{eqn:increment-state-linear} and~\eqref{eqn:increment-adjoint}, respectively. For a given variation $\delta c_\omega$, the second variation is
\begin{equation}
  \mathrm D^2j(c_\omega)[\Delta c_\omega,\delta c_\omega] = \mathrm D\mathcal T_\omega [\delta c_\omega;y_\omega,\Delta p_\omega] + \mathrm D\mathcal T_\omega [\delta c_\omega;\Delta y_\omega,p_\omega] + \mathrm D^2\mathcal T_\omega [\Delta c_\omega,\delta c_\omega;y_\omega,p_\omega].
\label{eqn:objective-hvp}
\end{equation}
The third term in \eqref{eqn:objective-hvp} accounts for the second derivative of the bilinear form with respect to the corresponding design realization. 
Substituting Eqs.~\eqref{eqn:objective-hvp} and~\eqref{eqn:realization-second-order} into the chain rule~\eqref{eqn:chain-manu-func} yields the Hessian-vector product for the objective functional with respect to the original design variable $c$
\begin{equation}
  H_{j,\omega}\Delta c
  =\mathrm D\widetilde c(c)
   \Bigl[\,h_{\beta,\eta_\omega}'(\widetilde c)\,H_j(c_\omega)\Delta c_\omega
    +h_{\beta,\eta_\omega}''(\widetilde c)\,
     \bigl(\mathrm D\widetilde c(c)[\Delta c]\bigr)\,
     \nabla_{c_\omega}j(c_\omega)\Bigr],
\label{eqn:objective-hvp-chain}
\end{equation}
where $\nabla_{c_\omega}j(c_\omega)$ is the sensitivity obtained from Eq.~\eqref{eqn:first-order-gradient-obj}, and $H_j(c_\omega)$ is determined from Eq.~\eqref{eqn:objective-hvp} through
\begin{equation}
  \delta c_\omega H_j(c_\omega)\Delta c_\omega = \mathrm D^2 j(c_\omega) [\Delta c_\omega,\delta c_\omega].
\end{equation}

\textbf{b. Spectral connectivity constraint.} 
For a given $\omega$, let $(\lambda_1,\vartheta)$ denote the eigenpair associated with the smallest eigenvalue. 
Assume that the spectral coefficients are twice continuously differentiable with respect to the design realization and that $\lambda_1$ is a simple eigenvalue.
To remove the arbitrary scaling of the eigenfunction, the following normalization is imposed with the mass weight $M_S(c_\omega)$
\begin{equation}
  \int_{\Omega_d}M_S(c_\omega)\,\vartheta^2\,\mathrm d\Omega=1.
\label{eqn:spec-mass-norm}
\end{equation}
Differentiating the eigenvalue equation~\eqref{eqn:spectral_conn_weak} and the normalization condition~\eqref{eqn:spec-mass-norm} gives the first directional derivative of the smallest eigenvalue with respect to the design realization
\begin{equation}
  \mathrm D\lambda_1(c_\omega)[\delta c_\omega]
  =\int_{\Omega_d}K_S'(c_\omega)\,\delta c_\omega\,|\nabla\vartheta|^2\,\mathrm d\Omega
  -\lambda_1\int_{\Omega_d}M_S'(c_\omega)\,\delta c_\omega\,\vartheta^2\,\mathrm d\Omega.
\label{eqn:spectral-first-order}
\end{equation}
The eigenpair increments $(\Delta\vartheta,\Delta\lambda_1)$ in the direction $\Delta c_\omega$ are required for the second-order sensitivity analysis,
with $\Delta\vartheta=\mathrm{D}\vartheta(c_\omega)[\Delta c_\omega]$ and $\Delta\lambda_1=\mathrm{D}\lambda_1(c_\omega)[\Delta c_\omega]$.
These increments are obtained from the linearization of the eigenvalue problem~\eqref{eqn:spectral_conn_weak} and the normalization condition~\eqref{eqn:spec-mass-norm} in the direction $\Delta c_\omega$.
Find $(\Delta\vartheta,\Delta\lambda_1)\in\mathcal X_S\times\mathbb{R}$ such that, for all $\delta\vartheta\in\mathcal X_S$,
\begin{equation}
  \begin{aligned}
    &\int_{\Omega_d}\Bigl[K_S(c_\omega)\,\nabla(\Delta\vartheta) \cdot\nabla\delta\vartheta
     -\lambda_1M_S(c_\omega)\,\Delta\vartheta\,\delta\vartheta
     -\Delta\lambda_1M_S(c_\omega)\,\vartheta\,\delta\vartheta\Bigr]\,\mathrm d\Omega\\
    =&-\int_{\Omega_d}\Bigl[K_S'(c_\omega)\,\Delta c_\omega\,\nabla\vartheta\cdot\nabla\delta\vartheta
     -\lambda_1M_S'(c_\omega)\,\Delta c_\omega\,\vartheta\,\delta\vartheta\Bigr]\,\mathrm d\Omega,
  \end{aligned}
\label{eqn:spectral-tangent}
\end{equation}
together with the linearized normalization condition
\begin{equation}
  2\int_{\Omega_d}M_S(c_\omega)\,\vartheta\,\Delta\vartheta\,\mathrm d\Omega
  +\int_{\Omega_d}M_S'(c_\omega)\,\Delta c_\omega\,\vartheta^2\,\mathrm d\Omega=0.
\label{eqn:spectral-normalization}
\end{equation}
For a given $\delta c_\omega$, differentiating Eq.~\eqref{eqn:spectral-first-order} in the direction $\Delta c_\omega$ yields the second directional derivative
\begin{equation}
  \begin{aligned}
    \mathrm D^2\lambda_1(c_\omega)[\Delta c_\omega,\delta c_\omega]
    ={}&2\int_{\Omega_d}\Bigl[K_S'(c_\omega)\,\delta c_\omega\,\nabla\vartheta\cdot\nabla(\Delta\vartheta)
      - \lambda_1M_S'(c_\omega)\,\delta c_\omega\,\vartheta\,\Delta\vartheta\Bigr]\,\mathrm d\Omega\\
     +&\int_{\Omega_d} {K_S''(c_\omega)\,\Delta c_\omega\,\delta c_\omega\,|\nabla\vartheta|^2}\,\mathrm d\Omega
     - \Delta\lambda_1\int_{\Omega_d}M_S'(c_\omega)\,\delta c_\omega\,\vartheta^2\,\mathrm d\Omega,
  \end{aligned}
\label{eqn:spectral-second-order}
\end{equation}
in which $M_S''\equiv0$ is used. 
The Hessian-vector product for the spectral connectivity constraint with respect to the original design variable is obtained by applying the chain rule to the realization mapping
\begin{equation}
  H_{\mathrm{s},\omega}\Delta c
  =-\frac{1}{s_{\lambda,\omega}} \mathrm D\widetilde c(c) \Bigl[\,
    h_{\beta,\eta_\omega}'(\widetilde{c})(H_{\lambda_1}(c_\omega)\Delta c_\omega)
    + h_{\beta,\eta_\omega}''(\widetilde c)\,
    \mathrm D\widetilde c(c)[\Delta c]\,
    \nabla_{c_\omega}\lambda_1\Bigr],
\label{eqn:constraint-hvp-chain}
\end{equation}
where $\nabla_{c_\omega}\lambda_1=K_S'(c_\omega)|\nabla\vartheta|^2-\lambda_1M_S'(c_\omega)\vartheta^2$ follows from Eq.~\eqref{eqn:spectral-first-order}.
The operator $H_{\lambda_1}(c_\omega)$ is determined from Eq.~\eqref{eqn:spectral-second-order} through
\begin{equation}
  \delta c_\omega H_{\lambda_1}(c_\omega)\Delta c_\omega = \mathrm D^2 \lambda_1 (c_\omega) [\Delta c_\omega,\delta c_\omega].
\end{equation}

\textbf{c. Phase-field energy.} 
The phase-field energy is defined directly on the original design variable $c$.
For a given $\delta c$, differentiation of its first variation in the direction $\Delta c$ yields
\begin{equation}
  \mathrm D^2\mathcal E(c)[\Delta c,\delta c]
  =\int_{\Omega_d}\Bigl[\Psi_B''(c)\,\Delta c\,\delta c
   +l_c^2\,\nabla\Delta c\cdot\nabla\delta c\Bigr]\,\mathrm d\Omega,
\label{eqn:phase-field-second-order}
\end{equation}
where $\Psi_B''(c)=2-12c+12c^2$. 
Thus, the Hessian-vector product for the total Lagrangian with respect to the design variable is obtained by substituting \eqref{eqn:objective-hvp-chain}, \eqref{eqn:constraint-hvp-chain} and \eqref{eqn:phase-field-second-order} into Eq.~\eqref{eqn:second-order-formula}.

\section{Constrained Newton-Krylov method}
\label{app:iga-newton-pdas}

Let $\mathcal{A}_j$ and $\mathcal{A}_\mathrm{s}$ denote the current active sets of the performance and spectral connectivity constraints, respectively. The residual equation follows from the variational derivative of the incremental Lagrangian~\eqref{eqn:incremental-lagrangian}
\begin{equation}
  \mathcal R(x)[\delta x] \coloneqq
  \mathcal R_c[\delta c] + \mathcal R_{\bar{J}}\delta \bar{J} + 
  \sum_{\omega\in\mathcal A_j}{\mathcal R_{j,\omega}\delta\alpha_{\omega}} + \sum_{\omega\in\mathcal A_{\mathrm{s}}}{\mathcal R_{\mathrm s,\omega}\delta\alpha_{\mathrm s,\omega} }
  + \mathcal R_{\mathrm v}\delta\alpha_{\mathrm v}  = 0, \quad \forall \delta x,
\label{eqn:incremental-residual-branch}
\end{equation}
with residual components
\begin{equation}
  \begin{aligned}
  \mathcal R_c[\delta c] & \coloneqq
  \frac{1}{M_0\Delta\tau} \int_{\Omega_d}
  (c-c^n)\delta c\,\mathrm d\Omega + \mathrm{D}\mathcal L (c, \bar{J}, \boldsymbol{\alpha})[\delta c], \\
  \mathcal R_{\bar{J}} &\coloneqq 1 - \sum_{\omega\in\mathcal{A}_j}{\alpha_\omega},\\
  \mathcal R_{j,\omega} &\coloneqq j_{\omega}(c) - \bar{J}, \qquad \omega\in\mathcal{A}_j, \\
  \mathcal R_{\mathrm{s},\omega} &\coloneqq g_{\mathrm{s},\omega}(c), \qquad \omega\in\mathcal{A}_{\mathrm{s}}, \\
  \mathcal R_{\mathrm v} &\coloneqq g_{\mathrm v}(c).
  \end{aligned}
\label{eqn:incremental-residual-components}
\end{equation}
The box constraints on the design variable are treated by an active-set method~\citep{hintermullerPrimalDualActiveSet2002}.
Once the current active sets are identified, the active variables are fixed at their bounds and the corresponding unknowns are eliminated from the linearized system. Hence, the variations and increments of the box-constraint multipliers $\alpha_c^{\pm}$ 
do not appear explicitly in Eq.~\eqref{eqn:incremental-residual-branch}.

The design-variable block of the linearized system is obtained by differentiating $\mathcal R_c[\delta c]$ in the direction $\Delta c$.
With $\bar{J}$, the multipliers $\boldsymbol{\alpha}$, and the test function $\delta c$ held fixed, the bilinear form is
\begin{equation}
  B_n[\Delta c, \delta c] \coloneqq
  \frac{1}{M_0\Delta\tau}\int_{\Omega_d}{\Delta c\,\delta c\,\mathrm d\Omega}
  + \mathrm{D}^2\mathcal L[\Delta c, \delta c],
\label{eqn:proximal-hessian}
\end{equation}
where $\mathrm{D}^2\mathcal L[\Delta c, \delta c]$ is derived in Appendix~\ref{app:2nd-sens}. Therefore, the total linearization is
\begin{equation}
  \begin{aligned}
  \mathrm{D}\mathcal R(x)[\Delta x, \delta x]
  ={}& B_n[\Delta c, \delta c]
  + \sum_{\omega\in\mathcal A_j}{\Delta\alpha_\omega \mathrm{D}j_\omega(c)[\delta c]}\\
  &+ \sum_{\omega\in\mathcal A_{\mathrm s}}{\Delta\alpha_{\mathrm{s},\omega} \mathrm{D}g_{\mathrm{s},\omega}(c)[\delta c]}
  + \Delta\alpha_{\mathrm v} \mathrm{D}g_{\mathrm v}(c)[\delta c]\\
  &- \bigl(\sum_{\omega\in\mathcal A_j}{\Delta\alpha_\omega}\bigr)\delta\bar{J}
  + \sum_{\omega\in\mathcal A_j}{\bigl(\mathrm{D}j_\omega(c)[\Delta c] - \Delta\bar{J}\bigr)\delta\alpha_\omega}\\
  &+ \sum_{\omega\in\mathcal A_{\mathrm s}}{\mathrm{D}g_{\mathrm{s},\omega}(c)[\Delta c]\,\delta\alpha_{\mathrm{s},\omega}}
  + \mathrm{D}g_{\mathrm v}(c)[\Delta c]\,\delta\alpha_{\mathrm v}.
  \end{aligned}
\label{eqn:consistent-linearization}
\end{equation}

\section*{Data availability}
Data will be made available on request.

\renewcommand{\refname}{References}
\bibliographystyle{latex/elsarticle-num-hyphen}
\bibliography{bib/references}

\end{document}